\documentclass[
superscriptaddress,
reprint,
nofootinbib,
 amsmath,amssymb,
altaffilletter,
aps,
prc,
]{revtex4-2}
\usepackage{layouts}
\usepackage[caption=false]{subfig}
\usepackage{graphicx}
\usepackage{hyperref}
\setpdflinkmargin{5pt}
\usepackage{dcolumn}
\usepackage{siunitx}
\DeclareSIUnit\evpercsq{\eV\per\clight\squared}
\DeclareSIUnit\clight{\text{\ensuremath{c}}}

\usepackage[dvipsnames]{xcolor}
\definecolor{cmntgreen}{rgb}{0.0, 0.8, 0.21}
\definecolor{cmntblue}{rgb}{0.0, 0.2, 0.7}
\definecolor{cmntbrown}{rgb}{0.6, 0.3, 0.2}

\usepackage[draft,inline,nomargin]{fixme}
\fxsetup{theme=color, mode=multiuser}
\FXRegisterAuthor{pts}{1}{\color{red}PTS}
\FXRegisterAuthor{g}{2}{\color{brown}\textbf{Info}}
\FXRegisterAuthor{adz}{3}{\color{pink}ADZ}
\FXRegisterAuthor{jkg}{4}{\color{teal}JKG}
\FXRegisterAuthor{ft}{5}{\color{blue}FT}
\FXRegisterAuthor{abt}{6}{\color{ForestGreen}ABT}
\FXRegisterAuthor{rr}{7}{\color{orange}RR}
\FXRegisterAuthor{ss}{8}{\color{purple}Space-saving comment}

\usepackage{subfiles}

\usepackage{svg}
\usepackage[ISO]{diffcoeff}

\usepackage[abbreviations]{glossaries-extra} 
\glssetcategoryattribute{abbreviation}{nohyper}{true}
\newabbreviation{cres}{CRES}{Cyclotron Radiation Emission Spectroscopy}
\newabbreviation{am}{AM}{amplitude modulation}
\newabbreviation{fm}{FM}{frequency modulation}
\newabbreviation{lw}{LW}{Liénard–Wiechert}
\newabbreviation{em}{EM}{electromagnetic}
\newabbreviation{tf}{TF}{transfer function}
\newabbreviation{wg}{WG}{waveguide}
\newabbreviation{ft}{FT}{Fourier transform}
\newabbreviation{awgn}{AWGN}{additive white Gaussian noise}
\newabbreviation{mle}{MLE}{maximum likelihood estimation}
\newabbreviation{crlb}{CRLB}{Cramér–Rao lower bound}
\newabbreviation{hfss}{HFSS}{High Frequency Structure Simulator}
\newabbreviation{snr}{SNR}{signal-to-noise ratio}
\newabbreviation{fir}{FIR}{finite impulse response}
\newabbreviation{rf}{RF}{radio frequency}

\makeatletter\let\myfsize\f@size\makeatother

\begin{document}

\newcommand{\iu}{\mathrm{i}} 
\newcommand{\euler}{\mathrm{e}} 
\newcommand{\bvec}[1]{\boldsymbol{#1}} 
\newcommand{\bmat}[1]{\boldsymbol{\mathrm{#1}}} 
\newcommand{\uvec}[1]{\boldsymbol{\hat{#1}}} 
\newcommand{\abs}[1]{ \left | #1 \right |} 
\newcommand{\vdot}{ \boldsymbol{\cdot} } 
\newcommand{\cross}{ \boldsymbol{ \times } } 

\newcommand{\pitcht}{\alpha} 
\newcommand{\pitch}{\alpha_0} 
\newcommand{\pitchb}{\alpha_*} 
\newcommand{\parameters}{\bvec{\theta}} 
\newcommand{\starttime}{t_0} 
\newcommand{\tracklength}{\tau} 
\newcommand{\ekin}{E_{\mathrm{kin}}} 
\newcommand{\azimuth}{\phi} 
\newcommand{\radial}{\rho} 
\newcommand{\phasecyc}{\varphi_c} 
\newcommand{\phaseax}{\varphi_{\mathrm{axial}}} 
\newcommand{\slope}{\delta \omega} 
\newcommand{\fcyclotron}{\omega_{c}} 
\newcommand{\faxial}{f_\mathrm{axial}} 
\newcommand{\polar}{\theta} 
\newcommand{\density}{\varrho} 

\newcommand{\ToDo}[2][]{{\color{blue} ToDo[#1]: #2 } }
\newcommand{\Comment}[2][]{{\color{blue} Comment[#1]: #2 } }

\newcommand{\subfigureautorefname}{\figureautorefname}

\bibliographystyle{apsrev4-2}

\title{Antenna Arrays for CRES-based Neutrino Mass Measurement}

\newcommand{\Case}{\affiliation{Department of Physics, Case Western Reserve University, Cleveland, OH 44106, USA}}
\newcommand{\Ghent}{\affiliation{Department of Physics and Astronomy, Ghent University, 9000 Ghent, Belgium}}
\newcommand{\Heidelberg}{\affiliation{Institute for Theoretical Astrophysics, Heidelberg University, 69120 Heidelberg, Germany}}
\newcommand{\Illinois}{\affiliation{Department of Physics, University of Illinois Urbana-Champaign, Urbana, IL 61801, USA}}
\newcommand{\Indiana}{\affiliation{Center for Exploration of Energy and Matter and Department of Physics, Indiana University, Bloomington, IN, 47405, USA}}
\newcommand{\Mainz}{\affiliation{Institute for Physics, Johannes Gutenberg University Mainz, 55128 Mainz, Germany}}
\newcommand{\KIT}{\affiliation{Institute of Astroparticle Physics, Karlsruhe Institute of Technology, 76021 Karlsruhe, Germany}}
\newcommand{\Berkeley}{\affiliation{Nuclear Science Division, Lawrence Berkeley National Laboratory, Berkeley, CA 94720, USA}}
\newcommand{\Livermore}{\affiliation{Lawrence Livermore National Laboratory, Livermore, CA 94550, USA}}
\newcommand{\MIT}{\affiliation{Laboratory for Nuclear Science, Massachusetts Institute of Technology, Cambridge, MA 02139, USA}}
\newcommand{\PNNL}{\affiliation{Pacific Northwest National Laboratory, Richland, WA 99354, USA}}
\newcommand{\PennState}{\affiliation{Department of Physics, Pennsylvania State University, University Park, PA 16802, USA}}
\newcommand{\PennStateARL}{\affiliation{Applied Research Laboratory, Pennsylvania State University, University Park, PA 16802, USA}}
\newcommand{\Pittsburgh}{\affiliation{Department of Physics \& Astronomy, University of Pittsburgh, Pittsburgh, PA 15260, USA}}
\newcommand{\Arlington}{\affiliation{Department of Physics, University of Texas at Arlington, Arlington, TX 76019, USA}}
\newcommand{\Washington}{\affiliation{Center for Experimental Nuclear Physics and Astrophysics and Department of Physics, University of Washington, Seattle, WA 98195, USA}}
\newcommand{\Yale}{\affiliation{Wright Laboratory and Department of Physics, Yale University, New Haven, CT 06520, USA}}

\author{A.~Ashtari~Esfahani} \altaffiliation{Present Address: Department of Physics, Sharif University of Technology, P.O. Box 11155-9161, Tehran, Iran} \Washington
\author{S. Bhagvati} \PennState
\author{S.~B\"oser} \Mainz
\author{M.~J.~Brandsema} \PennStateARL
\author{N.~Buzinsky} \altaffiliation{Present Address: Center for Experimental Nuclear Physics and Astrophysics and Department of Physics, University of Washington, Seattle, WA 98195, USA} \MIT
\author{R.~Cabral} \Indiana
\author{C.~Claessens} \Washington
\author{L.~de~Viveiros} \PennState
\author{A.~El~Boustani} \Mainz
\author{M.~G.~Elliott} \Arlington
\author{M.~Fertl} \Mainz
\author{J.~A.~Formaggio} \MIT
\author{B.~T.~Foust} \PNNL
\author{J.~K.~Gaison} \PNNL
\author{M.~Gödel} \Mainz
\author{M.~Grando} \PNNL
\author{P.~Harmston} \Illinois
\author{J.~Hartse} \Washington
\author{K.~M.~Heeger} \Yale
\author{X.~Huyan} \altaffiliation{Present Address: LeoLabs, Menlo Park, CA 94025, USA} \PNNL
\author{A.~M.~Jones} \altaffiliation{Present Address: Ozen Engineering, Sunnyvale, CA 94085, USA} \PNNL
\author{B.~J.~P.~Jones} \Arlington
\author{E.~Karim} \Pittsburgh
\author{K.~Kazkaz} \Livermore
\author{P.~T.~Kolbeck} \Washington
\author{B.~H.~LaRoque} \PNNL
\author{M.~Li} \MIT
\author{A.~Lindman} \Mainz
\author{C.-Y.~Liu} \Illinois
\author{E.~Machado} \Washington
\author{C.~Matth\'e} \Mainz
\author{R.~Mohiuddin} \Case
\author{B.~Monreal} \Case
\author{B. Mucogllava} \Mainz
\author{R.~Mueller} \PennState
\author{A.~Negi} \Arlington
\author{J.~A.~Nikkel} \altaffiliation{Present Address: Bamfield Marine Sciences Centre, Bamfield, British Columbia, Canada} \Yale
\author{E.~Novitski} \Washington
\author{N.~S.~Oblath} \PNNL
\author{M.~Oueslati} \Indiana
\author{J.~I.~Pe\~na} \MIT
\author{W.~Pettus} \Indiana
\author{V.~S.~Ranatunga} \Case
\author{R.~Reimann} \Mainz
\author{R.~G.~H.~Robertson} \Washington
\author{D.~Rosa~De~Jes\'us} \PNNL
\author{L.~Salda\~na} \Yale
\author{V.~Sharma} \Pittsburgh
\author{P.~L.~Slocum} \Yale
\author{F.~Spanier} \Heidelberg
\author{J.~Stachurska} \Ghent
\author{Y.-H.~Sun} \Washington
\author{P.~T.~Surukuchi} \Pittsburgh
\author{J.~R.~Tedeschi} \PNNL
\author{A.~B.~Telles} \email{Corresponding author: arina.telles@yale.edu} \Yale
\author{F.~Thomas} \Mainz
\author{M.~Thomas} \altaffiliation{Present Address: Booz Allen Hamilton, San Antonio, TX, 78226, USA} \PNNL
\author{L.~A.~Thorne} \Mainz
\author{T.~Th\"ummler} \KIT
\author{L.~Tvrznikova} \altaffiliation{Present Address: Waymo, Mountain View, CA 94043, USA} \Livermore
\author{W.~Van~De~Pontseele} \MIT
\author{B.~A.~VanDevender} \Washington \PNNL
\author{T.~E.~Weiss} \Yale
\author{T.~Wendler} \altaffiliation{Present Address: Pacific Northwest National Laboratory, Richland, WA 99354, USA} \PennState
\author{M.~Wynne}\Washington
\author{K.~Young} \Illinois
\author{E.~Zayas} \MIT
\author{A.~Ziegler} \altaffiliation{Present Address: HRL Laboratories LLC, Malibu, CA 90265, USA} \PennState

\collaboration{Project 8 Collaboration}
\date{\today}

\begin{abstract}
\gls{cres} is a technique for precision measurements of kinetic energies of charged particles, pioneered by the Project 8 experiment to measure the neutrino mass using the tritium endpoint method.
It was recently employed for the first time to measure the molecular tritium spectrum and place a limit on the neutrino mass using a \si{\centi \meter \cubed}-scale detector.
Future direct neutrino mass experiments are developing the technique to overcome the systematic and statistical limitations of current detectors.
This paper describes one such approach, namely the use of antenna arrays for \gls{cres} in free space.
Phenomenology, detector design, simulation, and performance estimates are discussed, culminating with an example design with a projected sensitivity of $m_{\beta} < $ \SI{0.04}{\evpercsq}.
Prototype antenna array measurements are also shown for a demonstrator-scale setup as a benchmark for the simulation.
By consolidating these results, this paper serves as a comprehensive reference for the development and performance of antenna arrays for \gls{cres}.
\end{abstract}
\newpage

\maketitle


\clearpage

\section{Introduction}
\label{sec:intro}
Antennas are ubiquitous in daily life and radio frequency techniques have been honed by humanity for well over a century.
In this paper, we describe the design and projected performance of a detector that employs this widely used technology for the purpose of fundamental physics---namely, the measurement of the neutrino mass.
The approach explored here uses the tritium endpoint method and a technique called \glsxtrfull{cres} with a new detection strategy consisting of antenna arrays.

\subsection{Neutrino Mass Measurement with CRES}
\label{sec:intro:CRES}
The current best limit on neutrino mass ($m_{\beta} < $ \SI{0.45}{\evpercsq}) has been reached by the KATRIN experiment using a MAC-E filter~\cite{Katrin:2024tvg, KATRIN:2022ayy}.
In hopes of surpassing KATRIN's ultimate projected upper limit of $m_{\beta} < $ \SI{0.3}{\evpercsq}, Project 8 has developed \gls{cres}, a technique designed for precise energy measurements of charged particles~\cite{originalcres}, with the ultimate goal of a mass measurement or exclusion of the inverted hierarchy at $m_{\beta} < $ \SI{0.04}{\evpercsq}~\cite{Project8:2017nal}.
In \gls{cres}, electrons from nuclear decays are emitted into a uniform magnetic field, causing them to undergo cyclotron motion and radiate with a frequency related to their energy.
For magnetic fields that can be reasonably achieved (\SIrange{0.01}{10}{\tesla}), that frequency falls within the same range used in the well-equipped fields of telecommunications, internet, and radar ($\sim$~\SI{250}{\mega \hertz}--\SI{250}{\giga \hertz}).
The cyclotron radiation can be collected with conventional \gls{rf} devices such as waveguides, antennas, or resonant cavities.
High precision in frequency translates to high precision in energy, which is the motivation for using this technique for increasing neutrino mass sensitivity.

Precise frequency measurements require the ability to observe the signal for a sufficiently long time, so one more element is necessary for \gls{cres} to succeed: a magnetic trap.
Without any confinement, the helical trajectories of the electrons would quickly propel them along the field lines and require observation over unreasonably long distances. 
A purely magnetic trap causes the electrons to be reflected at magnetic barriers and confined to a manageable volume without changing their energies. 
This oscillatory motion also introduces modulation into the radiation emitted by the electron. 
Simulating and analyzing the resulting complicated signal structure is challenging.
Nonetheless, \gls{cres} has been demonstrated by the Project 8 collaboration in a waveguide in two experiments.
The first proof-of-concept experimental phase measured the conversion lines of $^{83\mathrm{m}}$Kr in a WR-42 waveguide \cite{phaseIpaper} and the next phase expanded to study molecular tritium in a circular waveguide \cite{phaseIIprl, phaseIIprc}.
Both were limited to a few cubic millimeters of active volume, significantly limiting their statistical power. 
Ambitions of a future discovery-level neutrino mass measurement with CRES motivate extending the detector volume of Project 8 to the cubic-meter scale.

\subsection{CRES with Antenna Arrays in Free Space}

\begin{figure}
    \centering
    \includegraphics[width=.47\textwidth]{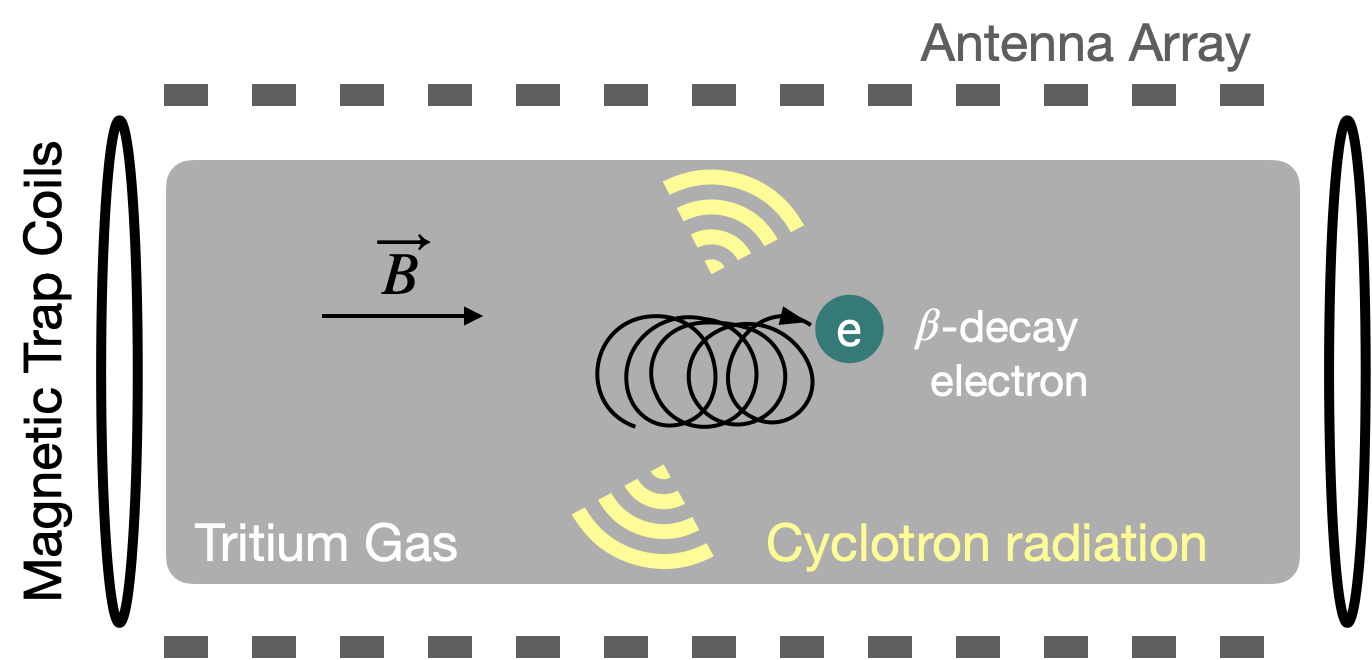}
    \caption{A conceptual sketch of an antenna-based \gls{cres} experiment. The view shown here is a slice along the length of a cylindrical detector. The axial motion is bounded in the horizontal direction by magnetic barriers formed by the current-carrying coils on either side of the tritium volume. The $z$ axis referenced in later sections is parallel to $\bvec{B}$.}
    \label{fig:antennacartoon}
\end{figure}

One natural way to expand the \gls{cres} volume is to move out of a waveguide and into a free-space environment, since waveguide sizes are limited by their operating frequencies.
Collecting the cyclotron radiation with antennas in free space would in principle allow for the experiment to be any size.
A conceptual sketch of \gls{cres} with antennas is shown in \autoref{fig:antennacartoon}.
Attempting \gls{cres} in free space does, however, pose several challenges.
First, the power from a single electron's cyclotron radiation is very small at the magnetic fields and energies relevant for Project~8.
For reference, a tritium endpoint \SI{18.6}{\kilo \electronvolt} electron in a \SI{1}{\tesla} magnetic field radiates $\sim \SI{1}{\femto \watt}$ of total power.
Only a portion of this power can be collected, since antennas cannot provide full solid angle coverage around the electron.
The experimental design becomes a delicate balance of trade-offs, requiring a full simulation to gauge feasibility.
In this paper we address this issue with a conceptual detector design developed using a custom simulation and antenna measurements, for which our performance estimates reach the Project 8 target sensitivity of \SI{40}{\milli \evpercsq}.
The result is still dependent on several idealizations so we present it here as a reference point for future \gls{cres} efforts, rather than a proposed experiment.

The organization of the paper is as follows: \autoref{sec:pheno} describes the phenomenology of a \gls{cres} electron in free space.
\autoref{sec:design} shows the experimental design to detect the signal of \gls{cres} electrons using antenna arrays.
Sections \ref{sec:sims} and \ref{sec:simvalid} are about the simulation and simulation validation of such an experiment.
\autoref{sec:recon} discusses the detection limits on signal reconstruction. 
\autoref{sec:large_detectors} shows how the conceived detector would perform.
Finally, \autoref{sec:sensitivity} evaluates its sensitivity reach to the absolute neutrino mass scale.

\section{Phenomenology of CRES Electrons in Free Space}

\label{sec:pheno}
First we describe the phenomenology of electrons undergoing cyclotron motion in free space as relevant to a \gls{cres} experiment.
We start with the electron trajectory and then derive a mathematical description for the emitted electric field that drives the antennas.

\subsection{Electron Kinematics in a Magnetic Trap}
\label{sec:electron_motion}
In an external magnetic field $\bvec{B}$ an electron with kinetic energy $\ekin$ is in helical motion with its cyclotron frequency given by
\begin{equation}
    \label{eq:cycltron_frequency}
    \fcyclotron = \frac{e B}{\gamma m_0} = \frac{e B}{m_0 + \ekin/c^2}
\end{equation}
and its gyroradius given by
\begin{equation}
    \label{eq:gyroradius}
    R_{g} = \frac{m_0 \gamma v_{\perp}}{e B} \, ,
\end{equation}
where $B=\abs{\bvec{B}}$ is the magnetic field strength, $m_0$ is the electron rest mass, $e$ is the elementary charge, $c$ is the speed of light, $\gamma$ is the electron’s Lorentz factor, and $v_{\perp}$ is the velocity component in the plane of the cyclotron orbit, perpendicular to the magnetic field $\bvec{B}$. 

As a consequence of the motion along the magnetic field direction, which we refer to as axial motion, $B$ needs a local minimum $B_0$ along the magnetic field direction to create a magnetic trap that increases the observation time~\cite{pheno_paper}.
If the electron's instantaneous pitch angle $\pitcht(t)$ is defined as the angle between electron momentum and the local magnetic field, then $v_{\perp}= v \sin\left(\pitcht(t) \right)$ and $v_{||} = v \cos(\pitcht(t))$. 
For a pitch angle of $\pitcht=90^{\circ}$ the electron has no velocity component parallel to $\bvec{B}$ and the electron motion is restricted to circular motion in a plane perpendicular to $\bvec{B}$. 
Conversely, a pitch angle of $\pitcht=0^{\circ}$ means that no cyclotron motion occurs at all. 
Under the assumptions of adiabatic invariance and rotational symmetry of the magnetic field, $v_{\perp}^2/B$ is a constant of motion \cite{Jackson1998-bb} and therefore the instantaneous pitch angle $\pitcht$ changes during the motion according to
\begin{equation}
\label{eq:adiabatic_invariance}
    \frac{\sin^2(\pitcht)}{B} = \frac{\sin^2(\pitch)}{B_0} \, ,
\end{equation}
where $\pitch$ is the pitch angle at the minimum $B_0$.
Electrons are trapped if they reach $\pitcht=90^{\circ}$ (i.e. $v_{||} = 0$) at some point in the magnetic field. For a field with a maximum $B_{\mathrm{peak}}$ and $\pitch$ constrained to $[0, \frac{\pi}{2}]$, the condition for trapping follows from \autoref{eq:adiabatic_invariance}
\begin{equation}
    \label{eq:trapped}
      \pitch \geq \arcsin\left(\sqrt{\frac{B_0}{B_{\mathrm{peak}}}} \right) \, .
\end{equation}
Setting the trap depth, the difference between $B_0$ and $B_{\mathrm{peak}}$, is thus equivalent to setting a lower bound on the pitch angles of the electrons available in a \gls{cres} experiment.
We assume the total $\bvec{B}$ is composed of a background field $\bvec{B}_{\mathrm{bkg}}$ aligned with $z$, combined with a trapping field on the few percent level of $\bvec{B}_{\mathrm{bkg}}$, both rotationally symmetric around the $z$ axis.
The equation of axial motion is \cite{Jackson1998-bb}
\begin{equation}
    v_{||}^2(z) = v_0^2 - \sin^2(\pitch) v_0^2 \frac{B(z)}{B_0} \, .
\end{equation}
The solution is found by integrating both sides and using \autoref{eq:adiabatic_invariance}
\begin{equation}
    \label{eq:axial_solution}
    t(z) = \frac{\sqrt{B_{\mathrm{max}}}}{v_0} \int_{z_0}^{z} \frac{\dl z'}{\sqrt{B_{\mathrm{max}} - B(z')}} \, .
\end{equation}
Where $z_0$ is the position where $B(z) = B_0$ and $B_{\mathrm{max}}$ is the maximum field experienced by a particular electron ($B_{\mathrm{max}} \leq B_{\mathrm{peak}})$.
This describes a periodic motion with frequency 
\begin{equation}
\label{eq:axial_frequency}
 \omega_a=\frac{\pi}{t(z_{\mathrm{max1}}) - t(z_{\mathrm{max0}})} \, ,   
\end{equation}
where $z_{\mathrm{max0}}$ and $z_{\mathrm{max1}}$ are the two solutions of $B_{\mathrm{max}} - B(z)=0$ to either side of the local minimum.

In addition to the cyclotron motion and the axial motion, the electron experiences slow drift motions due to non-uniformity of the magnetic field. 
A gradient of $B$ in the plane of the cyclotron orbit causes variations of the field experienced by the electron over a single orbit. 
This introduces a drift velocity which is perpendicular to both the magnetic field and its gradient. 
The drift velocity of this \textit{grad-B} motion is given by
\begin{equation}
    \label{eq:grad_b_v}
    \bvec{v}_{\nabla} = \frac{v_{\perp}^2}{2 B^2 \fcyclotron} \bvec{B} \cross \nabla_{\perp} B \, ,
\end{equation}
where $\nabla_{\perp} B$ is the gradient of $B$ in the plane orthogonal to $\bvec{B}$~\cite{Jackson1998-bb}.

If the field lines are curved with a curvature radius $R \gg R_g$ the motion along the curved lines introduces another drift motion with its velocity given by~\cite{Jackson1998-bb}
\begin{equation}
    \label{eq:curvature_drift}
    \bvec{v}_c = \frac{v_{||}^2}{\fcyclotron B^3} \bvec{B} \cross (\bvec{B} \cdot \nabla) \bvec{B} \, .
\end{equation}
This curvature is present in a magnetic trap, due to the required gradient along the symmetry axis, resulting in small radial field components. It can be shown that for a rotationally symmetric magnetic field both drift motions are such that they force the guiding center of the cyclotron motion in a circular motion around the symmetry axis~\cite{florianPhd}.
The electron motion described here is known as the guiding center approximation, which is discussed in detail in~\cite{Vandervoort1960,northrop1963,Jackson1998-bb}.

\subsection{Electromagnetic Fields from CRES Electrons}
\label{sec:em_fields}
Accelerated charges emit electromagnetic radiation which can be described by the Liénard-Wiechert potentials~\cite{Jackson1998-bb}. 
From these potentials it is possible to derive the electric and magnetic fields $\bvec{E}(\bvec{r},t)$ and $\bvec{B}(\bvec{r},t)$ at any time $t$ and position $\bvec{r}=R \, \uvec{n}$ generated by a point charge in arbitrary motion:
\begin{align}
    \label{eq:lw_solution_e}
    \bvec{E}(\bvec{r},t) &= \frac{1}{4 \pi \epsilon_0} \left( \frac{ q \left(\uvec{n} - \bvec{\beta} \right)}{\gamma^2 \left( 1 - \uvec{n} \vdot \bvec{\beta} \right)^3 R^2} \right. \nonumber \\
     &+ \left. \frac{q \uvec{n} \cross \left( \left (\uvec{n} - \bvec{\beta} \right ) \cross \dot{\bvec{\beta}} \right )}{c \left (1 - \uvec{n} \vdot \bvec{\beta} \right )^3 R } \right) \Bigg\rvert_{t_\mathrm{r}} \, ,
\end{align}
\begin{equation}
    \label{eq:lw_solution_b}
    \bvec{B}(\bvec{r},t) = \frac{1}{c} \, \uvec{n}(t_r) \cross \bvec{E}(\bvec{r},t) \, ,
\end{equation}
where $q$ denotes the magnitude of the charge, $\epsilon_0$ the permittivity of free space, $\bvec{\beta} = \frac{\bvec{v}}{c}$ the ratio of velocity to speed of light in vector form, and $\dot{\bvec{\beta}}$ its time derivative. 
\autoref{eq:lw_solution_e} is evaluated at the retarded time $t_r$ to account for the propagation delay between the source and the observer, using the implicit equation:
\begin{equation}
    \label{eq:retarded_time}
    c \abs{t - t_r} = \abs{\bvec{r} - \bvec{r_s}(t_r)} \, ,
\end{equation}
where $\bvec{r}_s$ is the source location. 
In \autoref{eq:lw_solution_e} the first term that only depends on velocity is the static component of the electric field and drops quickly with distance due to the $\frac{1}{R^2}$ dependence. 
The second term, which depends on the acceleration, is the dominating contribution at large distances and the relevant component for radiation. 

\begin{figure}[tbh]
\centering
 \subfloat[Tritium endpoint electron $\beta \approx 0.26$.]{
    \includegraphics[width=0.47\textwidth]{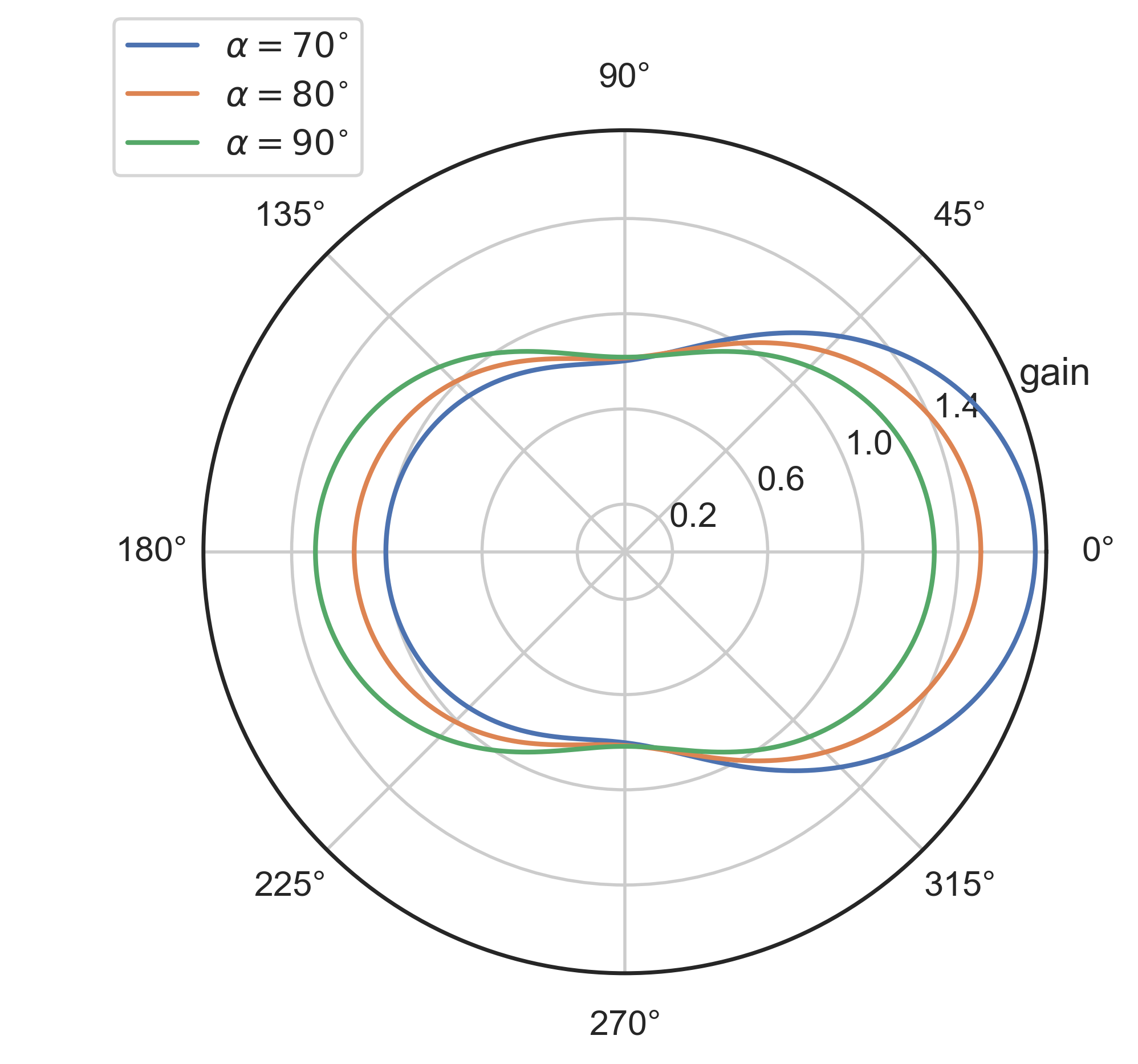}%
   \label{fig:electron_gain_endpoint}%
 }\\
 \centering
 \subfloat[High energy electron $\beta=0.75$.]{
    \includegraphics[width=0.47\textwidth]{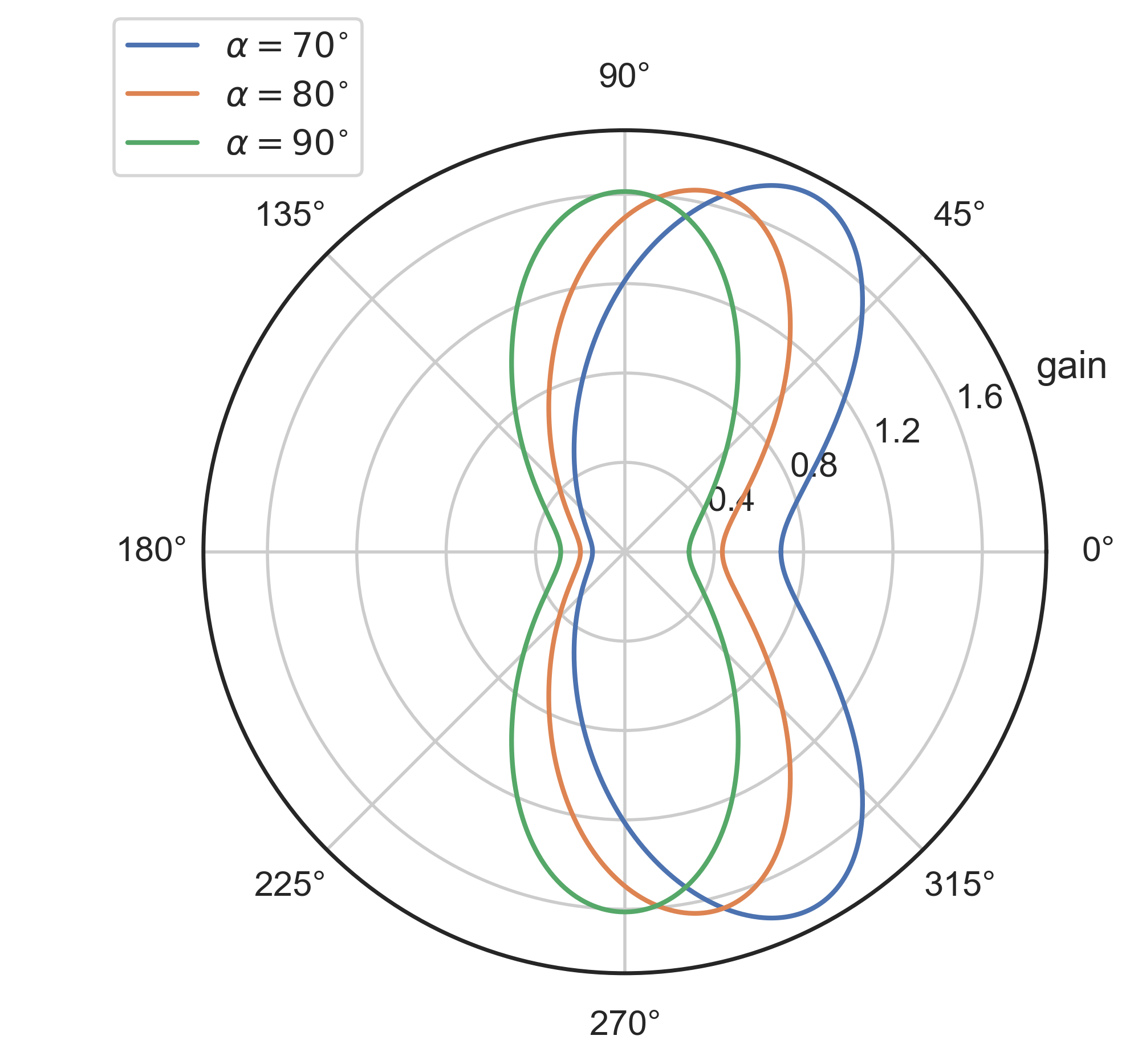}%
   \label{fig:electron_gain_high}%
 }
 \caption{Angular distribution of radiated power of electrons undergoing cyclotron motion. Power normalized to total radiated power such that it is represented as linear antenna gain. Plots depict the gain as it depends on polar angle $\polar$ between observer and magnetic field while it is symmetric for the azimuth. Distributions change shape depending on $\beta$ and $\pitcht$. Adapted from \cite{florianPhd}.}
 \label{fig:electron_gain}
\end{figure}
The total radiated power of the charge can be calculated with the relativistic Larmor formula as
\begin{equation}
    \label{eq:larmor}
    P_{\mathrm{Larmor}} = \frac{1}{4 \pi \epsilon_0} \frac{2 q^2 \omega_0^2}{3 c} \frac{\beta^2 \sin^2(\pitcht)}{1 - \beta^2} \, ,
\end{equation}
where $\omega_0$ is the non-relativistic cyclotron frequency. 
The charge radiates this power non-isotropically and the angular distribution of the radiated power $\diff{P}{\Omega}(\uvec{n})$ is given in ~\cite{Johner87}. 
It depends on the angle $\polar$ between $\uvec{n}$ and $\bvec{B}$.
The dependence on $\polar$ changes shape significantly with energy $\ekin$ and $\pitcht$. 
In the case of tritium beta decay electrons close to the endpoint with $\beta\approx 0.26$, the radiated power has a slight preference for directions parallel to $\bvec{B}$ as opposed to highly relativistic cases where the radiated power has a strong preference for directions orthogonal to $\bvec{B}$ (see \autoref{fig:electron_gain}). 
For pitch angles $\pitcht<90^{\circ}$ the distribution increases in the direction of axial motion.

\begin{figure}[tbh]
\centering
 \subfloat[Tritium endpoint electron $\beta \approx 0.26$.]{%
   \includegraphics[width=0.47\textwidth]{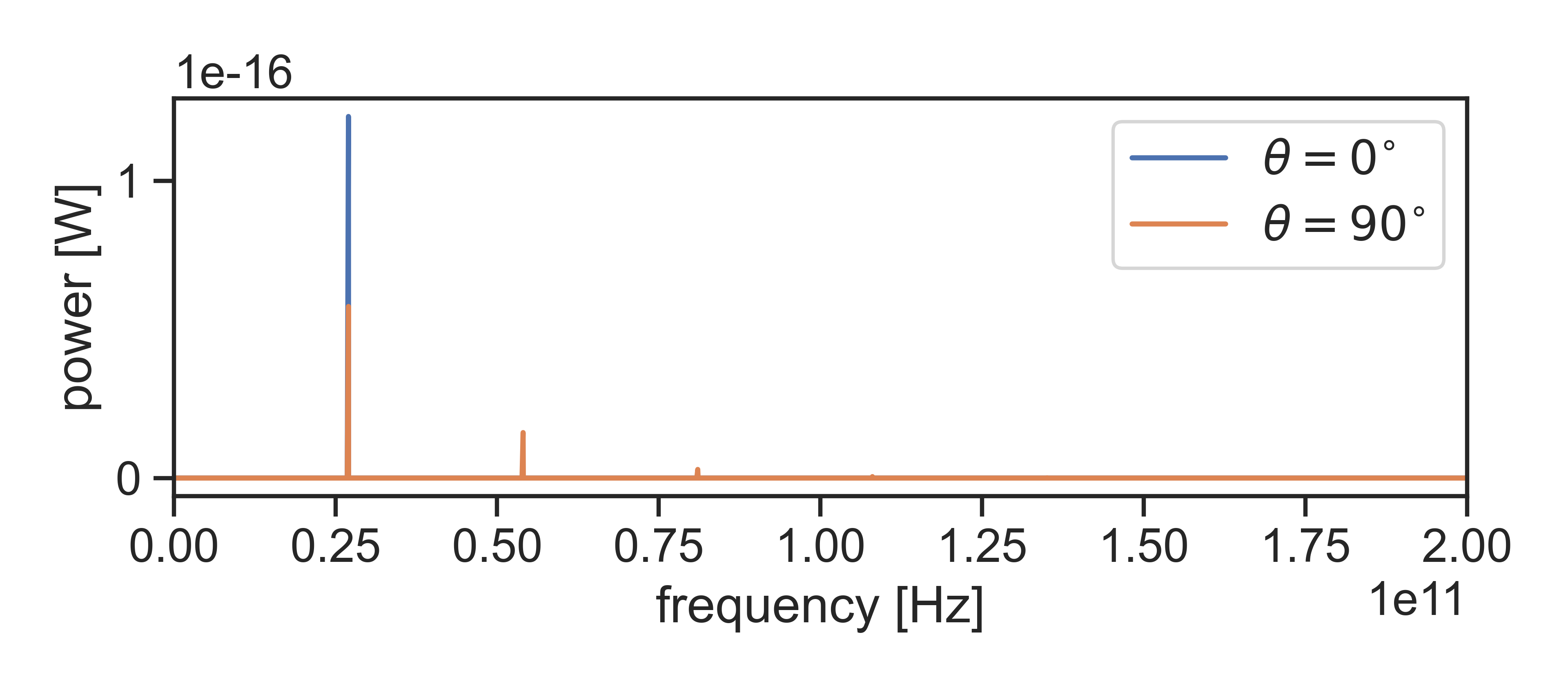}%
   \label{fig:electron_spectrum_endpoint}%
 }\\
 \centering
 \subfloat[High energy electron $\beta \approx 0.75$.]{%
   \includegraphics[width=0.47\textwidth]{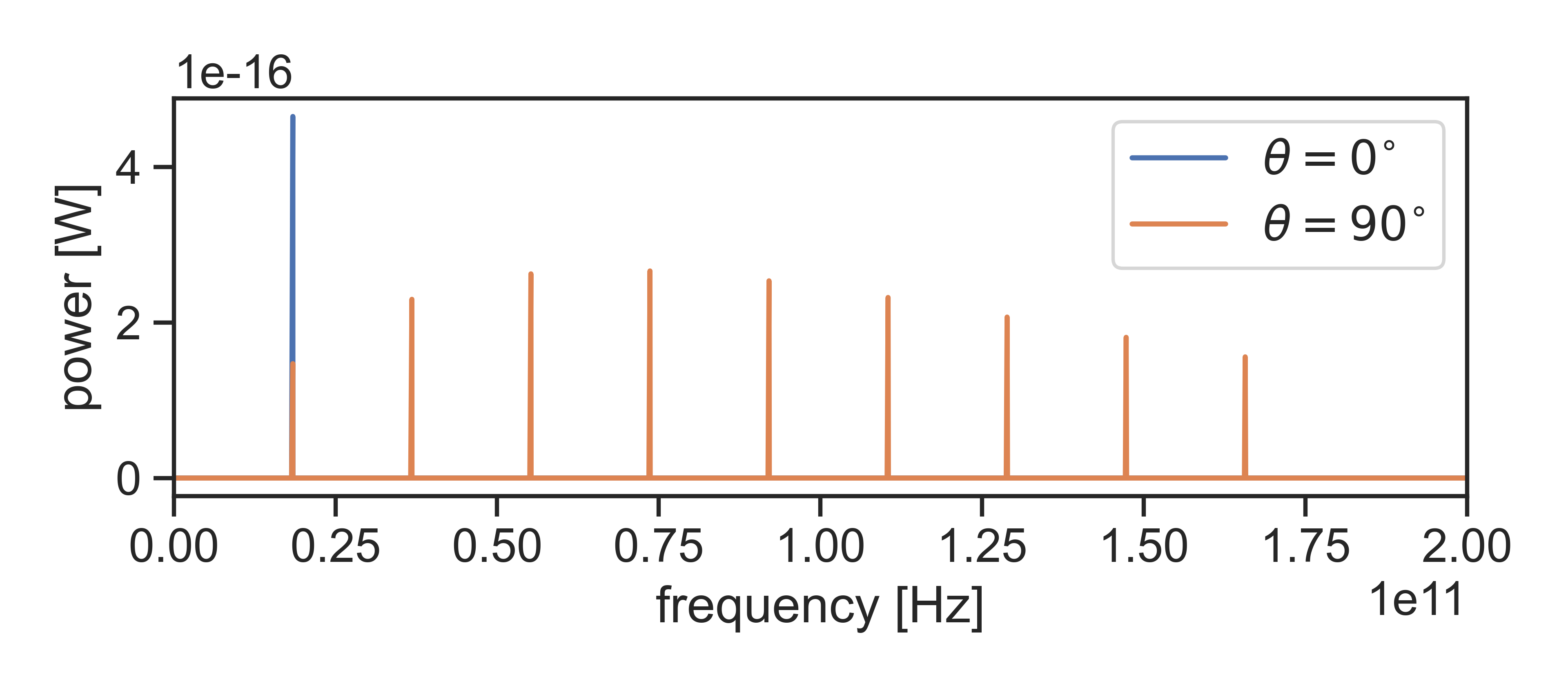}%
   \label{fig:electron_spectrum_high}%
 }
 \caption{Power spectra of electrons in \qty{1}{\tesla} magnetic field with $\pitcht=\qty{90}{\degree}$.
 Adapted from \cite{florianPhd}.}
 \label{fig:electron_spectrum}
\end{figure}
The frequency spectrum of the radiated power also depends on $\ekin$ and $\polar$~\cite{Tru58}.
In the non-relativistic case it consists of just a single peak at the cyclotron frequency. 
Going to relativistic energies, radiation contributions in the direction of motion create additional peaks at harmonics of the cyclotron frequency for $\polar>\qty{0}{\degree}$ as seen in \autoref{fig:electron_spectrum}. Thus the spectrum is given by
\begin{equation}
\label{eq:spectral_power_dist}
    \diffp{P}{\Omega, \omega}(\uvec{n}, \omega) = \sum_{n=1}^\infty \diff{P_n}{\Omega}(\uvec{n}) \delta(\omega - n \fcyclotron) \, ,
\end{equation}
where $\diff{P_n}{\Omega}(\uvec{n})$ is the angular power distribution for harmonic $n$ given in~\cite{Tru58}. 
For tritium endpoint electrons (\autoref{fig:electron_spectrum_endpoint}) the first harmonic is the most powerful for all observer angles $\polar$. 
Practical limitations on bandwidth prevent collection of power in higher harmonics. 
This results in an up to $\sim$\qty{25}{\percent} reduction in detectable power, depending on $\polar$.
\autoref{fig:electron_spectrum_high} shows that for higher energy electrons at $\polar=\qty{90}{\degree}$ the power is distributed into many peaks, with the maximum power shifting to higher harmonics.
Nevertheless, this is compensated by a $\sim20$ times higher Larmor power than a tritium endpoint electron.
In addition, if the power is integrated over all $\polar$ in the case of \autoref{fig:electron_spectrum_high}, the first harmonic is still the overall highest power peak.
Designing a \gls{cres} experiment sensitive only to the first harmonic is thus a viable option even at higher energies, though the feasibility changes with $\beta$ and $\bvec{B}$.

While the full expression for \autoref{eq:spectral_power_dist} derived in \cite{Tru58,Johner87} accounts for axial motion, it does not include drift motion. 
This is acceptable because the fraction of kinetic energy in the drift motion is insignificant compared to that in the cyclotron and axial motions~\cite{florianPhd}.

In addition to the power distributions, we also need the explicit vector form of the electric field to account for its phase and polarization. To first order in $\beta$, the radiation component of \autoref{eq:lw_solution_e} is
\begin{equation}
    \label{eq:lw_solution_non_relativistic}
    \bvec{E}(\bvec{r},t) \approx \frac{q}{4 \pi \epsilon_0 c} \left( \frac{1}{R} \, \uvec{n} \times \left( \uvec{n} \times \dot{\bvec{\beta}} \right) \right)\Bigg\rvert_{t_\mathrm{r}} \, .
\end{equation}
%
In a coordinate system where $\uvec{B}$ and the $z$-axis are aligned, the helical motion results in
\begin{equation}
\label{eq:beta_dot_helix}
    \dot{\bvec{\beta}} = \frac{1}{c}
    \begin{pmatrix}
        \fcyclotron^2 R_g \cos(\fcyclotron t + \phasecyc) \\
        -\fcyclotron^2 R_g \sin(\fcyclotron t + \phasecyc) \\
        \dot{v}_{||}
    \end{pmatrix} \, .
\end{equation}
For the relevant pitch angles $\dot{v}_{||}$ can be neglected.
If we transform to (right-handed) spherical coordinates with $\uvec{e}_r$ pointing along $\uvec{n}$ from the electron to the observer and $\theta$ the angle between between $\uvec{B}$ and $\uvec{n}$, we can substitute \autoref{eq:beta_dot_helix} into \autoref{eq:lw_solution_non_relativistic} to yield~\cite{florianPhd}
\begin{align}
    \label{eq:transformed_e}
    \bvec{E}_r(\bvec{r},t) = & \frac{q \fcyclotron^2 R_g}{4 \pi \epsilon_0 R c^2} \Bigl( \sin ( \zeta ) \uvec{e}_{\azimuth} - \cos(\polar) \cos(\zeta) \uvec{e}_{\polar} \Bigr) \Bigg\rvert_{t_\mathrm{r}} \ ,
    \nonumber \\ &
    \mathrm{for \ } \zeta = \fcyclotron t + \phasecyc + \azimuth .
\end{align}
This solution shows that the electric field is on the plane perpendicular to the direction $\uvec{n}$, with a phase shift of $-\pi/2$ between its two components.
This field is restricted to the fundamental frequency $\fcyclotron$ that we aim to detect for \gls{cres} experiments. 
The phases of both components depend on the azimuthal angle $\azimuth$ of the observer, which means that any two observers at the same polar elevation and at the same distance to the electron guiding center position will observe radiation with a phase shift of $\Delta \azimuth$ equal to their azimuthal angular distance. 
With this relation the amplitudes of $\bvec{E}_r$ follow an Archimedean spiral in the original $x$-$y$-plane as shown in~\cite{synca_paper}.
With the amplitude of the $\uvec{e}_{\polar}$-component decreasing with $\cos(\polar)$ for higher polar angles, this vector form represents the general case of elliptical polarization.

In conclusion, we use \autoref{eq:transformed_e} to model the explicit vector equation of the first harmonic of the electric field that drives the antennas. 
Due to the small $\beta$ approximation in \autoref{eq:lw_solution_non_relativistic} we apply relativistic corrections to the amplitude that account for the angular power distribution and power spectrum in \autoref{fig:electron_gain_endpoint} and \autoref{fig:electron_spectrum_endpoint}.
The characteristic frequency content, polarization, phase, and angular power distribution of this field are the primary considerations in the detector design discussed in the next section.

\section{Detector Design for Free-Space CRES Detection}
\label{sec:design} 
\begin{figure*}[htbp]
 \subfloat[A \SI{26}{\giga \hertz} ring of antennas in an MRI magnet with conceptual layout of vacuum and cryogenic system. Inset shows detailed view of slotted waveguide antennas.]{\includegraphics[width=0.48\textwidth]{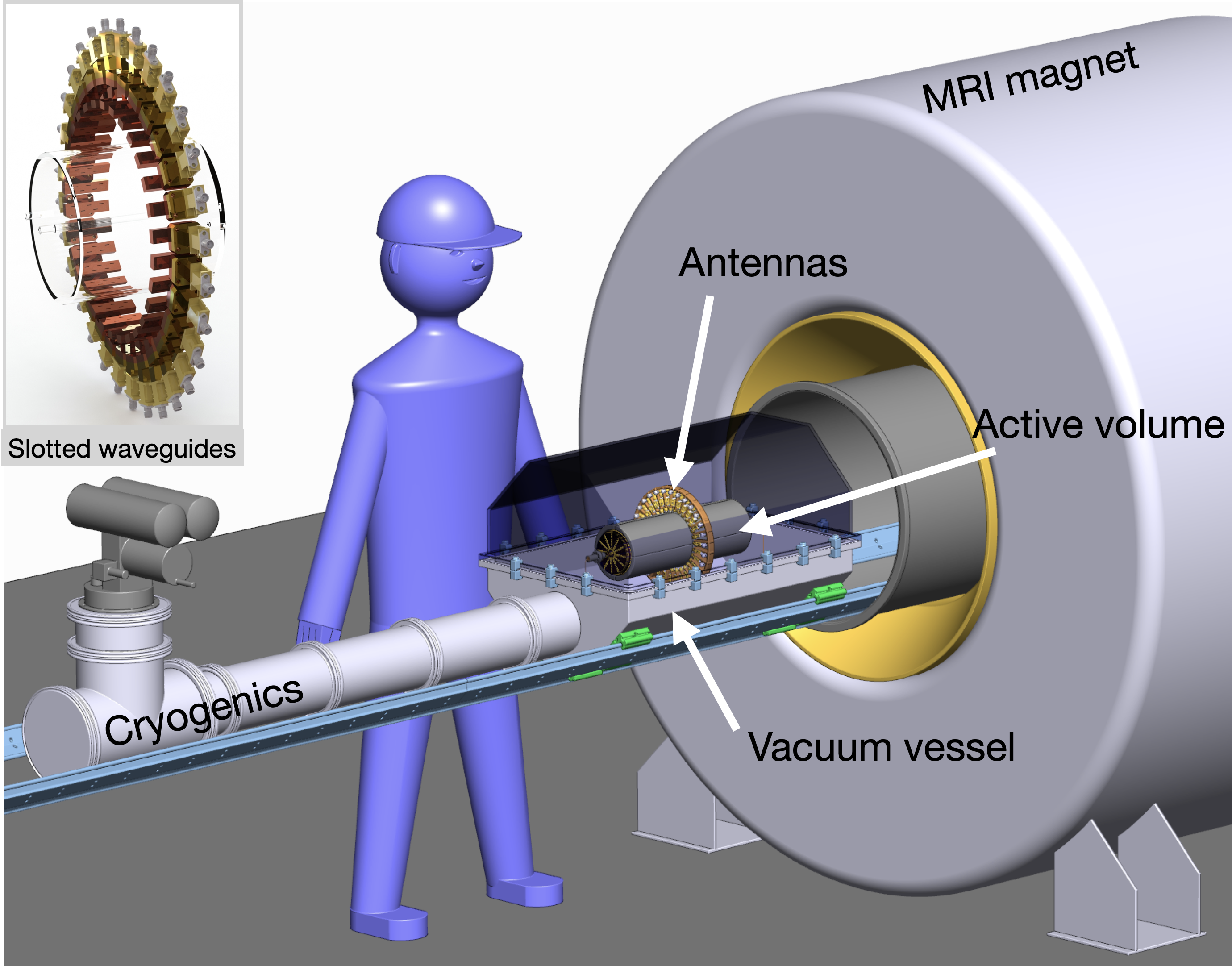}}
 \subfloat[A conceptual sketch of the large \SI{1.3}{\giga \hertz} experiment. Inset shows representative dipole antennas tiling the inner surface of the cylinder. The active volume highlights the acceptable field region of the antennas.]{\includegraphics[width=0.48\textwidth]{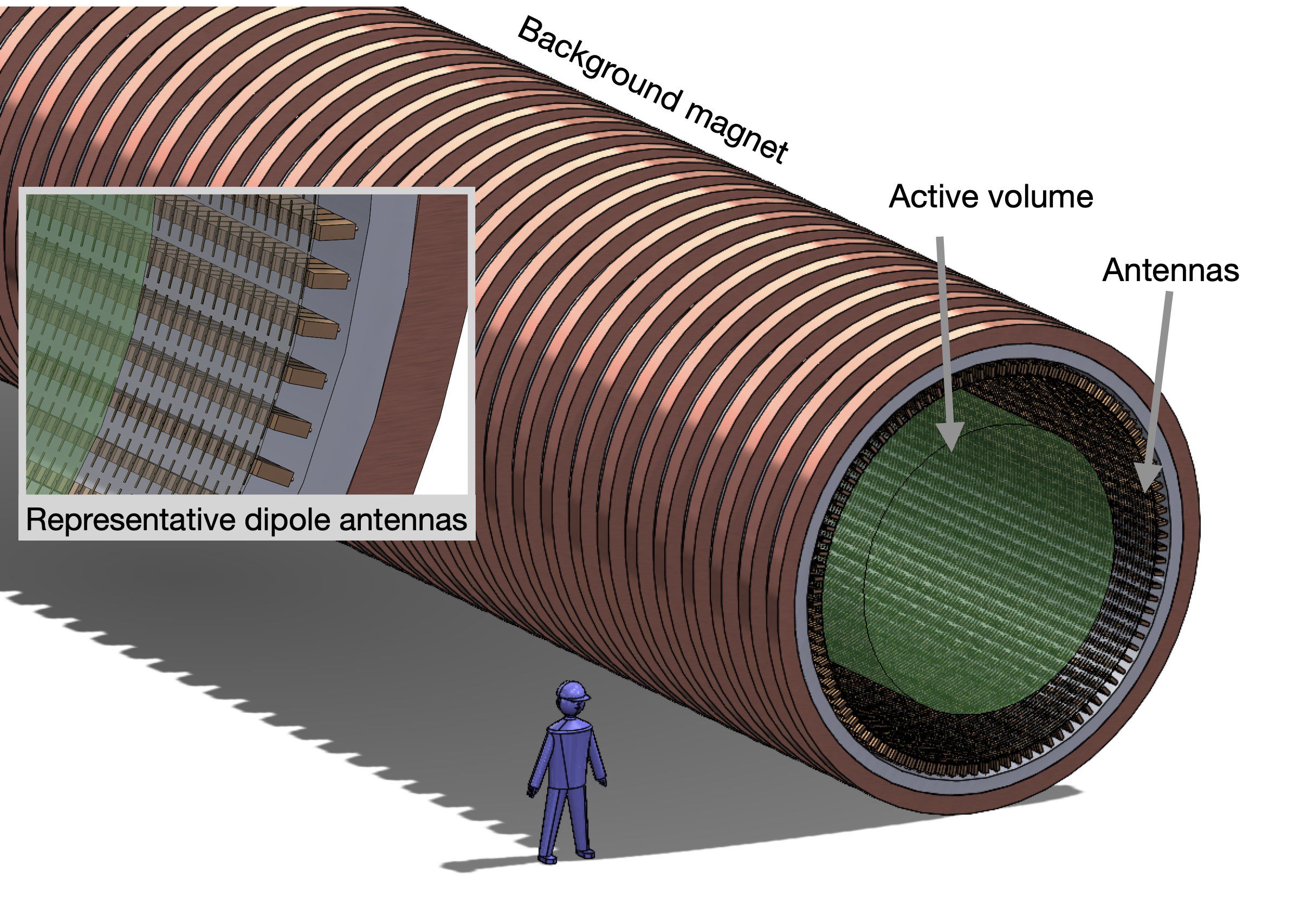}}
 \caption{Two examples of possible antenna-based \gls{cres} designs. Active volume denotes the radioactive source gas visible by the antennas. The purple person is the same size in both figures, to give a sense of scale.}
 \label{fig:fscd}
\end{figure*}
Having chosen the first harmonic as described above, the primary settable parameter for \gls{cres} detector design is the cyclotron frequency.
This frequency is set by the strength of the background magnetic field and is not intrinsically linked to the performance of the detector.
Practical considerations like cost and feasibility of fabrication, rather than physics reasons, set the bounds on magnetic fields that can be reasonably achieved.

Two detectors will be explored here as examples, differing mainly in their magnetic field and size.
One is designed for a medical MRI magnet at \SI{1}{\tesla} and the other is envisioned for a large custom magnet at \SI{0.05}{\tesla}, corresponding to cyclotron frequencies of $\sim$\SI{26}{\giga \hertz} and $\sim$\SI{1.3}{\giga \hertz}, respectively.
The \SI{1}{\tesla} detector is a convenient scale for prototyping, since high frequency antenna arrays are small enough to be tested on a lab bench.
Its active volume would be approximately \SI{0.001}{\meter \cubed} per antenna array ring shown in \autoref{fig:fscd}.
The \SI{0.05}{\tesla} case, with a $\sim$~\SI{250}{\meter \cubed} active volume, is used as the reference design in neutrino mass sensitivity estimates both for its larger size and because lower magnetic fields are important for atomic trapping efficiency in future Project 8 phases~\cite{snowmass}.
Together, these two magnetic field regimes allow us to describe multiple aspects of antenna \gls{cres} detectors.

We broadly conceptualize \gls{cres} designs as a set of nested cylinders.
The trajectory of the electron motivates this geometry, with the cylinder axis aligned with $\bvec{B}_{\mathrm{bkg}}$ field.
The innermost cylinder houses the tritium gas, which is surrounded by the antenna array. 
The cryogenics and vacuum vessel are next, followed by a set of current-carrying coils that generate the magnetic trap, all housed inside the large background field magnet.
Some of the layers may be rearranged, but all of them must be present to perform \gls{cres}.
Here we discuss the two \gls{cres}-specific layers in detail: the antennas and the magnetic trap. 

\subsection{Design of Antenna Arrays for CRES}
\label{sec:antenna_arrays}
The antenna arrangement is dictated by the electron fields and trajectories described in \autoref{sec:pheno}, briefly summarized here.
Following the established coordinate system, the $z$-axis is aligned with the background magnetic field.
An electron born at the origin with a pitch angle of $\pitch=90^{\circ}$ traces a circle in the $x$-$y$-plane, and electrons of lower pitch angles follow helical trajectories up and down the $z$-axis.
Magnetic non-uniformity adds a circular drift motion of the guiding center about the $z$-axis.
Resolving the guiding center position of the electron in the $x$-$y$-plane is important to distinguish between multiple electrons, correct for drift motions, and account for magnetic field variations over detector radius.
The motion along $z$ does not need to be resolved because it is parametrized entirely by the pitch angle, which is encoded in the frequency spectrum (described in \autoref{sec:sims}).
Completely tiling the cylinder with independent antennas is the best solution, but passively combining antennas where possible is desirable to reduce cost and complexity.
For these reasons, independently instrumented antennas must be placed along the circumference of the cylinder to enable digital beamforming~\cite{balanis01} in the $x$-$y$-plane, though they can be passively combined into phased sub-arrays along the $z$-axis to reduce the number of DAQ channels and amplifiers.

The number of elements in the phased sub-arrays must be chosen with caution.
Consider a $1 \times N$ sub-array sitting at the position~$(x, y, z) = (R, 0, 0)$: it is at a radius $R$ from the center, facing inward, with its $N$ elements centered axially along $z$.
The number $N$ is closely tied to the gain pattern and the field regions of the sub-array.
Increasing $N$ will narrow its beam in the $x$-$z$-plane (the H-plane).
An example of a pattern in this plane is the multi-lobed plot in \autoref{fig:five_slot_pattern}.
Larger $N$ will also increase overall antenna size and extend the reactive near-field boundary further from the sub-array. 
Inside this region the fields do not propagate power and are thus not usable for signal detection~\cite{pozar}.
In the next region, the radiative near-field, power does propagate but the wave-fronts incident on a receiver are spherical, which can cause destructive interference between elements. 
Due to the nature of the \gls{cres} detector as an antenna array facing inward toward a dynamic point source, it is impractical to avoid the radiative near-field region entirely. 
$N$ must be small enough such that the sub-array is excited mostly in phase by the radiation coming from the electron, or else the power loss between elements is intolerable.
In practice, the acceptable region boundaries (and thus maximum $N$) are also limited by the overall radius of the experiment.

In their typical uses for communication and radar, phased antenna arrays commonly consist of electric dipole wire antennas, microstrip patch antennas, or slotted waveguides~\cite{pozar}.
Wire antennas are simple to fabricate and have broad spatial coverage, though poor radiation efficiency.
Patch antennas are a viable candidate given their low cost and flat physical shape, lending themselves well to a layered structure.
Slotted waveguides are attractive because of their very low ohmic losses.
Here slotted waveguides are used in the \SI{1}{\tesla} (\SI{26}{\giga \hertz}) case for their superior high frequency performance and dipole antennas are chosen for \SI{0.05}{\tesla} (\SI{1.3}{\giga \hertz}) due to their simplicity.
Future designs could consider patch antennas, but they are not studied here.

\begin{figure}[tb]
    \centering
    \subfloat[Photo of one antenna. The tabs are for mounting.]
    {\includegraphics[width=0.8\linewidth]{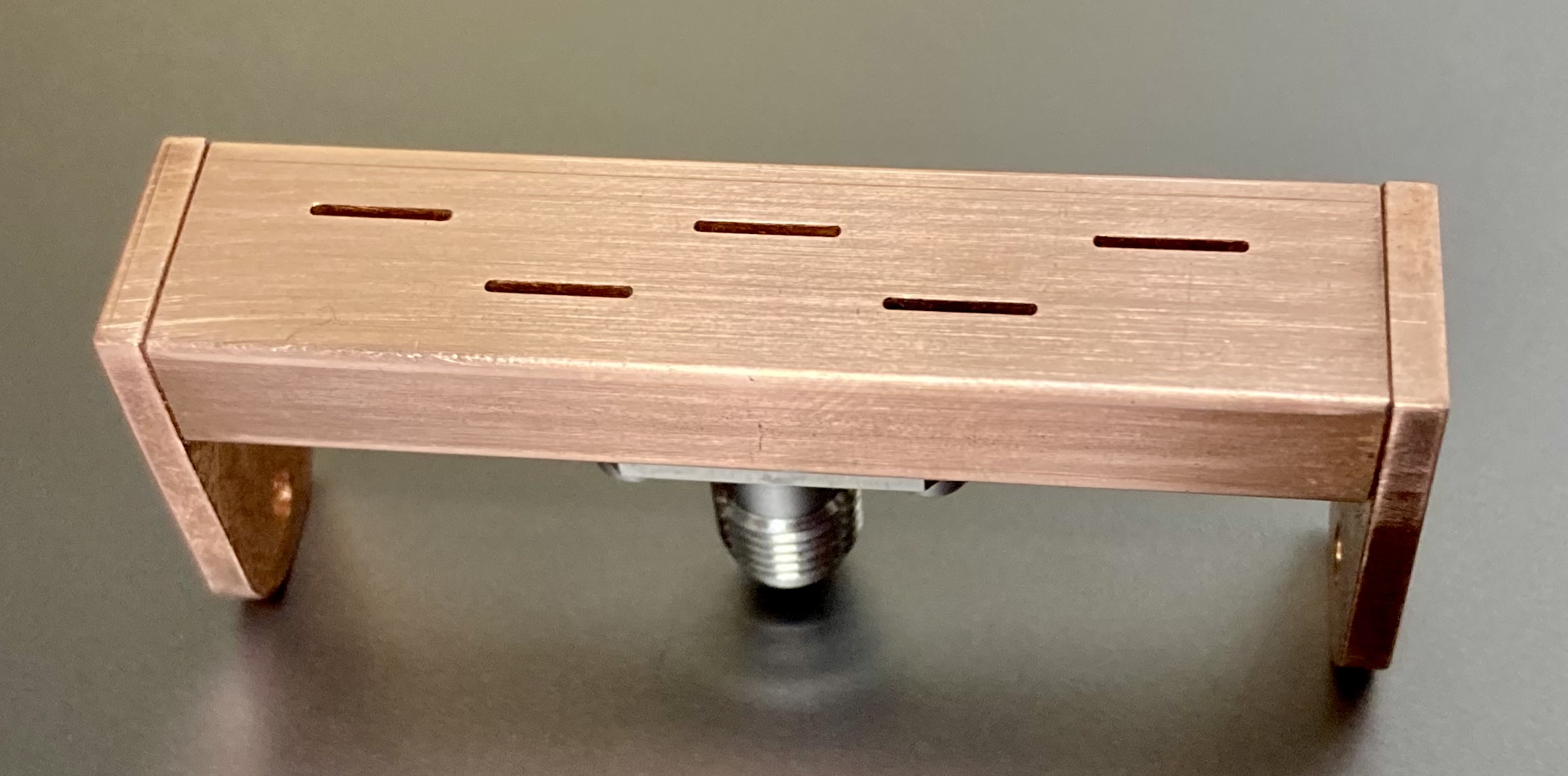}}
    \\
    \subfloat[All dimensions in millimeters. Polarization and normal vectors are shown in blue and green. Section on the right is taken through the center.]
    {\includegraphics[width=0.95\linewidth]{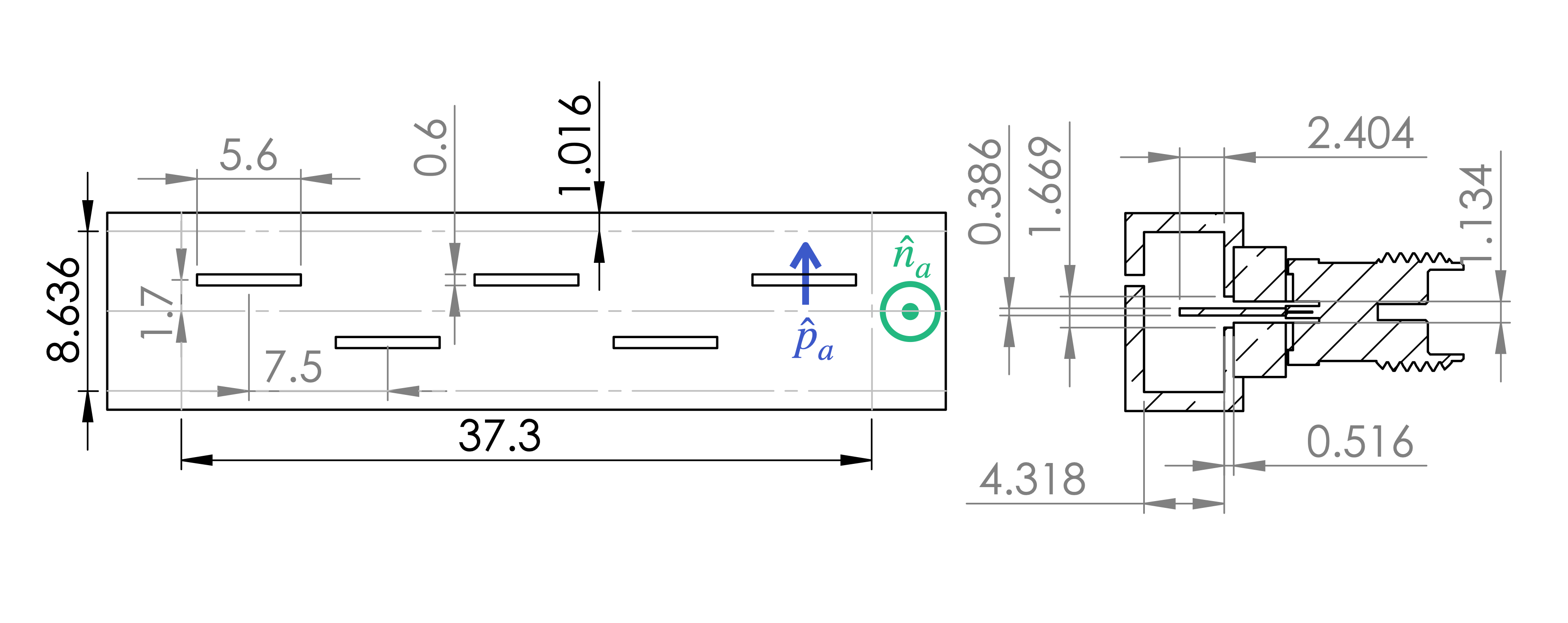}}
    \caption{Center-fed five-slot waveguide antenna designed for the \SI{1}{\tesla} experiment and used for simulation validation in \autoref{sec:simvalid}.}
    \label{fig:fiveslot}
\end{figure}

Optimizations were conducted in Ansys \gls{hfss} to settle on the final \SI{26}{\giga \hertz} antenna design for the constraints of a MRI magnet, resulting in a center-fed, five-slot waveguide antenna.
A photo and technical drawing of a slotted waveguide antenna used for this work is shown in \autoref{fig:fiveslot}.
When an electromagnetic field is incident on the face of the slotted waveguide antenna, a voltage potential is induced across the slots.
This in turn induces a current that flows around the slots along the inside of the waveguide.
The current generates an internal field in the waveguide's fundamental mode, which is picked up by a pin attached to a coax adapter in the back center of the antenna.
The antenna body is made out of copper WR-34 waveguide, the slots are machined, and the coax-to-waveguide adapter is a \SI{2.92}{\milli \meter} field replaceable connector, press-fit with a beryllium-copper pin.
Note that the polarization of the waves that can be received and transmitted by the antenna is perpendicular to the slots themselves.
When placed along the inner surface of the cylinder as described above, this polarization aligns with the azimuthal polarization of the \gls{cres} fields.
The size and spacing of the slots are determined by optimizing for central frequency, bandwidth, and gain pattern ~\cite{elliott_slot_arrays}.
Since we consider a relatively small bandwidth for \gls{cres} signals, the resulting design is fairly resonant, allowing it to have a high gain (\autoref{fig:five_tf}).
The performance of the antenna prototypes are shown in \autoref{fig:individual_antenna_gain}.

In contrast, half-wave electric dipole antennas were chosen for the \SI{1.3}{\giga \hertz} experiment, mostly for their simplicity as a representative antenna.
This is considered sufficient for the scope of the study, since meter-scale \gls{cres} physical prototypes are not yet feasible.
A more in-depth design study using the optimizations in \autoref{sec:sensitivity} would determine the best antenna for the \SI{0.05}{\tesla} case.

One aspect of the antenna arrangement in \autoref{fig:fscd} appears at odds with the \gls{cres} radiation pattern shown in \autoref{fig:electron_gain} -- no antennas are placed at the maxima of the radiated power, i.e. at the end-caps of the cylinder.
Though it would be blind to the $x$-$y$ position of the electron, a circularly polarized antenna placed here would be beneficial for \gls{snr}.
In the final Project~8 experiment, however, the \gls{cres} source will be a beam of tritium atoms aligned with the $z$-axis. 
Therefore, demonstrator detectors keep that area free from instrumentation to remain consistent with the future visions of the experiment.

\subsection{Magnetic Trap Design}
\label{sec:trap}

An electron trap is necessary for sufficiently long observation times and good energy resolution, as discussed in \autoref{sec:intro:CRES}.
Preserving electron kinetic energy from the beta decay requires a purely magnetic bottle trap. 
The trap shape determines the signal structure through the electron trajectory, making it a key aspect of detector design.
Generally, the trap is formed by current-carrying coils of wire placed on either end of the cylindrical detector as indicated in \autoref{fig:antennacartoon}.

Since the trapping volume is a current-free space, the magnetic vector potential is a solution of the Laplace equation and the field can be conveniently expressed as a multipole expansion. 
The expansion coefficients are fully defined by the magnetic field profile along the symmetry axis for a cylindrically symmetric setup.
For \gls{cres} we assume adiabatic motion within the trap, therefore during trap design we require that magnetic field gradients are small, i.e. $\frac{|\nabla \bvec{B}|}{|\bvec{B}|} R_g \ll 1$.
The quantity in \autoref{eq:adiabatic_invariance} is thus conserved and the trapping condition can be given by \autoref{eq:trapped}.

The trap shape has a direct impact on statistics because the trap depth determines the fraction of decay electrons that are confined and could potentially be measured with \gls{cres}.
This fraction is called the trapping efficiency and is given by
\begin{equation}
    \label{eq:trapping_efficiency}
    \epsilon_\mathrm{trap}(\radial,z) = \sqrt{1 - \frac{|\bvec{B}(\radial,z)|}{|\bvec{B}_\mathrm{max}(\radial)|}},
\end{equation}
where $|\bvec{B}_\mathrm{max}(\radial)|$ is the magnetic field maximum along the trajectory of an electron with radial position $\radial$. 
For averaged trapping efficiencies of a few percent, the trap depth  is at the percent level of the background field. 

For the purposes of trap design, it is helpful to consider the ideal trap for antenna-based \gls{cres} in neutrino mass measurements. 
This is a box trap that is perfectly flat within the active volume and has infinitely sharp walls at its boundaries.
Electrons would not experience magnetic field values that differ based on pitch angle and the trajectory-averaged cyclotron frequency would be unique for a given electron energy.
(In contrast, a perfectly harmonic trap profile yields a degeneracy between pitch angle and energy.)
Furthermore in a trap that is perfectly flat radially, the electron signal structure would be independent of radial position and \gls{cres} event reconstruction would be greatly simplified.
However, from Maxwell's equations it follows that the box trap is not physically realizable, and in general the axial gradients which are needed to form a trap imply the existence of radial gradients.
All realistic traps must increase smoothly in the axial direction, which causes some radial and pitch angle dependence of the cyclotron frequency, worsening the energy resolution. 
Therefore, increasing trap depth is favorable for a higher event count, but it must be balanced against its impact on energy resolution.

The following design considerations are used while designing the magnetic trap:
\begin{itemize}
    \item Due to the rotational symmetry of the setup, we generate the magnetic field by circular current-carrying coils.
    \item We keep the trap $\pm z$ symmetric. 
    \item Coils cannot be placed within the tritium gas volume, nor should the coils intersect the field of view of the antennas. Otherwise reflections on the coils distort the radiation observed by the antennas. 
    \item Coil positions and radii are used to set the spatial extent of the trap, coil currents to control the depth, and both to determine the overall shape. 
    \item Because $z$ position is not tracked in \gls{cres}, we require that the trap does not have any side minima in the field profile, so that all electrons traverse the trap symmetrically and do not get trapped locally. 
\end{itemize}
The impact of the magnetic field shape on the trapping efficiency and especially the energy resolution is not straightforward in general. 
The design for any given \gls{cres} prototype requires detailed event simulation, which is discussed in the next section.

\section{Simulation of a Free-Space CRES Experiment}
\label{sec:sims}
To simulate the voltage time series produced by an antenna array in \gls{cres} experiments, we have developed a new simulation package, CRESana~\cite{cresana_archive,florianPhd}.
In the following sections we discuss how the electron motion in the trap, the electric fields, and the antenna response are combined to generate the simulated signal.
We finish by highlighting several of its main spectral features.

\subsection{Simulating Electron Motion}
\label{sec:sims:electron}
The initial momentum and the magnetic field fully determine the electron trajectory as described in \autoref{sec:pheno}.
We restrict the magnetic field $\bvec{B}(\radial, z)$ to be rotationally symmetric around the $z$-axis. 
CRESana allows for generating the field through three options: direct input of a field map, defining a polynomial function $B_z(0, z)$ for the $z$-component of the field along the cylinder axis, or defining an assembly of electromagnetic coils.
In the case of the polynomial function, $\bvec{B}(\radial, z)$ is calculated with the multipole expansion, while the coils are implemented based on analytic field solutions of current loops from \cite{nasaBField}.

The electron trajectory is calculated by solving the axial and drift motions separately, while the actual cyclotron motion is not resolved due to the guiding center approximation. 
The trajectory of axial motion is found by solving the integral in \autoref{eq:axial_solution} along the electron's magnetic field line $(z, \radial(z))$ with $B(z) = \abs{\bvec{B}(\radial(z),z)}$. 
We evaluate the integral numerically for a number of evenly spaced points $z_i \in [z_{\mathrm{max0}}, z_{\mathrm{max1}}]$ yielding $t_i=t(z_i)$. $z_{\mathrm{max}0/1}$ are the two roots of the denominator of the integrand on either side of the minimum $B_0$. 
$B_{\mathrm{max}}$ is calculated using \autoref{eq:adiabatic_invariance} without knowledge of $z_{\mathrm{max}}$. 
Interpolation of the points $(t_i, z_i)$ yields the inverse function $\Bar{z}(t)$, which describes the path $z_{\mathrm{max0}} \rightarrow z_{\mathrm{max1}}$ and is only valid for the first half of the axial period. 
For the full axial trajectory $z(t)$ we first extend this solution to the full axial period by exploiting the symmetry of the motion for the reversed path $z_{\mathrm{max1}} \rightarrow z_{\mathrm{max0}}$ and subsequently the full trajectory length by periodic summation. 
Note that in \autoref{eq:axial_solution} $v_0$ is assumed constant, which is a very good approximation over a single axial period, but for long simulation times the axial frequency slowly decreases due to the electron radiating energy. 
While only a minor effect, we account for it by applying a first order correction to $z(t)$ using the time dependent energy. 

\begin{figure}[tbh]
\centering
   \includegraphics[width=0.5\textwidth]{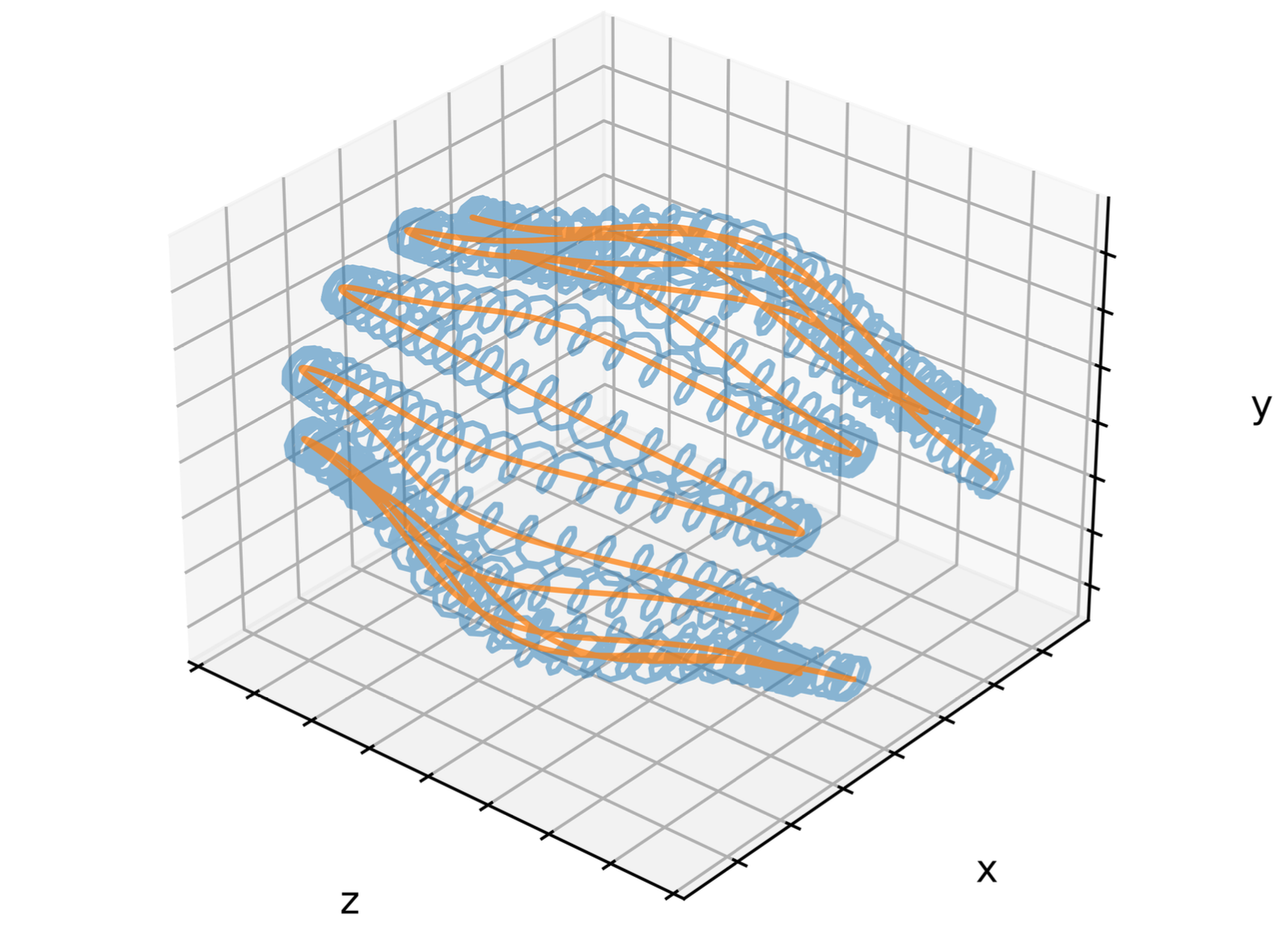}
 \caption{3D electron motion with an orange line representing the guiding center and a blue line indicating the electron’s actual position. An exaggerated synthetic motion is depicted aimed at improving visualization of key features rather than replicating the exact physical simulation. Adapted from~\cite{florianPhd}.}
 \label{fig:electron_motion}
\end{figure}
We calculate the instantaneous drift velocity $v_D(t)$ as the sum of \autoref{eq:grad_b_v} and \autoref{eq:curvature_drift}, which depend on the axial position of the electron. 
We only calculate the absolute value, since it is known that the drift motion forces the guiding center into a circular motion along $\uvec{\azimuth}$ in a rotationally symmetric field~\cite{florianPhd}.
Therefore, we only need the instantaneous drift phase of that circular motion, given as the accumulated angular position found by numerical integration of the instantaneous angular frequencies:
\begin{equation}
\label{eq:azimuth_phase}
    \azimuth(t)=\azimuth_0 + \int_{0}^{t} \frac{v_D(z(t'))}{\rho(z(t'))}dt' \, 
\end{equation}
where $\azimuth_0$ is determined by the electron's initial azimuthal position.
The combined 3-dimensional trajectory of the guiding center is then given by
\begin{equation}
    \bvec{r}_s(t)  =
    \begin{pmatrix}
                \rho(z(t)) \cos(\azimuth(t)) \\
                \rho(z(t)) \sin(\azimuth(t)) \\
                z(t)
    \end{pmatrix} \, .
\end{equation}
Using a radius $\radial(z(t))$ for the circular drift motion, we account for variations in radial position as the electron follows the magnetic field line. This motion is visualized in \autoref{fig:electron_motion}.

The trajectory $\bvec{r}_s(t)$ leads to all the other parameters of interest for \gls{cres}.
We track the magnetic field values $B(t)$ along the trajectory and then determine the instantaneous pitch angle $\pitcht(t)$ using \autoref{eq:adiabatic_invariance}. 
The instantaneous kinetic energy $\ekin(t)$ is the solution of the differential equation $\diff{\ekin}{t}=-P_{\mathrm{Larmor}}(t)$, where the total radiated relativistic power $P_{\mathrm{Larmor}}(t)$ itself depends on $\ekin(t)$ (\autoref{eq:larmor}).
Finally, with $B(t)$ and $\ekin(t)$ we calculate the instantaneous cyclotron frequency $\fcyclotron(t)$ with \autoref{eq:cycltron_frequency}.

\subsection{Electric Field at Antenna}
\label{sec:sims:field}
The electric field at the antennas also follows from the electron motion. 
We obtain the vector form $\bvec{E}_r(t)$ of the field from the approximation of the fundamental frequency in \autoref{eq:transformed_e}. 
It is also useful to characterize the field by an instantaneous power $P_E(t)$ and phase $\varphi_E(t)$. 
The latter needs to take into account that the cyclotron frequency $\omega_c$ is time-dependent due to the Doppler effect and magnetic field variations along the electron trajectory.

\subsubsection{Retarded Time}
Since the electron is in motion, \autoref{eq:transformed_e} is evaluated at the retarded time $t_r$, defined in \autoref{eq:retarded_time} where $\bvec{r}$ is the antenna position and $\bvec{r}_s$ is the electron's position. 
To solve this equation we simulate the electron trajectory at twice the sampling rate and calculate the delay time to all antennas at each trajectory sample. 
The delay time is the time when radiation from that trajectory sample has propagated to an antenna. 
For the $j$-th antenna at position $\bvec{r}_j$, this delay is 
\begin{equation}
    \label{eq:delaytime}
    {t_D}_j(t_r)=\frac{\abs{\bvec{r}_j-\bvec{r}_s(t_r)}}{c} + t_r. 
\end{equation}
By interpolating the results we can evaluate the retarded time at an antenna's sampling time as ${t_r}_j(t)={t_D}_j^{-1}$. 
This approach assumes that for each sample time there is only a single path from the electron trajectory to the antenna and reflection effects are negligible.
This is validated by the measurements described in \autoref{sec:simvalid}.

Once we know the retarded time ${t_r}_j(t)$, we can use the trajectory parameters as described at the end of \autoref{sec:sims:electron} to calculate $\fcyclotron({t_r}_j(t))$, $P_{\mathrm{Larmor}}({t_r}_j(t))$, and the distance vector $\bvec{d}({t_r}_j(t))=\bvec{r}_s({t_r}_j(t)) - \bvec{r}_j$. 

\subsubsection{Power}
For determining the power incident on an antenna, we use the Friis transmission equation~\cite{antenna_engineering_friis}. Assuming unity receiver gain for now (the antenna response is treated in the next section), the power is
\begin{equation}
    \label{eq:friis}
    {P_{E}}_j(t) = \left. P_{\mathrm{Larmor}} \, G_e \, \left( \frac{c}{2 \fcyclotron \abs{\bvec{d}}} \right)^2 \right |_{{t_r}_j(t)} \, ,
\end{equation}
where $G_e$ denotes the electron's ``transmitter gain." 
All symbols are evaluated at the retarded time.
The gain $G_e$ implements the relativistic corrections for the field approximation from \autoref{eq:transformed_e}, which are given by $\diff{P_1}{\Omega}(\uvec{n})$ in \autoref{eq:spectral_power_dist}. 
In practice this means we implement the anisotropy seen in \autoref{fig:electron_gain} and reduce the Larmor power by the fraction lost to the higher harmonics as seen in \autoref{fig:electron_spectrum}. 
Both effects depend on the direction relative to the B-field direction $\uvec{n}=-\frac{\bvec{d}({t_r}_j(t))}{\abs{\bvec{d}({t_r}_j(t))}}$. 

\subsubsection{Phase}
Finally, the instantaneous field phase at each antenna is given as the integral over all past instantaneous field frequencies:
\begin{equation}
    \label{eq:phase_integrated}
    \varphi_j(t) = \int_0^{{t_r}_j(t)} \fcyclotron({t_r}_j(t')) \dl t' + \phasecyc + \phi_a({t_r}_j(t)) \, .
\end{equation}
The Doppler shift of the cyclotron frequency is included through a coordinate transformation to the retarded time~\cite{florianPhd}.
In addition to the frequency integral, \autoref{eq:phase_integrated} includes the initial phase of the cyclotron motion $\phasecyc$ and a phase $\phi_a$ that implements the characteristic Archimedean spiral described in \autoref{sec:em_fields}.

\subsection{Simulation of Antenna Response}
\label{sec:sims:antenna}
The antenna response function converts the electric field into a time-varying voltage, which we model as an arbitrary modulated cosine function
\begin{equation}
    \label{eq:antenna_voltage}
    U_{\mathrm{real}}(t) = A(t) \cos(\varphi(t)) \, ,
\end{equation}
where $A(t)$ and $\varphi(t)$ are the instantaneous amplitude and phase.
Using antenna impedance $Z$, the amplitude is a simple conversion of the instantaneous antenna output power: 
\begin{equation}
    \label{eq:antenna_amp}
    A(t) = \sqrt{2 {P_{\mathrm{out}}}(t) Z}.
\end{equation}
Therefore, the variables of interest are instantaneous power and phase at the antenna output after applying the response function to the incident field's power $P_E$ and phase $\varphi_E$. 

The response function depends on the frequency, the polarization, and the source direction of the radiation incident on the antenna, all of which can be treated separately with their respective effects on the frequency spectrum of the output voltage.
\subsubsection{Polarization Mismatch}
Antennas are only sensitive to radiation polarized in a fixed direction $\uvec{p}_a$, hence the instantaneous electric field $E(t)$ that drives the antenna is the component of $\bvec{E}_r(t)$ that is parallel to $\uvec{p}_a$ with $E(t) = \uvec{p}_a \cdot \bvec{E}_r(t)$. 
For example, the five-slot antenna polarization vector $\uvec{p}_a$ is in the same plane as the slots but oriented orthogonal to them as in \autoref{fig:fiveslot}. 
We implement this polarization mismatch effect with a power loss factor $M_{\mathrm{pol}}$ for the output voltage spectrum. 
For a general radiation source with elliptic polarization the mismatch factor is $M_{\mathrm{pol}}=A_x^2 (\uvec{x} \cdot \uvec{p}_a)^2 + A_y^2 (\uvec{y} \cdot \uvec{p}_a)^2$ where $\uvec{x}$ and $\uvec{y}$ are basis vectors aligned with the axes of the polarization ellipse and $A_x$ and $A_y$ are the amplitudes of the field in that respective direction. 
For an electron source radiating from a direction $\uvec{d}$ we find from \autoref{eq:transformed_e} $\uvec{x}=\uvec{e}_{\azimuth}$, with $A_x=1$ and $\uvec{y}=\uvec{e}_{\polar}$ with $A_y=\cos(\polar)$, where the spherical coordinate system is defined such that $\polar$ is the angle enclosed by $\uvec{e}_r=-\uvec{d}$ and the magnetic field direction $\uvec{B}$.
\subsubsection{Frequency response}
\begin{figure}[tb]
\centering
   \includegraphics[width=0.5\textwidth]{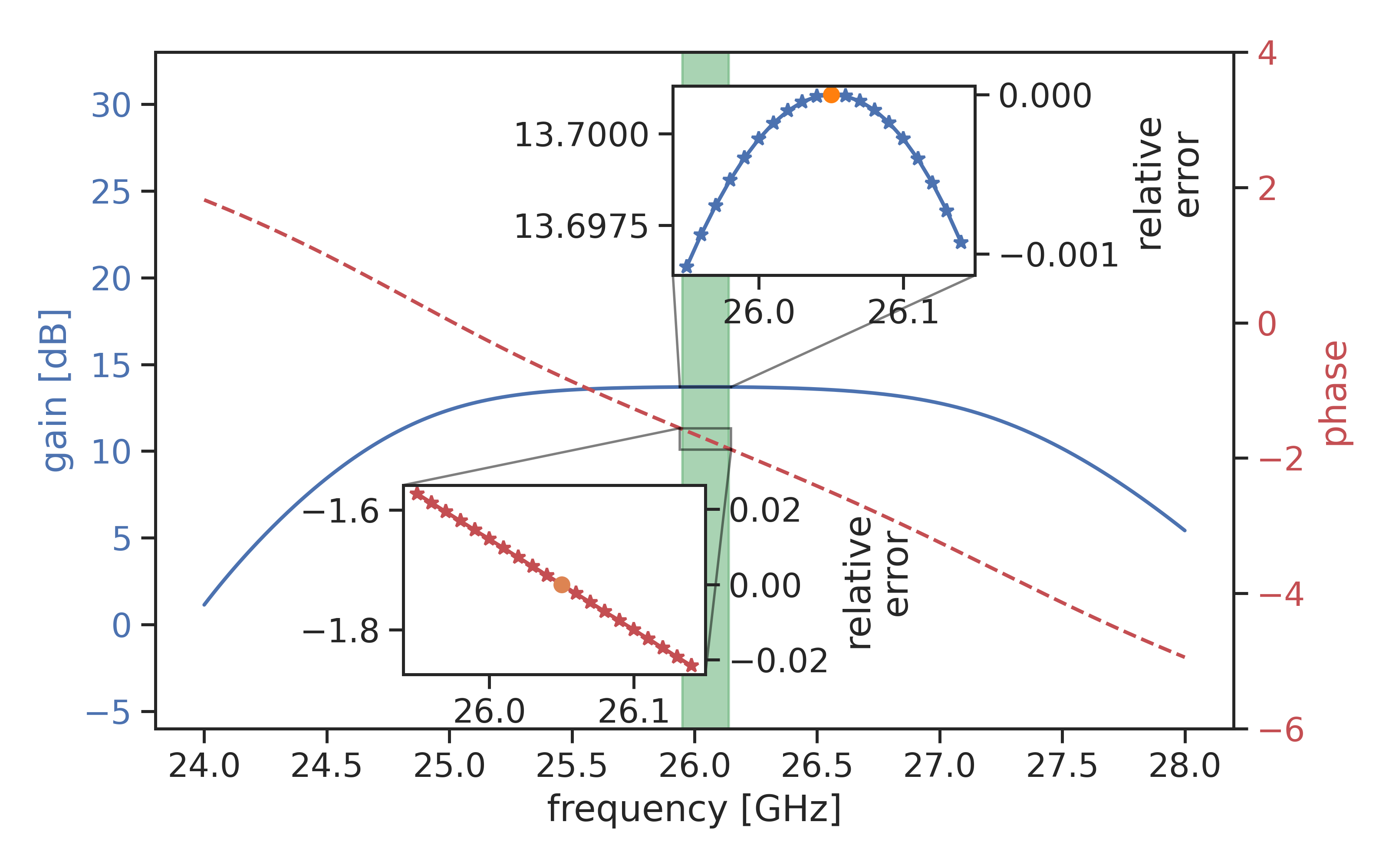}
 \caption{Gain (blue) and phase (red) of the five-slot antenna depending on frequency of the incoming radiation. Near the center of the antenna's frequency band there is a plateau with almost constant gain. The green region marks a \qty{200}{\mega \hertz} band for \gls{cres}. The insets show zooms into that region each with a secondary y-axis showing the respective relative error if only the value at its center (orange) is used. For the phase error this is relative to $2\pi$. Adapted from \cite{florianPhd}.}
 \label{fig:five_tf}
\end{figure}
The frequency response is given by the antenna's transfer function defined as
\begin{equation}
    \label{eq:transfer_function}
    H(\omega) = \frac{U(\omega)}{E(\omega)} \, ,
\end{equation}
which relates the input electric field $E$ to the output voltage $U$ at frequency $\omega$. 
We use \gls{hfss} to obtain the transfer function for use in the \gls{cres} simulation. 
By design, the bandwidth for the antennas under consideration is wider than the narrow bandwidth of interest for \gls{cres} in tritium beta spectroscopy, as can be seen in \autoref{fig:five_tf}. 
We can thus implement the frequency response as a constant gain $G_F=G(\omega_0)$, where $\omega_0$ is the central frequency of the \gls{cres} spectrum, and ignore the effect of the phase.
From the insets in \autoref{fig:five_tf} we observe that this approximation introduces errors of $\lesssim 0.1\%$ in the gain and $\lesssim 2\%$ in the phase.
\subsubsection{Directional response}
\begin{figure*}[!htbp]
   \subfloat[3D render from \gls{hfss} to show the physical orientation of the radiation pattern relative to the antenna.]{\includegraphics[width=0.3\textwidth]{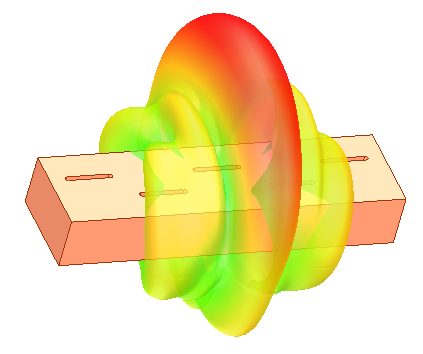}}
   \subfloat[We compare patterns from simulating the antenna in \gls{hfss} to the patterns produced by our integrated \gls{cres} simulation (CRESana) with the approach described in the text. Note that these are perpendicular slices of the 3D plot on the left. Adapted from \cite{florianPhd}.]{\includegraphics[width=0.7\textwidth]{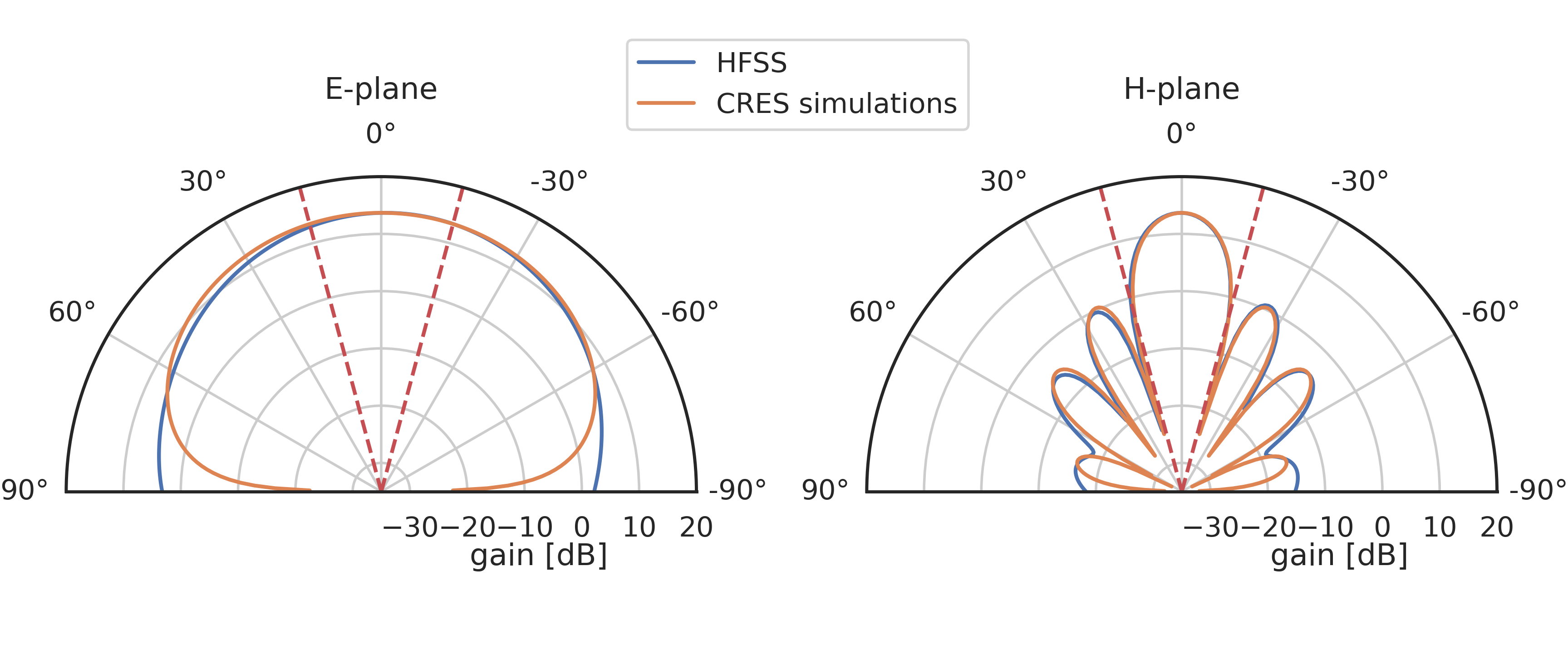}}
   \caption{Simulated gain patterns of the five-slot waveguide antenna in the E and H planes at a frequency of \qty{26}{\giga \hertz}.}
 \label{fig:five_slot_pattern}
\end{figure*}
The antenna's directivity $D(\uvec{d})$ describes its directional response in the form of a power damping factor. 
\autoref{fig:five_slot_pattern} shows how $D(\uvec{d})$ changes the overall antenna gain of the five-slot antenna in the E and H-plane at a frequency of \qty{26}{\giga \hertz}, depending on the angle to the normal.
The blue curves and 3D plot in \autoref{fig:five_slot_pattern} show a simulation of the gain pattern in \gls{hfss}. 
To reproduce this behavior in CRESana, we chose to sample the signals at each slot position individually and then sum those signals incoherently.
For a single slot of the waveguide antenna we approximate $D(\uvec{d})$ by the analytic directivity of an electric half-wave dipole antenna, where the dipole axis is aligned with the slot and the E and H-planes are swapped~\cite{Booker1946}. 
However, in the E-plane our approximation only gives good agreement with \gls{hfss} in the narrow range of $\pm$\qty{15}{\degree}. 
This is due to the fact that \gls{hfss} is a full electromagnetic simulation of the physical antenna with currents induced all over the outside surface, causing radiation in the $\pm \qty{90}{\degree}$ direction as well. 
We found that using the cosine of the incident angle for the directivity of the E-plane increases the range of agreement to $\pm$\qty{60}{\degree} at the price of worse agreement when going to $\pm$\qty{90}{\degree}.
The orange curves in \autoref{fig:five_slot_pattern} show the resulting gain pattern in CRESana using this approach. 
The structure with multiple side-lobes in the H-plane is the result of interference of the individual slots due to their displacement. 
These plots show that our approach reproduces the \gls{hfss} results well. 
\subsubsection{Combined antenna response}
Combining all the aspects discussed above, the output power of a single antenna element is calculated using the antenna polarization, directivity and the transfer function gain shape:
\begin{equation}
    \label{eq:output_power}
    P_{\mathrm{out}}(t) = M_{\mathrm{pol}}(t) \cdot D(\uvec{d}(t)) \cdot G(\omega_0) \cdot P_E(t) \, ,
\end{equation}
depending on the input radiation power at the antenna $P_E$, the direction of the source $\uvec{d}$, and the polarization of the radiation. 
The voltage time series amplitude is calculated as in \autoref{eq:antenna_amp} using $P_{\mathrm{out}}(t)$ from \autoref{eq:output_power} and $Z$ evaluated from \gls{hfss} at $\omega_0$.
Because we treat the frequency response as constant, the antennas do not alter the frequency content and the phase of the incident electric field $E(t)$ is preserved.

\subsection{Sampled Signal}
\label{sec:sims:signal_sampling}
Finally, CRESana directly takes digital IQ-samples for the output voltage time-series of the antenna array
\begin{equation}
    \label{eq:voltage_generic_sinusoid}
    {U_{I/Q}}_j(t_i) = \frac{1}{2} A_{\mathrm{LO}} A_j(t_i) \euler^{\iu \left(\frac{\pi}{2} + \varphi_j(t_i) - 2 \pi f_{\mathrm{LO}} t_i \right)} \, ,
\end{equation}
where $t_i$ are the sample times and $j$ denotes the antenna number. 
With the frequency $f_{\mathrm{LO}}$, we implement a local oscillator (LO) for down-conversion with an idealized low-pass filter. 
It shifts all frequencies to baseband, reducing the required sampling rate to the bandwidth necessary for capturing the full signal spectra at the relevant energies for tritium beta spectroscopy, which is typically \SIrange{200}{400}{\mega \hertz}. 
$A_j(t)$ and $\varphi_j(t)$ are the instantaneous amplitude and phase of the analog antenna voltages and are the result of all the simulation steps above. 
These are only evaluated for the sample times $t_i$ requested here. 

With this simulation approach, we are unable to resolve any effect on time scales shorter than the cyclotron period. 
In the electron motion simulation we chose the guiding center approximation, and thus the cyclotron frequency is treated as constant over a single period. 
In addition, most analytic expressions e.g. for power and angular power distribution are also averaged over a cyclotron orbit. 
These are reasonable assumptions considering that the spatial resolution of the antenna array cannot resolve the cyclotron orbit, and the sampling rate cannot resolve effects on these time scales.

\subsection{CRES Electron Signal Structure}
\label{sec:sims:signal_structure}
\begin{figure*}[tbh]
 \subfloat[Spectra without modulation using the average cyclotron frequency.]{%
   \includegraphics[width=0.5\textwidth]{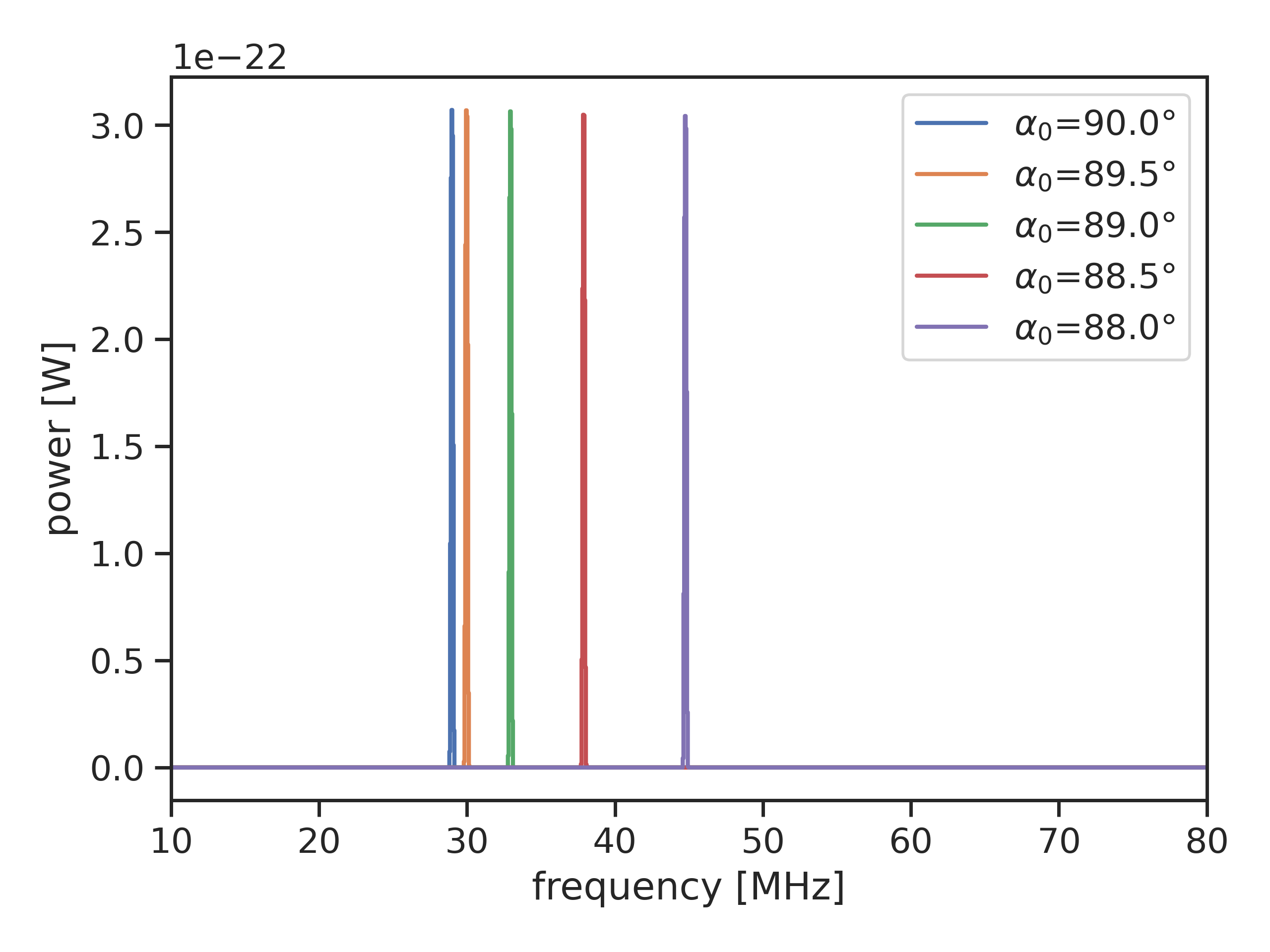}%
   \label{fig:spectra_harmonic_no_modulation}%
 }\hfill
  \subfloat[Spectra with all modulation effects.]{%
   \includegraphics[width=0.5\textwidth]{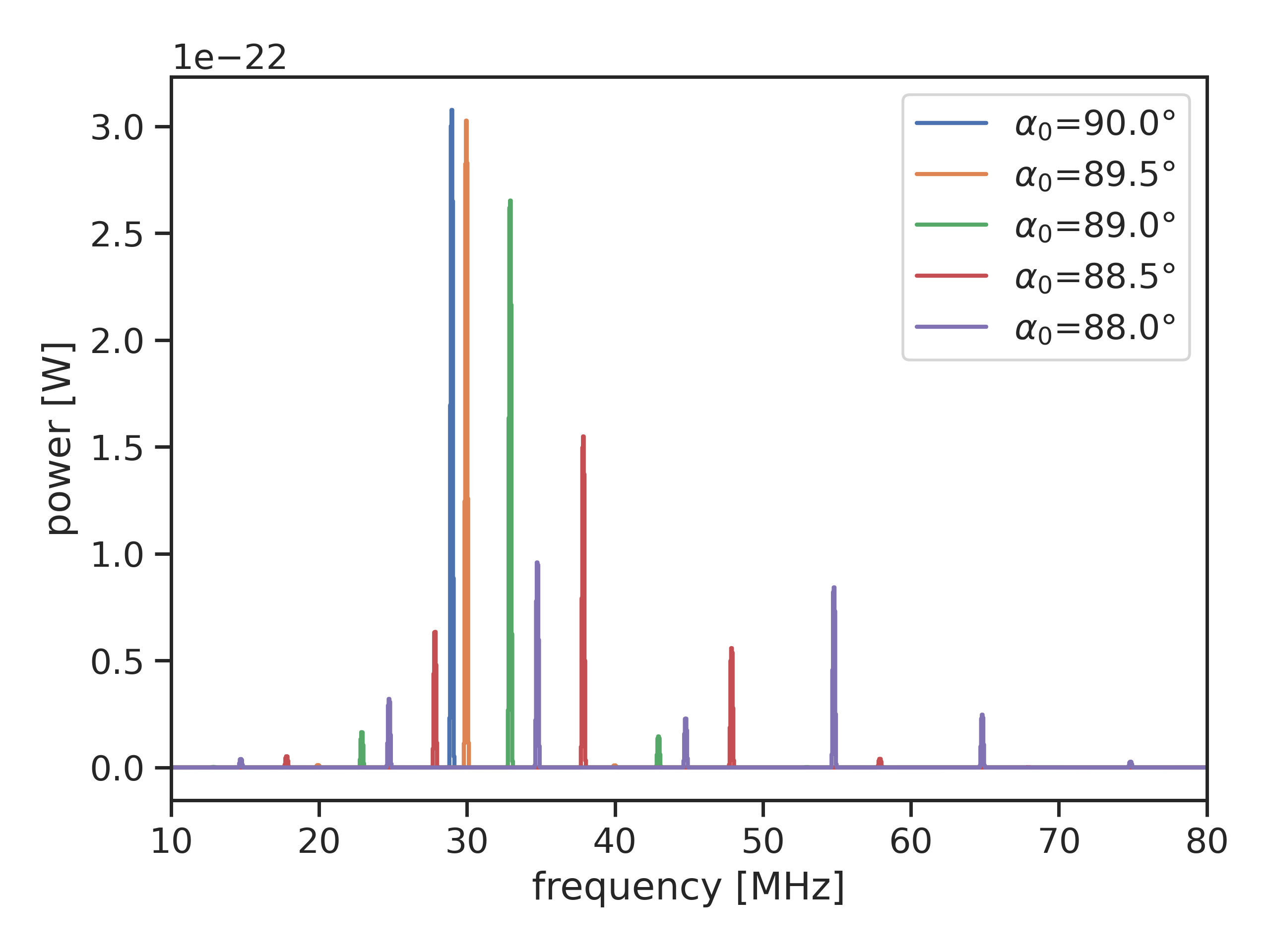}%
   \label{fig:spectra_harmonic_shift}%
 }
 \caption{Spectra of electron signals with $\ekin=\SI{18.6}{\kilo \electronvolt}$ in a harmonic trap. Adapted from \cite{florianPhd}. 
 }
 \label{fig:spectra_harmonic}
\end{figure*}
\begin{figure*}[tbhp]
\centering
   \includegraphics[width=0.75\textwidth]{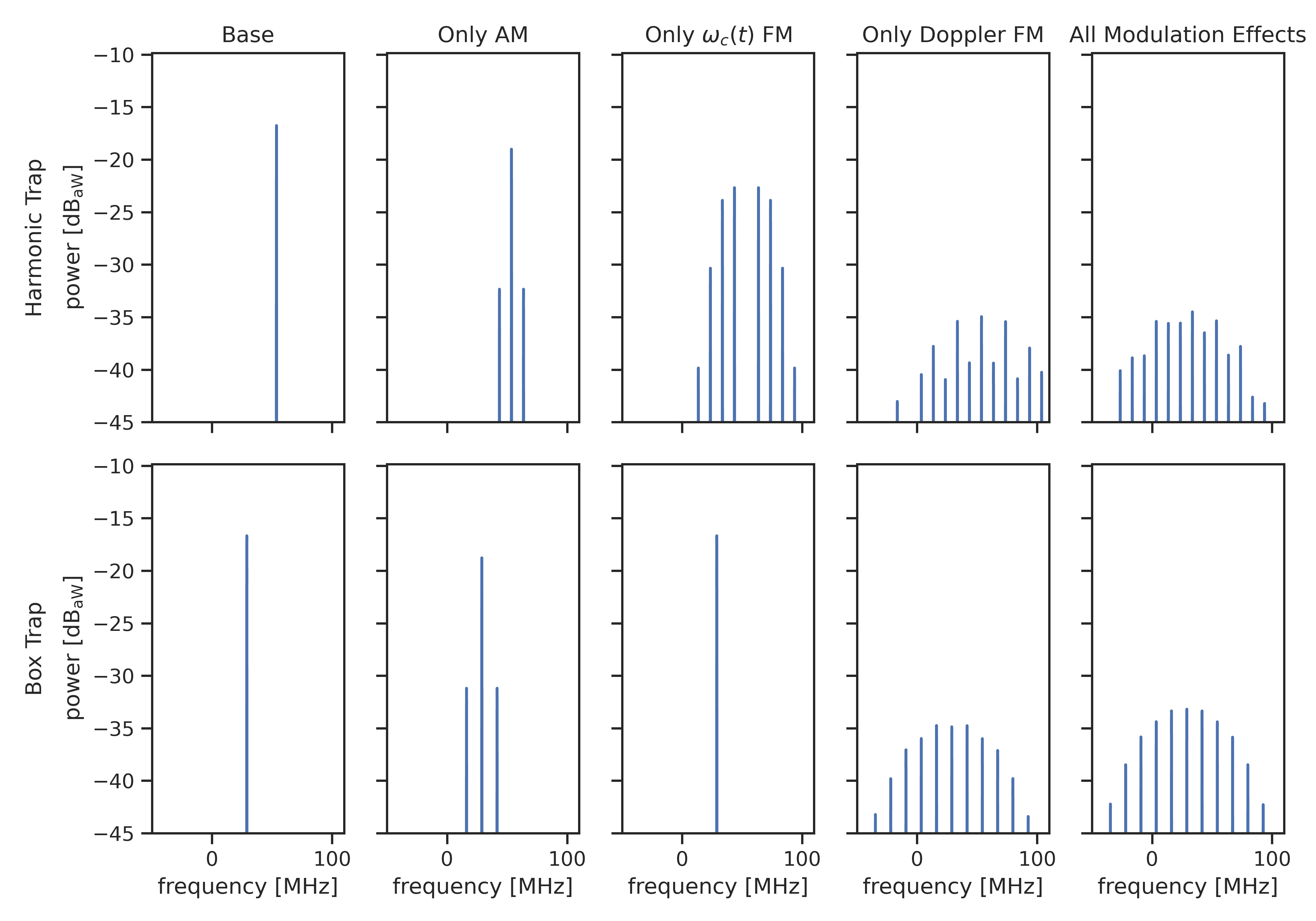}
\caption{Spectra for $\pitch=\SI{87.5}{\degree}$, $\radial=0$ and $\ekin=\SI{18.6}{\kilo \electronvolt}$ with the five-slot antenna at \qty{10}{\centi \meter} distance in a trap with a harmonic profile $B(z)$ and a box trap with infinitely sharp walls but otherwise constant along $z$. In the first column all modulation effects are disabled, the second enables only \gls{am}, the third only \gls{fm} due to changes of $B(z)$, the fourth enables only \gls{fm} due to Doppler shifts and the last enables all modulation effects. 
Adapted from~\cite{florianPhd}.}
 \label{fig:harmonic_flat_all_effects}
\end{figure*}

\gls{cres} electron signals exhibit several characteristic spectral features. 
While their specific manifestation is greatly influenced by the magnetic trap and the antenna array configuration, we can discuss these features generically.
For example, in any experimental configuration the spectra depend strongly on the electron's pitch angle $\pitch$, as can be seen in \autoref{fig:spectra_harmonic_shift} for spectra in a trap with a harmonic profile along the rotation axis.

Electrons with pitch angle $\pitch=\SI{90}{\degree}$ generate spectra with a single peak that corresponds to the downshifted cyclotron frequency. 
At different antenna positions this signal appears with different time delays, phase shifts and amplitudes according to the radiation's travel distance. 

For electrons with $\pitch<\SI{90}{\degree}$ the periodic axial motion adds additional features in the form of \glsxtrfull{fm} and \glsxtrfull{am} sidebands. 
\gls{am} and \gls{fm} are established techniques for encoding information in \gls{rf} signals and are mathematically well understood~\cite{Faruque2017}.
Sidebands appear at frequencies offset from the carrier frequency (the main peak) by integer multiples of the modulation frequency. For \gls{cres}, the carrier frequency is the average cyclotron frequency.  
In an antenna array centered in a $\pm z$-symmetric trap, the modulation frequency is typically twice the axial frequency.
This is because the amplitude and frequency variations repeat in the second half of the axial cycle as electrons pass twice in front of the antennas.

The sources of \gls{fm} are Doppler shifts and variations of the instantaneous cyclotron frequency due to variation of $B(z)$ along the cylinder axis~\cite{pheno_paper}. 
As lower pitch angle electrons explore higher magnetic field regions, the second effect also increases the average cyclotron frequency, thereby shifting the carrier position, as shown in \autoref{fig:spectra_harmonic_no_modulation}.

\gls{am} arises from the variation of the distance $\abs{\bvec{d}}$ in \autoref{eq:friis}, as well as the source direction relative to the antenna, which affects $G_e$ and the antenna response.
Additionally, the variations of the magnetic field lead to slight changes in the Larmor power (see \autoref{eq:larmor}), contributing marginally to the overall \gls{am}.

\autoref{fig:harmonic_flat_all_effects} shows a comparison of spectra for the same electron in two traps with different simulation conditions. 
Although both traps have similar magnetic field strengths and axial frequencies for this electron, their spectra differ due to the functional form of the magnetic trapping field. 
In the harmonic trap, the main peak shifts to the average cyclotron frequency, while it remains constant in the box trap due to the uniform background field, as seen in the first column. 
As expected for \gls{am}, the second column only adds two significant sidebands~\cite{Faruque2017}, which are stronger in the box trap since the electron's axial travel distance is about \qty{70}{\percent} larger. 
In the third column the carrier disappears in the harmonic trap---a well known feature of \gls{fm} with high modulation index---whereas no sidebands are added in the box trap, as the cyclotron frequency stays constant in the constant magnetic field. 
Finally, in both traps, Doppler-induced \gls{fm} plays the largest role in shaping the final spectrum as seen in the fourth and fifth columns.
In general we found this to be the dominant effect for most setups in our simulations~\cite{florianPhd}.
\begin{figure}[tbhp]
\centering
   \includegraphics[width=0.5\textwidth]{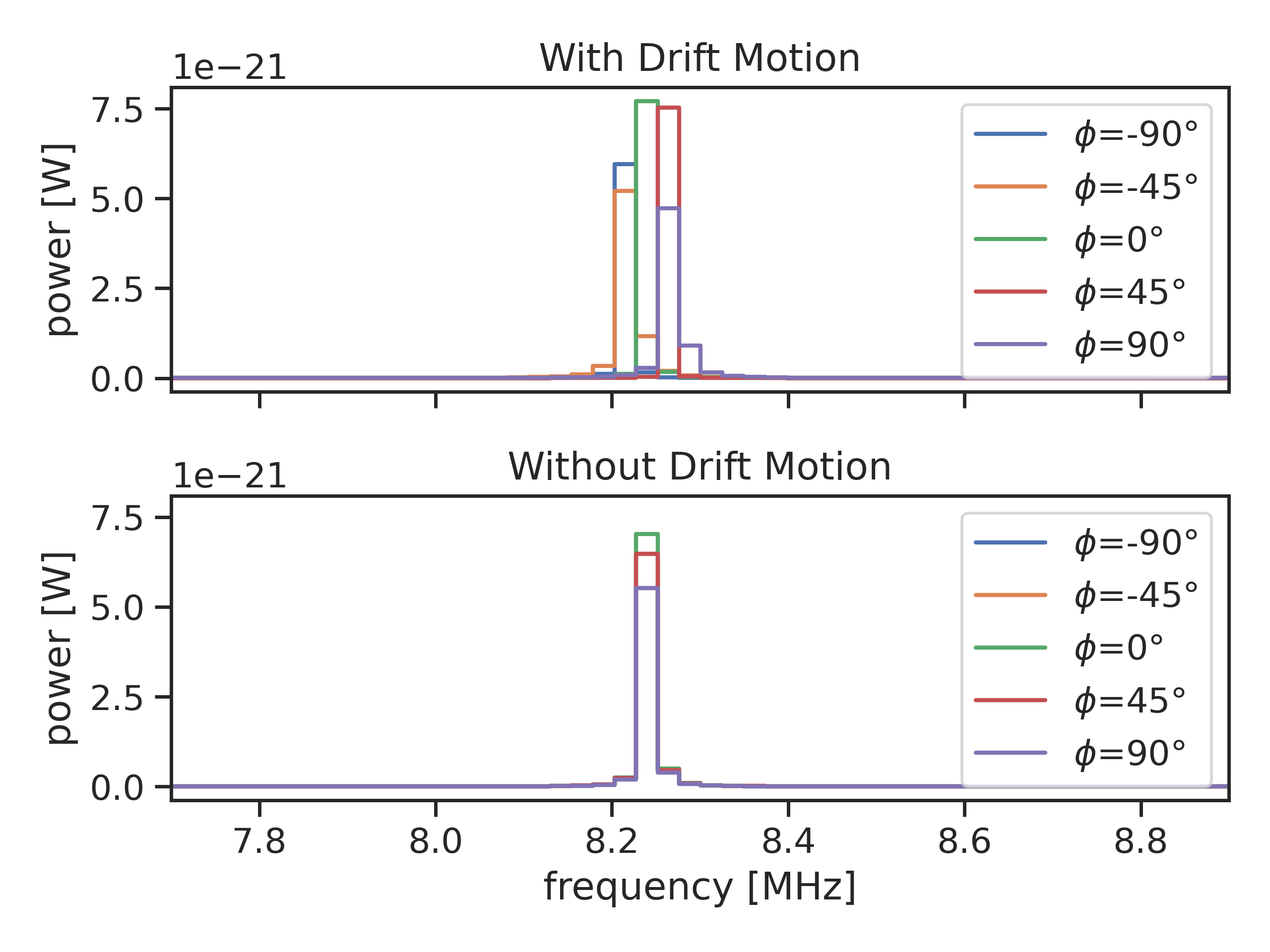}%
    \caption{Spectra of an electron with $\pitch=\SI{90}{\degree}$ and $\radial>0$ experiencing drift motion compared to spectra with drift motion disabled. The drift motion Doppler-shifts the cyclotron frequency for different azimuthal antenna positions $\azimuth$. Adapted from \cite{florianPhd}.}
   \label{fig:drift_motion_spectra}%
 \end{figure}
\begin{figure}
\centering
   \includegraphics[width=0.5\textwidth]{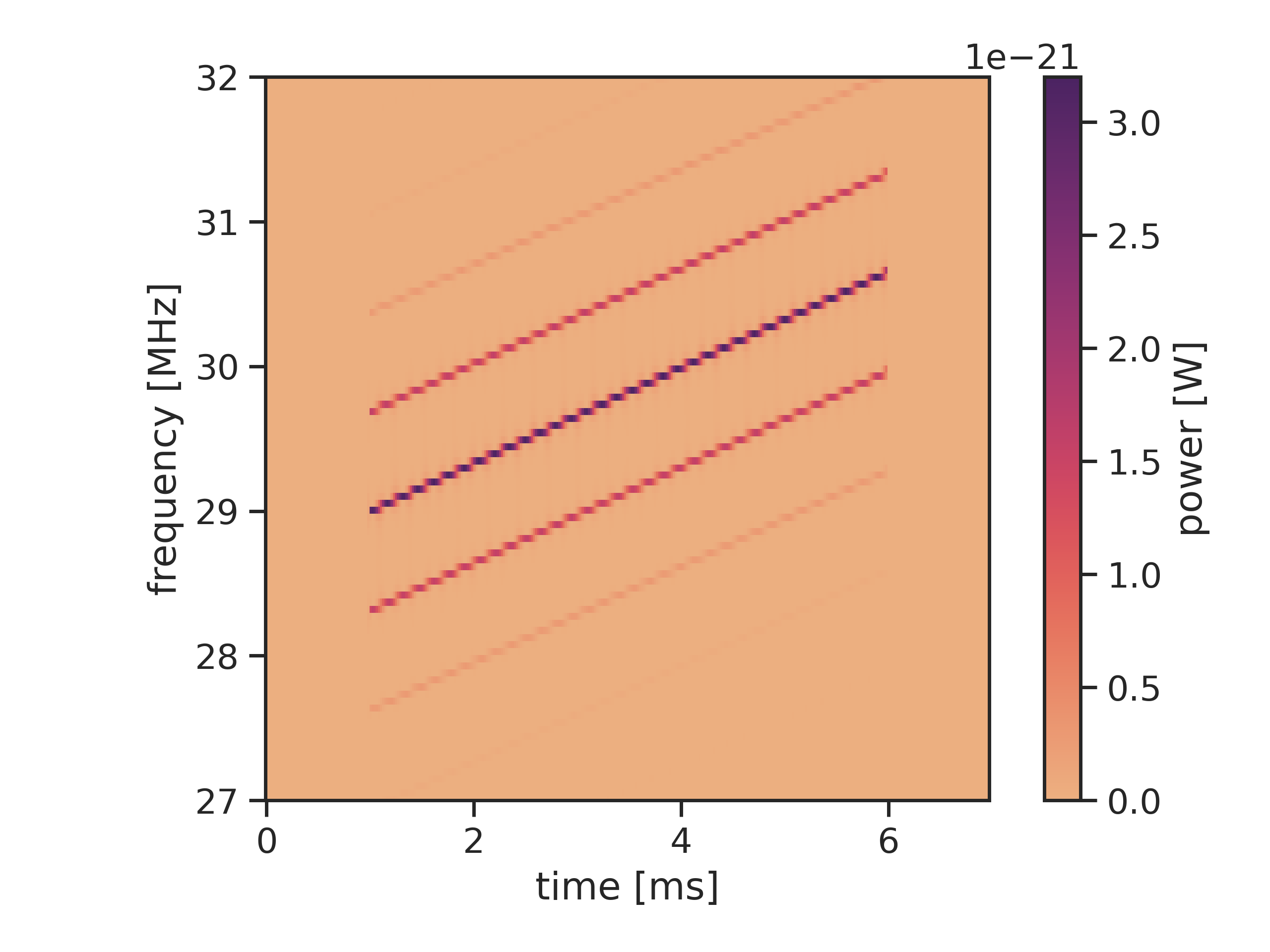}%
    \caption{Time-frequency spectrogram of a \SI{5}{\milli \second} duration CRES event observed by a single antenna. Multiple sidebands are visible that shift in frequency over time due to the energy loss. A setup with small axial frequency was chosen such that these lines are visible in a close-up. Adapted from \cite{florianPhd}.}
   \label{fig:spectrogram_sidebands}%
\end{figure}

Another notable feature is a small variation in the cyclotron frequency for different azimuthal antenna positions $\azimuth$ for $\radial>0$, as seen in \autoref{fig:drift_motion_spectra}.
This is caused by the Doppler shifts from the drift motion, since the relative velocities between electron and antenna vary with the antenna's azimuthal position. 
In most cases the drift motion is too slow to add visible \gls{am} and \gls{fm} sidebands.
While this effect might seem insignificant, if left unaccounted for the incoherence of antennas causes a noticeable reduction in \gls{snr} of simple trigger algorithms \cite{antenna_detection_paper}.

The last feature of interest in the signal spectra is caused by the energy loss of the electron due to the radiated power. 
For sufficiently long observation times the energy loss manifests as a linear frequency chirp, i.e. over time the cyclotron frequency shifts to higher values in a linear fashion (see \autoref{fig:spectrogram_sidebands}). 
The chirp rate, or slope, is the frequency change per time $\slope$~\cite{pheno_paper}:
\begin{equation}
    \label{eq:chirp_rate}
    \slope = \frac{P_{\mathrm{Larmor}} \fcyclotron}{m_0 c^2 + \ekin} . 
\end{equation}
It is mostly determined by the magnetic field strength, since that sets the the radiated power $P_{\mathrm{Larmor}}$ in \autoref{eq:larmor} and cyclotron frequency $\fcyclotron$ in \autoref{eq:cycltron_frequency}.

\section{Prototype Antenna Array Measurements}
\label{sec:simvalid}
In previous sections we have described antenna modeling in \gls{hfss} and electron signal simulation in CRESana. 
This section aims to cross-check those tools and quantify their imperfections towards a fuller, more realistic application.
A series of room temperature measurements were made using a synthetic \gls{cres} antenna (SYNCA) as a representation of a \gls{cres} electron \cite{synca_paper}.
The SYNCA is a static source and therefore does not address the aforementioned spectral features, but it is useful for quantifying signal losses due to multipath reflections and for directly measuring reconstruction accuracy.
The design goal was to reach a reflection-based signal loss small enough to allow event tracking and achieve millimeter-scale accuracy in position reconstruction. 
These design goals were met, and the details of these measurements are provided below.
As addressed in \autoref{sec:design}, the \SI{26}{\giga \hertz} scale is more feasible for lab testing than \SI{1.3}{\giga \hertz}.
All measurements described here were conducted near \SI{26}{\giga \hertz} and unless otherwise specified, ``antennas" refers to the center-fed, five-slot antennas of~\autoref{fig:fiveslot}.
\begin{figure}[htbp]
    \includegraphics[width=0.45\textwidth]{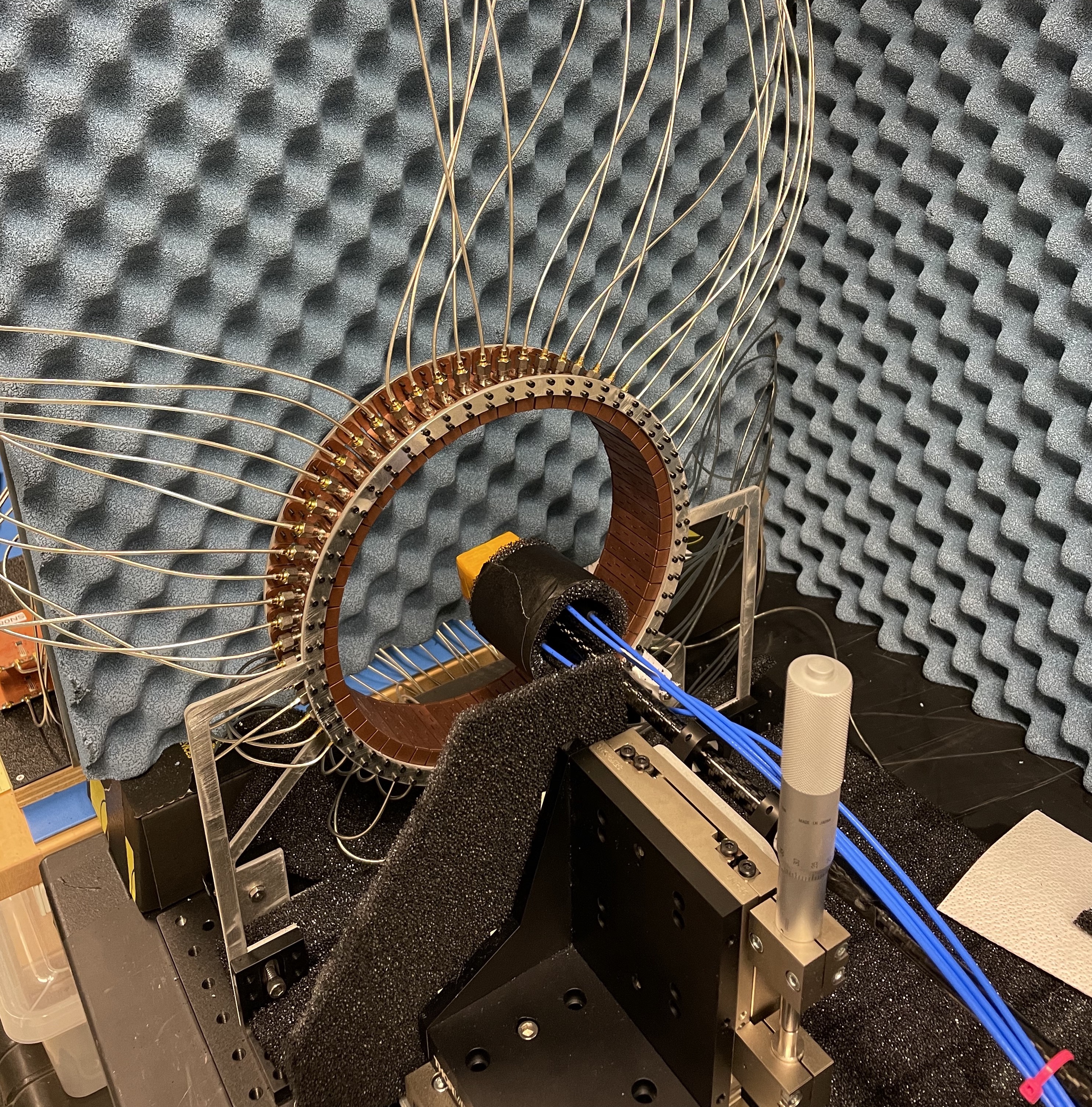}%
    \caption{Experimental setup of the full ring of sixty receiver antennas as well as the SYNCA source antenna in the center, mounted on a 3-axis stage.}
     \label{fig:jugaad_photo}
\end{figure}
\subsection{Characterization of Individual Antennas}
\label{sec:ind_measurements}
\begin{figure*}[tbh]
    \includegraphics[width=0.95\textwidth]{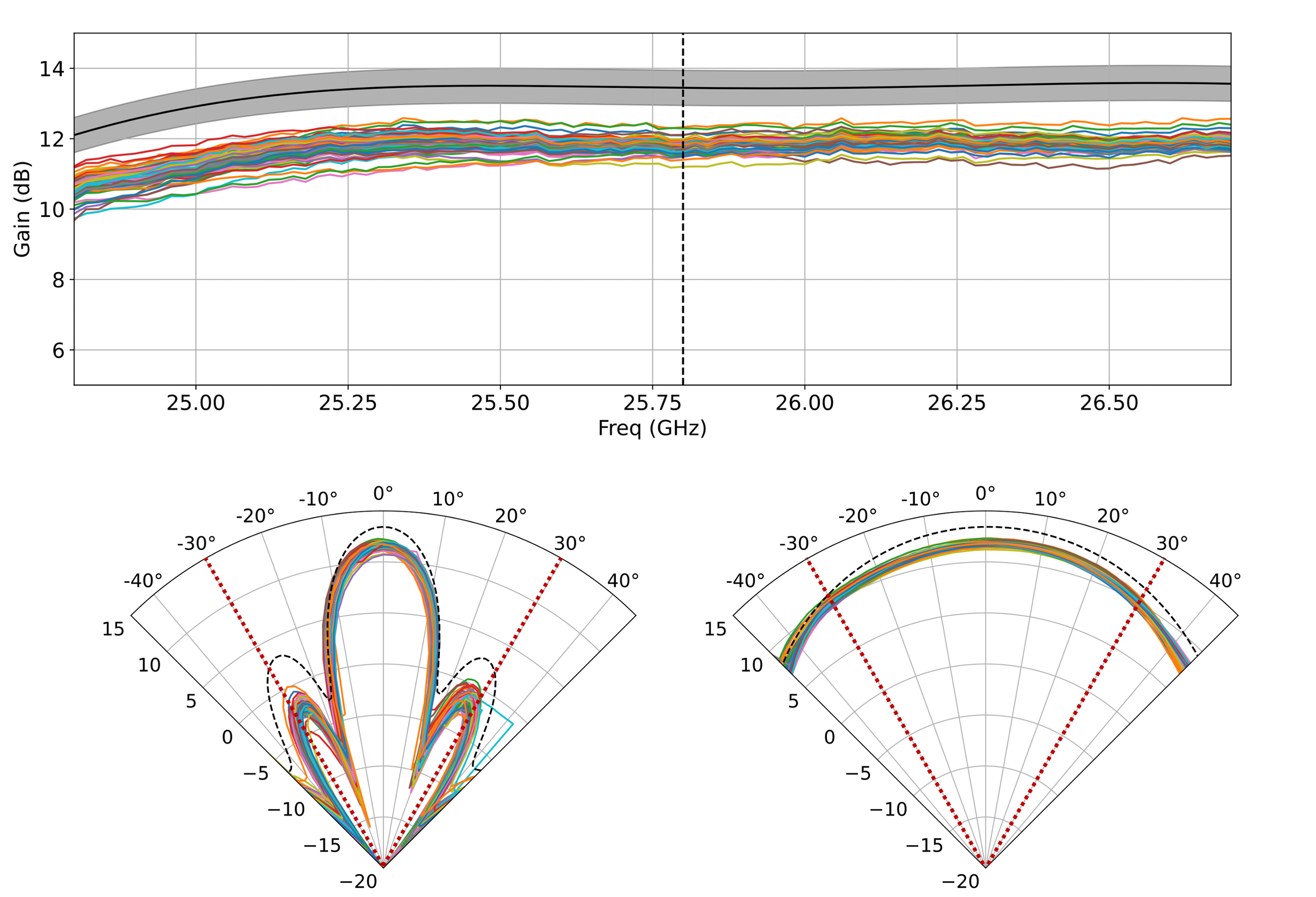}
    \caption{Boresight gain as a function of frequency as measured for all sixty antennas with \gls{hfss} simulated response in gray (top). Beam pattern as measured at the central frequency across the H-plane (left) and E-plane (right) of the antennas. Solid colored lines correspond to individual antenna measurements and the black dashed lines correspond to the ideal models from \gls{hfss} simulations. The red dashed line corresponds to the field of view characterized for the full array measurements in \autoref{sec:fullarray}.}
     \label{fig:individual_antenna_gain}
\end{figure*}

A standard gain horn antenna was used to characterize individual antennas and measure their transfer function and beam pattern.
The boresight gain was measured within a \qty{2}{\giga \hertz} bandwidth around the central frequency (\qty{25.8}{\giga \hertz}).
The beam pattern was also measured with this setup by rotating the antennas in the H- and E-planes.
Comparing the boresight gains to those modeled by \gls{hfss}, none of the $60$ antennas used for these measurements were more than \qty{2}{\decibel} below the ideal antenna gain.
Similarly the gain as a function of angle was consistently 1-\qty{2}{\decibel} below the ideal gain in the main lobe and approximately \qty{5}{\decibel} below the modeled gain in the side lobes (results shown in \autoref{fig:individual_antenna_gain}).
These results were sufficient to move forward with full array measurements, and showed that relative antenna differences were $<$\qty{2}{\decibel}.
\subsection{Antenna Array Measurement Setups}
\label{sec:benchmarking} 
In order to benchmark antenna simulations, two receiver antenna array configurations were used: a synthetic array generated from a single antenna, and a full array of $60$ antennas.
In both cases, the SYNCA was the source antenna.
\subsubsection{Synthetic Single-Antenna Array Setup}
First, a single receiver antenna was placed at a fixed distance from the SYNCA, which was mounted on a rotary and translation stage.
This single receiver antenna was used to simulate a full array by rotating the SYNCA through a full $360^\circ$ rotation at multiple off-axis radial positions (0 to \SI{35}{\milli \meter} from axial center in 5~mm steps). 
Measurements were taken to represent antenna positions at $6^\circ$ increments.
Phase-locked data runs were then digitally combined into a synthetic receiver array.
This imitates a ring of antennas while avoiding real, multipath reflections off of a physical array, as well as any relative antenna differences.
Vertical alignment of the SYNCA with the central plane of the synthetic array was accomplished with a manually controlled optical post mount (0 to \SI{25}{\milli \meter} in \SI{5}{\milli \meter} steps out of the plane of the synthetic antenna ring).
Continuous wave signals were delivered to the SYNCA by generating a \qty{64}{\mega \hertz} baseband sinusoid signal upconverted to \qty{25.864}{\giga \hertz} using a \qty{25.8}{\giga \hertz} LO and bandpass filter. 
The signals emitted by the SYNCA were received by the slotted-waveguide antenna and downconverted to baseband using the same LO. 
\subsubsection{Full Array Setup}
\label{sec:fullarray}
To measure physical array effects, bench-top measurements were then taken for a full ring of 60 antennas using a Keysight FieldFox vector network analyzer.
A photo of the experimental setup is shown in \autoref{fig:jugaad_photo}.
For a given position of the SYNCA source, each channel was measured sequentially with a phase-locked signal using a custom 1$\times$64 electronic switch.
S-parameters of the antennas were measured using a frequency sweep from \SI{25.1}{\giga \hertz} to \SI{26.5}{\giga \hertz} in \SI{0.01}{\giga \hertz} steps.
\gls{rf}-absorbing material was placed around the setup to mitigate any reflections from the surrounding environment.
A manually controlled 3-axis stage was used to take data for a variety of SYNCA positions.
The stage was initially located using a plastic jig fit to the antenna ring.
\begin{figure*}[!hbtp]
    \centering
    \subfloat[Synthetic array.]{
        \includegraphics[width=0.48\textwidth]{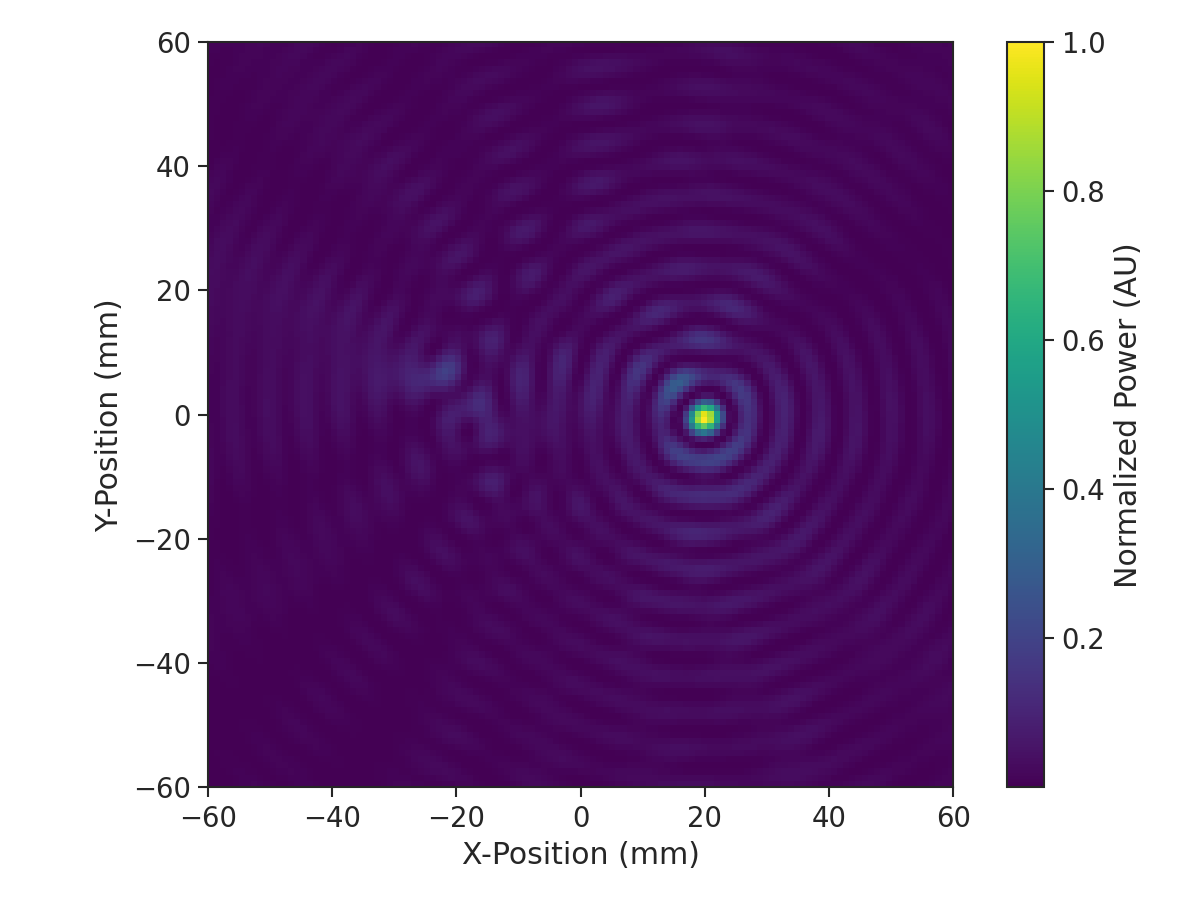}
    }
    \subfloat[Full array.]{
        \includegraphics[width=0.48\textwidth]{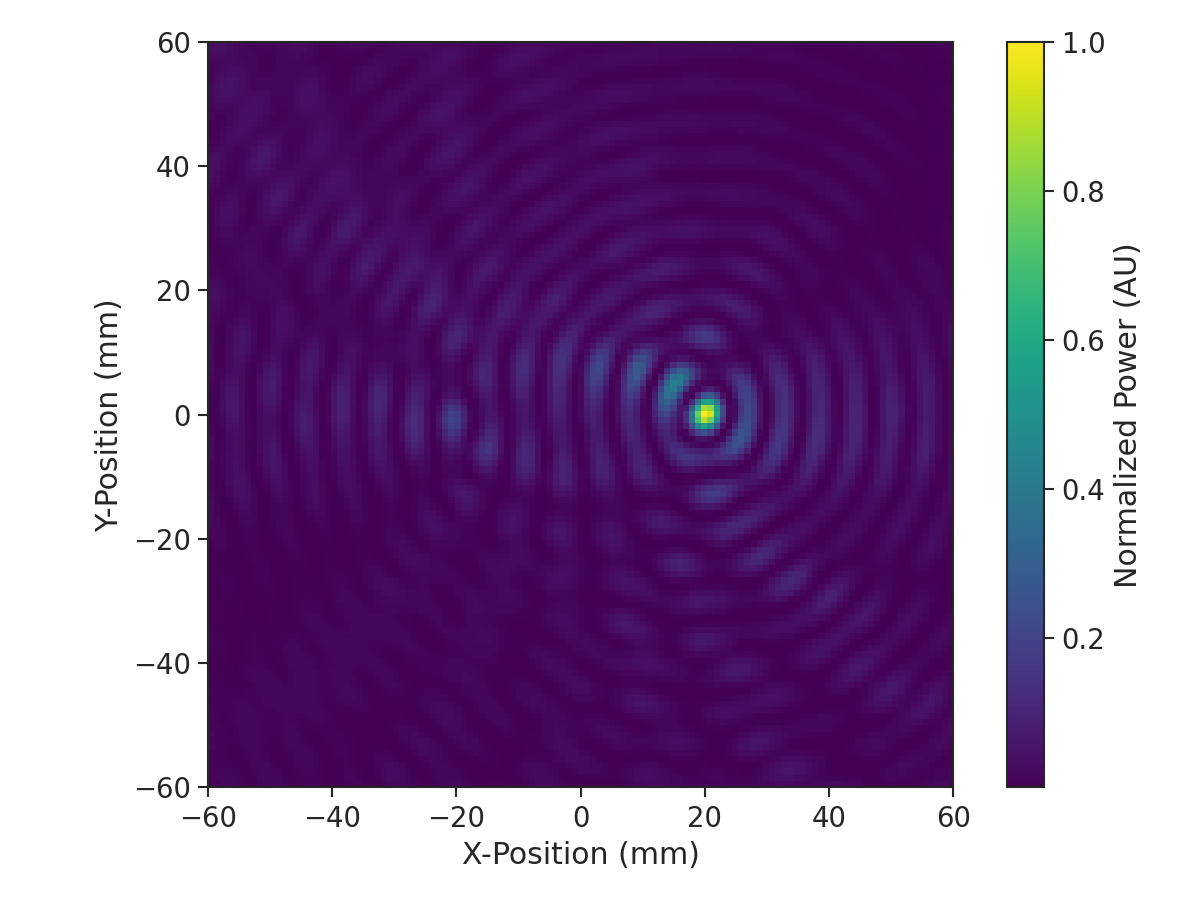}
    }
    \caption{Beamforming images obtained for the synthetic and full antenna arrays with the SYNCA located at $\radial{}$ = \qty{20}{\milli \meter} and $z=$~\qty{0}{\milli \meter}. The images are normalized such that the maximum beamformed power is equal to unity. The observed phase errors that lead to a reduction in the maximum beamformed power do not lead to a significant difference between the reconstructed and actual positions of the SYNCA.}
    \label{fig:beamforming_images}
\end{figure*}
\subsection{Validating Position Reconstruction Techniques}
\label{sec:position_recon}
\begin{figure*}[htbp!]
\subfloat{
   \includegraphics[width=0.47\textwidth]{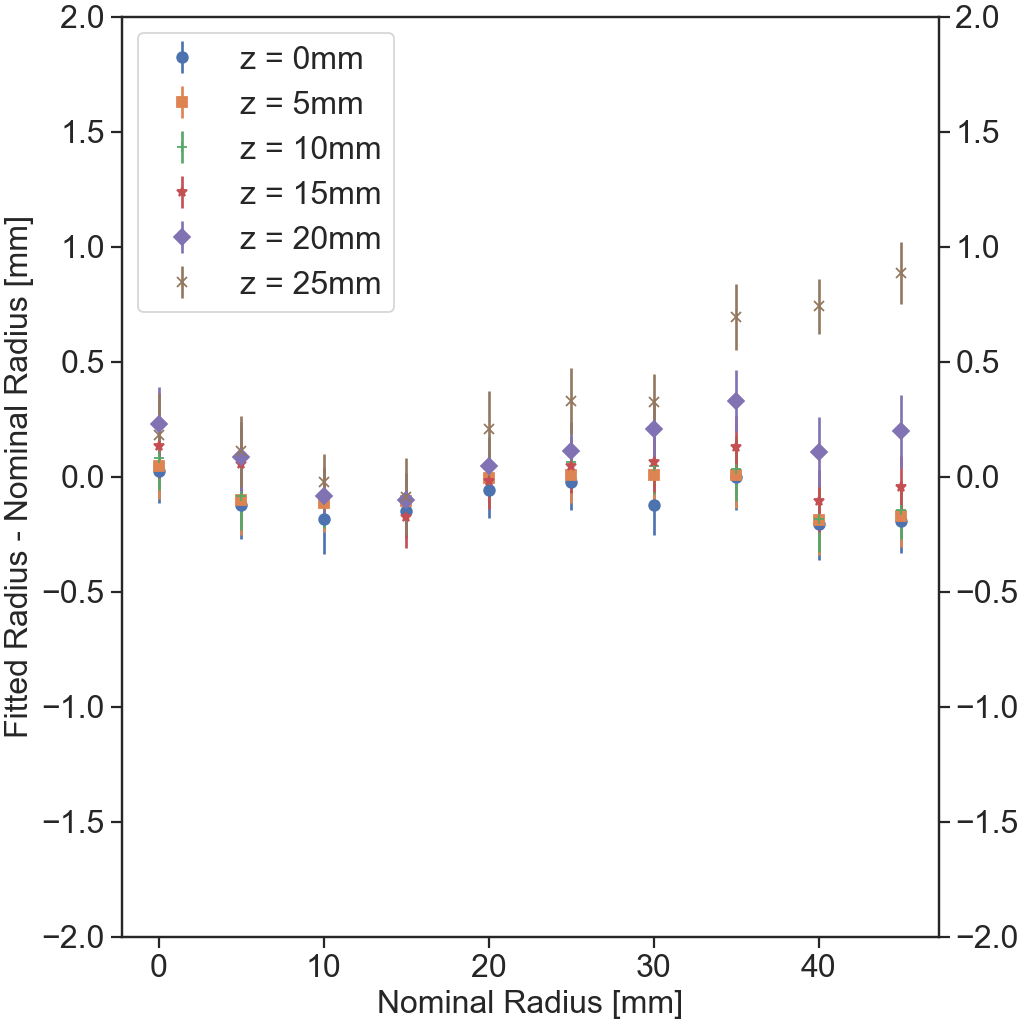}
 }
 \subfloat{
   \includegraphics[width=0.47\textwidth]{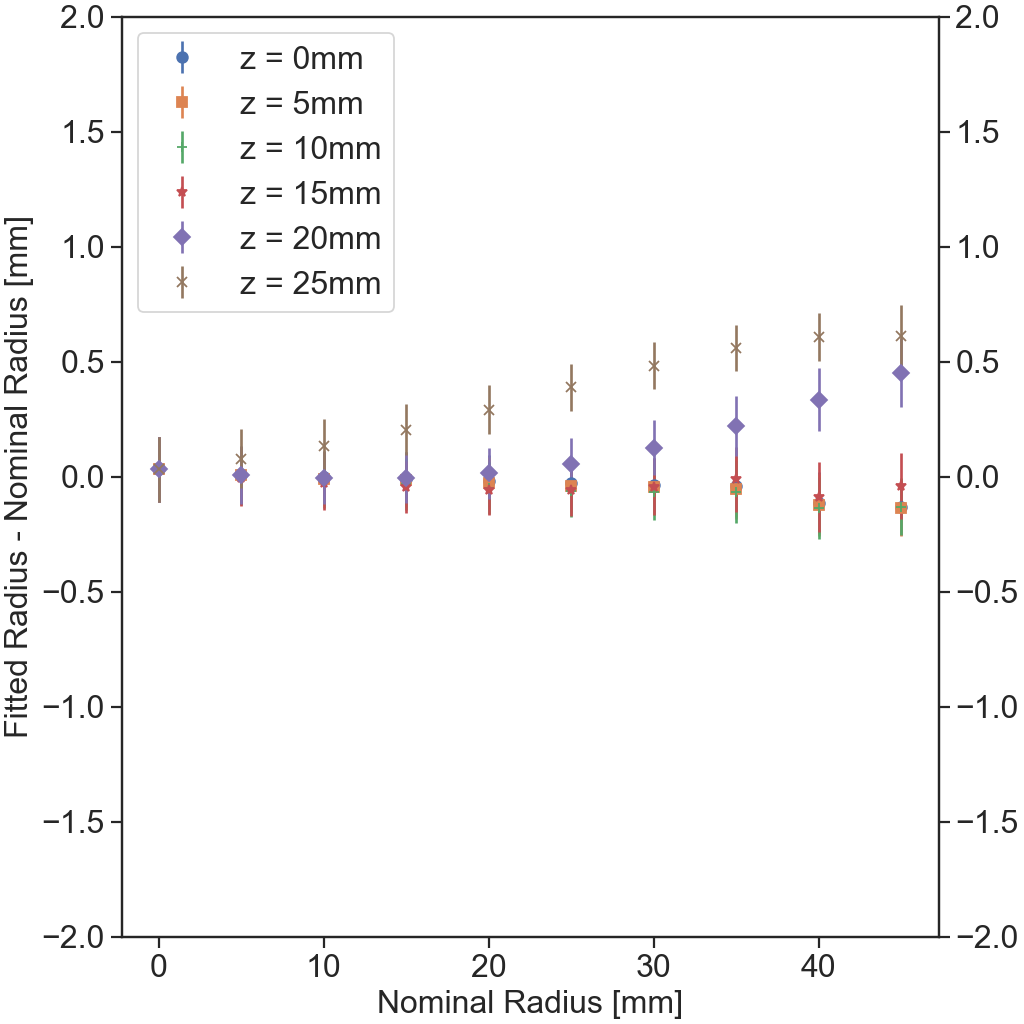}
 }
 \caption{Differences between fitted and nominal radial positions of a $60$-antenna ring for (left) data and (right) simulation. Different colors correspond to different $z$ positions of the synthetic \gls{cres} source. The measurements and simulations show sub-millimeter agreement, comparable to the physical positioning uncertainty.}
 \label{fig:reconstructed_position}
\end{figure*}

Data was digitally beamformed by doing a phased summation of the individual channels in the array, both for the synthetic and full arrays~\cite{antenna_detection_paper}. 
Representative images are presented in \autoref{fig:beamforming_images}.
For both measured and simulated data, the beamformed image at each nominal position is fitted with a 2D Gaussian at the central peak of the distribution.
The difference between the fitted and nominal radial positions is shown in \autoref{fig:reconstructed_position}.
Error bars are determined from the uncertainties on the means of the fitted Gaussian for each image, with widths of the Gaussians being a few millimeters.

The different z-positions of the SYNCA introduce a slight radially-dependent bias into the reconstructed position, since the SYNCA's phase response and pattern are not uniform out-of-plane.
In a real \gls{cres} experiment with a magnetic trap, the electron exhibits periodic axial motion as described in \autoref{sec:pheno}.
The beamformed position would be determined from radiation emitted within the plane of the antenna ring, greatly reducing the reconstruction bias seen in this setup.
We also see evidence of a coherent, sinusoidal error over radius, which is likely from phase uncertainties in the SYNCA source.
Both of these uncertainties could be reduced through a dedicated calibration in a larger full-scale experiment.
In this case, however, calibration is not required, as the total uncertainty is already below one millimeter, which is comparable to the scale of the uncertainty from the positioning system.
Overall, the data shows excellent agreement between simulation and data as well as sub-millimeter accuracy across the range of measured positions.
\begin{figure*}[htbp!]
    \includegraphics[width=.99\textwidth]{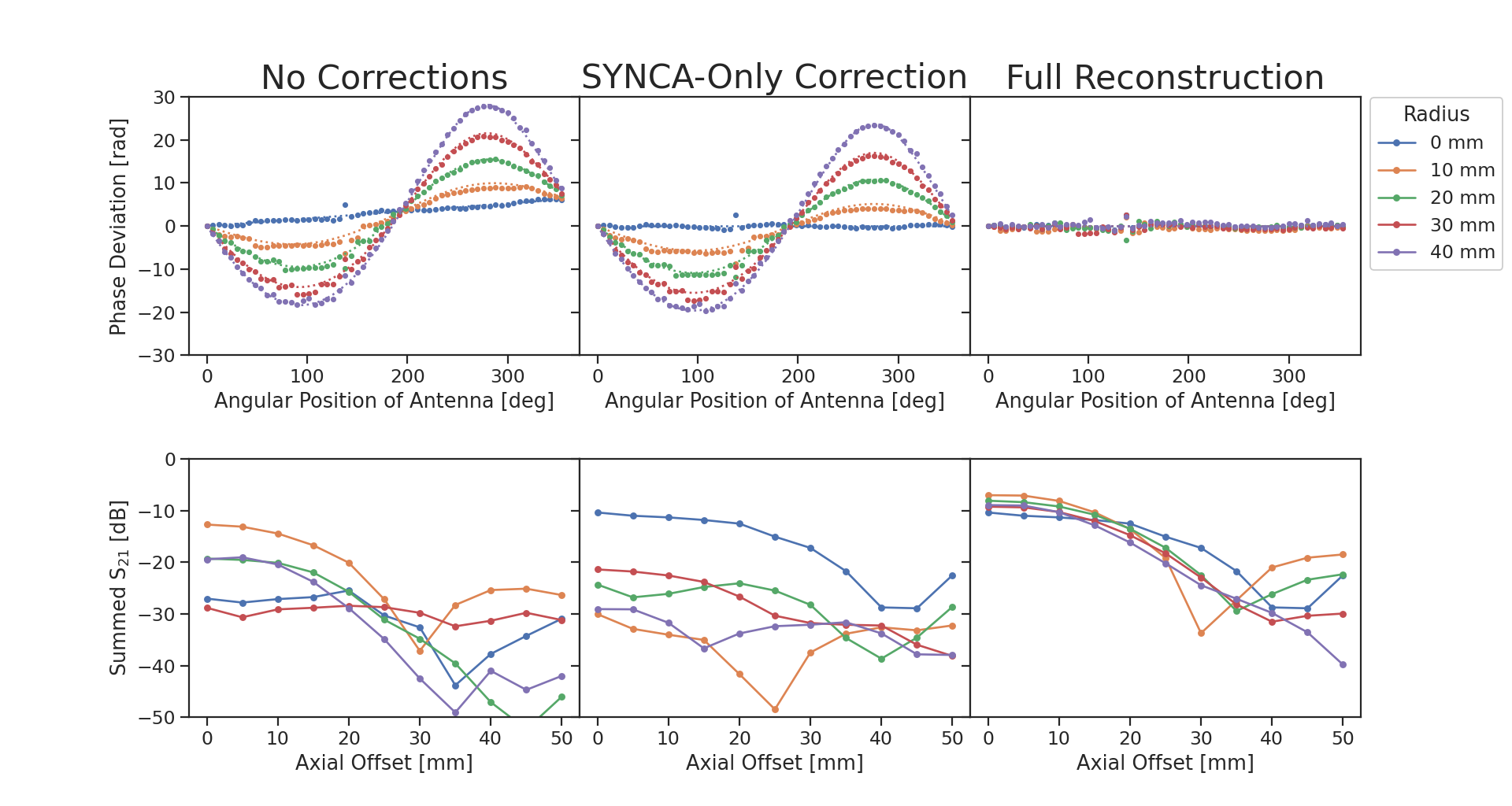}%
    \caption{ (top) Deviation in phase measured at different antennas as a function of angle across three levels of reconstruction as measured at the central frequency of the waveguide antennas. (bottom) Power of transmitted signal coherently summed across all sixty antennas as a function of frequency and at the corresponding levels of reconstruction of the above plots. From left to right these plots have no phase corrections, phase corrections accounting for the modeled SYNCA phase only, and full reconstruction including both the SYNCA phase and the SYNCA radial position. All plots include measurements with the SYNCA source located at the center of the array as well as at \qty{10}{\milli \meter}, \qty{20}{\milli \meter}, \qty{30}{\milli \meter}, and \qty{40}{\milli \meter} offsets. A significant increase in power is reconstructed when the phase offsets are completely accounted for.}
     \label{fig:phase_reconstruction}
\end{figure*}
\subsection{Quantifying Signal Losses}
The primary goal of the array measurements is to isolate and quantify the signal losses coming from SYNCA phase errors, receiver-to-receiver relative phase errors, and effects from multipath reflections.
A subset of these phase errors are handled by the reconstruction process illustrated in \autoref{fig:phase_reconstruction}.
The left-most plots have no phase reconstruction, and the central plots include a linear phase offset due to the Archimedean spiral pattern of the SYNCA source fields.
The right-most plots include all SYNCA-related phase adjustments as well as a spatial offset from the location of the point in the beamformed image, described in greater detail in~\cite{synca_paper}.
The lower plots show the total reconstructed power, via the S$_{21}$ parameter summed over all sixty channels, in a \qty{50}{\mega \hertz} wide window about the antenna central frequency.
The total reconstructed power increases substantially across all radii after the phase corrections are applied.
This holds true for axial offsets less than \qty{30}{\milli \meter}, beyond which the SYNCA moves out of the main lobe of the antenna array's radiation pattern.

However, phase errors from other sources persist, and can only be quantified by comparing results between the synthetic array, the full array, and simulations of both setups.
Due to the differences in receiver electronics for the synthetic and full arrays, the reconstructed signal power cannot be compared directly. 
Instead, the reconstructed power relative to the maximum simulated power for each setup is used for comparison.
This was done over a range of radial and angular positions for the SYNCA source.
The result was a mean power loss due to uncorrected phase errors of 15\% for the synthetic array and 23\% for the full array. 
We attribute the approximate 10\% increase in power loss to both antenna-to-antenna differences and multipath effects in the full array, which can qualitatively be seen in \autoref{fig:beamforming_images}.
These measurements place an upper bound on the scale of uncertainties in the full array setup.
In principle, effects from the SYNCA phase errors could be calibrated out, and specific contributions from receiver antenna phase mismatches could be measured directly in dedicated studies to improve performance.
While the specific contributions from receiver antenna phase mismatches and multipath effects have not yet been clearly separated, preliminary estimates of beamformed images suggest that the contributions from these two terms are comparable.
Regardless of the relative strength of these effects, the overall measured $\sim$10\% power loss due to the presence of the array allows us to proceed with using our simulations for evaluating the feasibility and performance of \gls{cres} experiments.

\section{Signal Detection and Parameter Estimation}
\label{sec:recon}

\gls{cres} event reconstruction is the procedure of using the acquired raw voltage time series to estimate the starting kinetic energies of electrons trapped in the detector. 
These energy estimates are then used to construct the tritium beta decay spectrum and measure the mass of the neutrino. 
Generally event reconstruction can be divided into signal detection and parameter estimation. 
Signal detection is the decision problem: determining whether the given data contain an electron signal or if they consist only of noise. 
For data which contain an electron signal with high statistical confidence, the parameter estimation problem is to obtain a value for the electron's kinetic energy and the associated uncertainty.

Both the detection and the parameter estimation problems have established general solutions in statistical signal processing literature, however, for specific problems these may be computationally infeasible. 
Computationally {\it feasible} solutions need to make use of the signal's characteristics to find a compromise between computational cost and estimation performance.
Often the signal detection and parameter estimation steps happen at the same time. 
For the remainder of this chapter we present the application of the general solutions of the two distinct problems to \gls{cres} event reconstruction, providing the upper bound for the detector performance parameters in \autoref{sec:large_detectors} while neglecting computational cost.

\subsection{Signal Detection with Matched Filtering}
\label{sec:reco:MF}

A fundamental difference between event reconstruction with an antenna array \gls{cres} detector and event reconstruction in previous experiments is the multi-channel nature of the data. 
The increase in the raw data generation rate as well as the reduction in the average signal power per channel requires an approach to triggering that can combine many weak signals to reconstruct \gls{cres} events~\cite{antenna_detection_paper}.
For the matched filtering approach described here, detection performance is not degraded for arbitrary numbers of channels.

The dominant source of electronic noise for an antenna array \gls{cres} experiment is assumed to be Nyquist-Johnson thermal noise, which is well-approximated by a complex \gls{awgn} distribution with variance
\begin{equation}
    \sigma^2 = k_B T \Delta f R \, ,
\end{equation}
where $k_B$ is Boltzmann's constant, $T$ is the system noise temperature, $\Delta f$ is the sampling rate and $R$ is the system impedance. 
The detector that maximizes the true detection probability (detection efficiency) for \gls{cres} signals is the matched filter \cite{kay1998fundamentals}. 
Since \gls{cres} signals have unknown parameters, a matched filter detector must employ the template bank approach (also used in gravitational wave detection \cite{LIGO_I, LIGO_II}). 
In this method, a set of pre-generated simulated signal templates are used to detect the presence of \gls{cres} signals buried in the antenna array time series data. 

The test statistic that describes the detection probability of a matched filter template bank can be obtained by posing the detection problem as a statistical hypothesis test between two alternate hypotheses,
%
\begin{align}
    \mathcal{H}_0\colon&\bvec{x}[n]=\bvec{\nu}[n]\\
    \mathcal{H}_1\colon&\bvec{x}[n]=\bvec{s}[n]+\bvec{\nu}[n].
\end{align}
%
Hypothesis $\mathcal{H}_0$, is the null hypothesis in which the data is composed purely of \gls{awgn}, denoted by the vector $\bvec{\nu}[n]$.
The alternative hypothesis, $\mathcal{H}_1$, is the signal hypothesis where both signal, denoted as $\bvec{s}[n]$, and noise are present in the data. 

To decide between the two hypotheses we calculate the likelihood ratio test prescribed by the Neyman-Pearson theorem \cite{kay1998fundamentals}. 
For $N_\mathrm{ch}$ the number of antennas and $N_\mathrm{sample}$ the number of samples in the data, define a matrix of array data in the shape $N_\mathrm{ch}\times N_\mathrm{sample}$.
The template bank matched filter test statistic is given by
%
\begin{equation}
    \mathcal{T}_i[m]=\left|\sum_{n=m}^{m+N_\mathrm{signal}-1}{\sum_{k=0}^{N_\mathrm{ch}-1}{\bmat{h}_i^\dagger[k, n-m]\bmat{x}[k, n]}}\right|^2,
    \label{eq:recon-mf-score}
\end{equation}
where $N_\mathrm{signal}$ is the duration of the signal template. 
In \autoref{eq:recon-mf-score}, we compute the complex cross-correlation between the array data matrix, $\bmat{x}$, and the matched filter template matrix, $\bmat{h}_i$, by calculating the separate cross-correlation for each antenna signal and then summing over all channels to obtain the matched filter test statistic as a function of the delay, $m$. 
The cross-correlation is computed for a range of delays $m\in[0,N_\mathrm{sample}-N_\mathrm{signal}]$, where $N_\mathrm{signal}$ is the duration of the signal template. 
To check if a data segment contains a signal that matches the template, we compare the maximum value of the cross-correlation to a threshold. 
We decide that the signal is present if
\begin{equation}
    \label{eq:detection_threshold}
    \max_m\mathcal{T}_i[m]>\gamma,
\end{equation}
where $\gamma$ is the decision threshold. 
This test is performed for each template $\bmat{h}_i$ until a signal is found or all templates are exhausted. 

In preceding Project~8 experiments~\cite{phaseIIprc}, the template model $\bmat{h}$ was a single channel sinusoid with unknown frequency, which is equivalent to setting a threshold on the time frequency spectrogram of the data according to \autoref{eq:recon-mf-score} and \autoref{eq:detection_threshold}~\cite{kay1998fundamentals}.
This led to a conventional naming of an electron signal as a ``track" due to its appearance on spectrograms like \autoref{fig:spectrogram_sidebands}.
A single track, or event, refers to the signal only between the decay and the first scatter off residual gas in this paper.
The scatter causes a discrete frequency change and jump in the spectrogram.
Note that subsequent signals are also called tracks in much of Project~8 literature, but here we limit our discussion to just the first track after the decay.

The noise distribution of a single matched filter template follows a $\chi^2$ distribution with two degrees of freedom.
For a number of tested templates $N_t$, a combinatorial factor must be taken into account. 
The signal matched filter distribution follows a noncentral $\chi^2$-distribution with two degrees of freedom, where the noncentrality parameter is given by the \gls{snr} of the signal, defined by
\begin{equation}
    \label{eq:SNR}
    \mathrm{\gls{snr}} = \frac{2 P_{\mathrm{det}} \tracklength}{k_B T} \, ,
\end{equation}
where $P_{\mathrm{det}}$ denotes the total detected signal power and $\tracklength$ denotes the duration of the electron track. 
The detection probability $P_\mathrm{D}$ of this detector is
\begin{equation}
    \label{eq:MF_performance}
    P_\mathrm{D}(P_{\mathrm{FP}}) = Q_{\chi'^2_2(\mathrm{\gls{snr}})}\left(-2 \log\left(1-(1-P_{\mathrm{FP}})^\frac{1}{N_{t}} \right) \right) \, ,
\end{equation}
where $P_{\mathrm{FP}}$ is the probability of a false positive detection, $Q_{\chi'^2_2(\mathrm{\gls{snr}})}$ is the survival function of the noncentral $\chi^2$ distribution with 2 degrees of freedom and \gls{snr} as the noncentrality parameter.
Equivalently, this can be expressed as $P_\mathrm{D} = Q_1(\sqrt{\mathrm{\gls{snr}} },\sqrt{\gamma})$ where $Q_m(\lambda, k)$ is the Marcum Q-function.
The decision threshold is selected based on the acceptable level of false positives at the signal detection stage. 
A higher decision threshold will result in fewer false positives at the cost of rejecting a larger proportion of real \gls{cres} signals. 
The best neutrino mass sensitivity is found by optimizing between the detection efficiency and probability of detecting a false event.
Using simulations one can directly study how changes in the decision threshold affects the sensitivity of the experiment. 

For our sensitivity studies we calculate \autoref{eq:SNR} by simulating a noiseless signal for an electron track as outlined in \autoref{sec:sims} and then calculating $P_{\mathrm{det}}$ as the root-mean-square (RMS) power of the signal summed over the whole array. 
With this simulated \gls{snr} we can calculate the upper bound on the detection probability for an electron and an acceptable false positive rate using \autoref{eq:MF_performance}.
The size of the matched filter template bank is a result of the number of parameters required to describe a \gls{cres} track and their spacing for statistically independent templates, which we further discuss in \autoref{sec:recon:estimation}. 
A high dimensional parameter space can easily result in a matched filter template bank that is too large to be checked in real time due to practical limits on the available computational power. 
Optimizations of this general approach and alternatives for real time computation to cope with the significant raw data rates of large scale antenna arrays can be found in~\cite{andrewPhd, antenna_detection_paper}.

\subsection{Parameter Estimation with the Maximum Likelihood Method}
\label{sec:recon:estimation}
An electron track has eight free parameters: the electron's initial position and momentum $(x,y,z)$ and $(p_x, p_y, p_z)$, the start time $\starttime$, and the track duration $\tracklength$ which is the time it takes before the electron scatters. 
An equivalent but more convenient representation of these parameters is $\parameters=(\ekin, \pitch, \radial, \azimuth, \phaseax, \phasecyc, \starttime, \tracklength)$, where $\pitch$ is the pitch angle at the trap minimum, $\radial$ is the radial position, $\azimuth$ is the azimuthal angle of the position, $\phaseax$ is the initial phase of the axial motion and $\phasecyc$ is the initial phase of the cyclotron motion. 
Ideally we would use \gls{mle} to get an estimate $\parameters$ for the parameters together with the profile likelihood approach and Wilks' theorem \cite{wilks_theorem} to construct parameter uncertainties $\Delta \parameters_i$ for 68\% confidence.
Of the eight parameters the only parameter of interest is $\ekin$, whereas the remaining ones are nuisance parameters that only need to be accounted for if they are correlated with $\ekin$.

Under the assumption of \gls{awgn} the log-likelihood function describing a single track in $N_{\mathrm{samples}}$ of noisy multi-channel data $\bmat{x}$ is
\begin{equation}
    \label{eq:loglikelihoodfunction}
    \ell\left(\parameters | \bvec{x} \right) = -\frac{1}{\sigma^2} \left( \abs{\bvec{x}}^2 + \abs{\bvec{s}(\parameters)}^2 - 2 \mathrm{Re}\left( \bvec{x}^H \bvec{s}(\parameters) \right) \right) + \mathrm{const} \, ,
\end{equation}
where $\bvec{x}$ and $\bvec{s}$ now denote the vectorization of the multi-channel matrices $\bmat{x}$ and $\bmat{s}$ respectively and $\bvec{x}^H \bvec{s}$ denotes the inner product of $\bvec{x}$ and $\bvec{s}$. 
Note that it is required to have the same number of samples from the signal model as the input data, and that every data point is complex due to the use of IQ-sampling. 

In preceding experiments, the slow linear frequency chirp of the tracks naturally led to line fits in time-frequency spectrograms~\cite{phaseIIprl,phaseIIprc}. 
Although it is straightforward to adjust this model to the multi-channel antenna case using static phase-shifts, this approach does not capture the modulated nature of the signal induced by the trap-dependent motion of the electron and is therefore only valid in a small region of the parameter space by design. 
With the ambition of increasing the sensitivity of the experiment, the use of a signal model with modulation based on the phenomenology in \autoref{sec:pheno} is essential. 
Nevertheless, the described procedure also suffers from the typical complications for \gls{mle}, namely side minima and poor choice of initial conditions for the minimizer.
For the specific signals and parameter space, initial conditions could be provided from the signal detection stage with a template bank. 
Further research is required to arrive at a simple model with modulation which is suitable for \gls{mle} under the conditions of real data taking.

Despite this, we want to obtain the expected spread of parameter estimates under idealized conditions for our study of the detector performance parameters in \autoref{sec:large_detectors}. 
Using Monte Carlo (MC) data with fixed parameters $\parameters_{\mathrm{true}}$ it is possible to analyze parameter resolutions $\Delta \parameters(\parameters_{\mathrm{true}})$ by repeating the same MC experiment many times and looking at the distribution of maximum likelihood estimates $\uvec{\theta}$. 
For these experiments we have $\bmat{x} = \bmat{s}(\parameters_{\mathrm{true}}) + \bmat{n}$, where $\bmat{s}(\parameters)$ in both the data and the likelihood function is provided by the CRESana simulation tool described in \autoref{sec:sims} and $n_{ij} \sim \mathcal{CN}(0, \sigma^2)$, a complex Gaussian.
To save computation time of sampling many experiments we follow the Asimov data set approach discussed in~\cite{cowan_asymptotic_2011} to estimate $\Delta \parameters$ based on the expectation value of $\bmat{x}$ for the given $\parameters_{\mathrm{true}}$. 
The expectation value of the signal is the signal itself, while the expectation value of the noise is zero. Thus we simulate signals without noise while still taking the noise scale into account in the likelihood function via $\sigma$.
As a consequence, we can neglect the issue of finding the minimum numerically for the goal of obtaining parameter resolutions. Instead we estimate the Gaussian covariance matrix from the likelihood landscape in the vicinity of $\parameters_{\mathrm{true}}$. 
With this approach we learn the true physical limits from the information that is available in the recorded signal, but not how difficult it may be to recover that information from the signal in practice.

\begin{figure}
    \centering
    \includegraphics[width=0.5\textwidth]{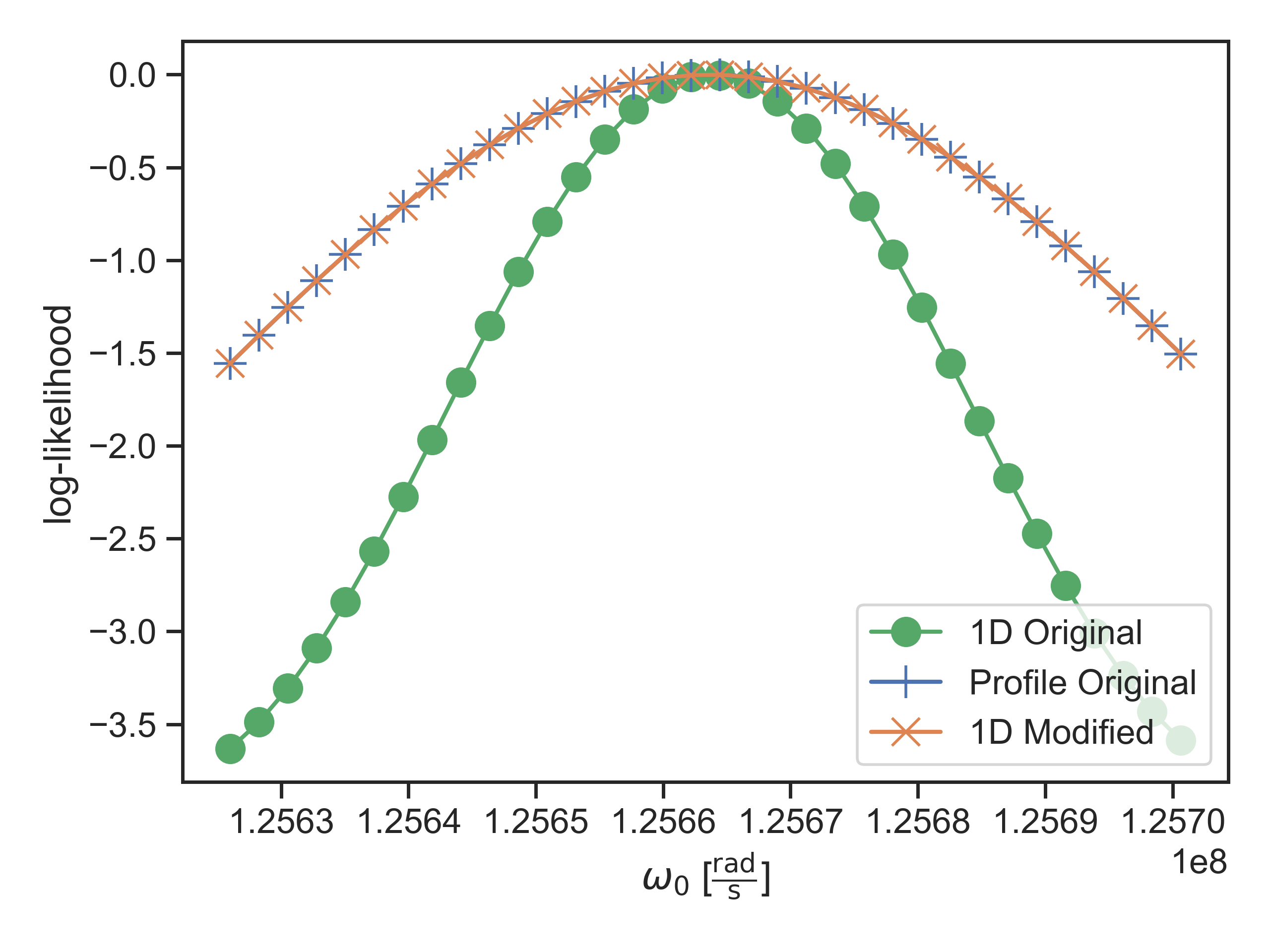}
    \caption{Comparison of original likelihood function (\autoref{eq:loglikelihoodfunction}), its profile likelihood, and the improved performance of the modified function (\autoref{eq:loglikelihood_final}) along the $\omega_0$ axis of a chirp.}
    \label{fig:invariant_cyclotron_phase_LLH}
\end{figure}

To reduce the computational cost of the analysis of the likelihood landscape we restrict the 8-dimensional parameter space to the most relevant parameters that exhibit strong correlations with each other.
First, $\starttime$ is correlated with the initial cyclotron frequency due to the frequency chirp. The degree of this correlation depends on the chirp rate~\cite{nickPhd}, which is defined in \autoref{eq:chirp_rate}.
This yields $\frac{\slope}{2 \pi} \sim \qty{370}{\mega \hertz \per \second}$ for the \qty{1}{\tesla} setup, and $\frac{\slope}{2 \pi} \sim \qty{0.05}{\mega \hertz \per \second}$ for the \qty{0.05}{\tesla} setup.
Considering these chirp rates, $\starttime$ must be included in the likelihood function for the analysis of the \qty{1}{\tesla} setup but can be neglected for the \qty{0.05}{\tesla} case. 
Next, we consider $\phasecyc$, which introduces a constant phase shift of the recorded signal, i.e. $\bmat{s}(\parameters) = \euler^{\iu \phasecyc}\bmat{s}(\parameters_{\phasecyc=0})$. Again, this results in a strong correlation with the initial signal frequency \cite{nickPhd}. 
To deal with this correlation, we construct a modified likelihood function. 
By modifying \autoref{eq:loglikelihoodfunction} to
\begin{equation}
    \label{eq:loglikelihood_final}
    \ell_{\mathrm{mod}}\left(\parameters | \bvec{x} \right) = -\frac{1}{\sigma^2} \left( \abs{\bvec{x}}^2 + \abs{\bvec{s}(\parameters)}^2 - 2 \abs{\bvec{x}^H \bvec{s}(\parameters)} \right) + \mathrm{const} \, ,
\end{equation}
the likelihood function becomes invariant under $\phasecyc$ with $\ell_{\mathrm{mod}}\left(\parameters | \bvec{x} \right) = \max_{\phasecyc} \left( \ell\left(\parameters | \bvec{x} \right) \right)$. Therefore, \autoref{eq:loglikelihood_final} is a profile likelihood of the remaining parameters, as demonstrated in \autoref{fig:invariant_cyclotron_phase_LLH}. Hence, using \autoref{eq:loglikelihood_final} correlations of $\phasecyc$ are eliminated.
It can be shown that the parameters $\tracklength$, and $\azimuth$ are not strongly correlated with the remaining parameters and can be neglected~\cite{florianPhd}.
Finally, from the remaining parameters, we chose to only consider ($\ekin, \pitch, \radial$) since these are strongly correlated with each other in many setups, while we neglect $\phaseax$ as we only expect a weak impact from potential correlations with the aforementioned parameters.

Minimization of the likelihood function with respect to these three parameters and determining the width of the profile likelihood yields a measure of the achievable event-wise energy resolution. 
The performance of this triggering and reconstruction on \gls{cres} events in an antenna array is discussed in the following section.

\section{Antenna Array CRES Detector Performance Parameters}
\label{sec:large_detectors}\label{sec:performance}

The sensitivity of a physics analysis is determined by the performance parameters of the detector. 
Although the goals of various physics analyses may differ, they are all influenced by the same performance parameters, which can be grouped into efficiency, resolution, and background. 
While these performance parameters are properties of the full ensemble of events, they come from the reconstruction parameters of individual events, which in turn depend on the trigger and reconstruction algorithms. 
The matched filter triggering and likelihood reconstruction approach developed in \autoref{sec:recon} is used here to evaluate the event-by-event parameters of \gls{snr} and event-wise energy resolution (see \autoref{sec:performance:individual}), as well as the ensemble parameters of background rate, efficiency, and ensemble energy resolution (see \autoref{sec:performance:ensemble}).
The performance parameters we discuss here are valid for low magnetic fields, since this is the target regime for future \gls{cres} experiments, as addressed in \autoref{sec:design}.

\subsection{Individual Event Performance Parameters}
\label{sec:performance:individual}

\subsubsection{Signal to Noise Ratio}

\begin{figure}[tb]
    \centering
    \includegraphics[width=0.5\textwidth]{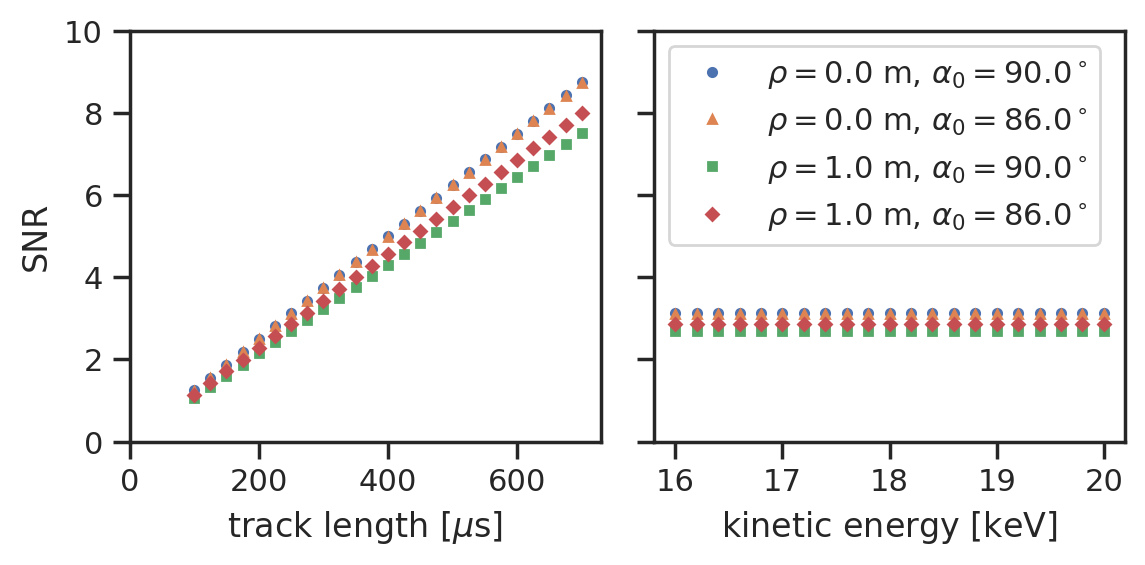}
    \caption{\gls{snr} as a function of track length (left) and electron kinetic energy (right) for different combinations of radial position and pitch angle. For the left plot a kinetic energy of \SI{18.6}{\kilo \electronvolt} and for the right plot a mean track length of \SI{250}{\micro\second} are used.}
    \label{fig:snr_dependence}
\end{figure}

The definition of \gls{snr} is given in \autoref{eq:SNR}, which shows that the \gls{snr} should scale linearly with track length. 
Since the \gls{snr} is also proportional to the detected power $P_\mathrm{det}$, it depends on the kinetic energy. 
The dependence on kinetic energy comes from the $\beta^2/(1-\beta^2)$ in \autoref{eq:larmor}, which is a very slowly-varying function.
Even within a range of \SI{1}{\kilo\electronvolt} around the tritium endpoint, the change in \gls{snr}  is $<5\%$. 
In addition, the transfer function in \autoref{fig:five_tf} is nearly constant, such that no additional energy dependence is introduced by the antenna response. 
This expectation is verified by full event simulation as shown in \autoref{fig:snr_dependence}. 
The linear \gls{snr} scaling with track length is confirmed and a very small dependence on the kinetic energy itself is observed. 
Therefore, the \gls{snr} needs to be explicitly computed only as a function of radius and pitch angle for one reference track length, $\tracklength_\mathrm{ref}$, after which the following equation can be applied:
\begin{equation}
    \mathrm{\gls{snr}}  (\ekin, \radial, \pitch, \tracklength) = 
    \mathrm{\gls{snr}}  (\radial, \pitch, \tracklength_\mathrm{ref}) \cdot \tracklength/\tracklength_\mathrm{ref} \,,
\end{equation}
which reduces the computational cost significantly. 
An example of $\mathrm{\gls{snr}} (\radial,\pitch)$ scan is shown in the top panel of \autoref{fig:snr_dE_scan} for the configuration described in \autoref{sec:sensitivity:setup}.

\subsubsection{Event-wise Energy Resolution}

\begin{figure}
    \centering
        \includegraphics[width=\linewidth]{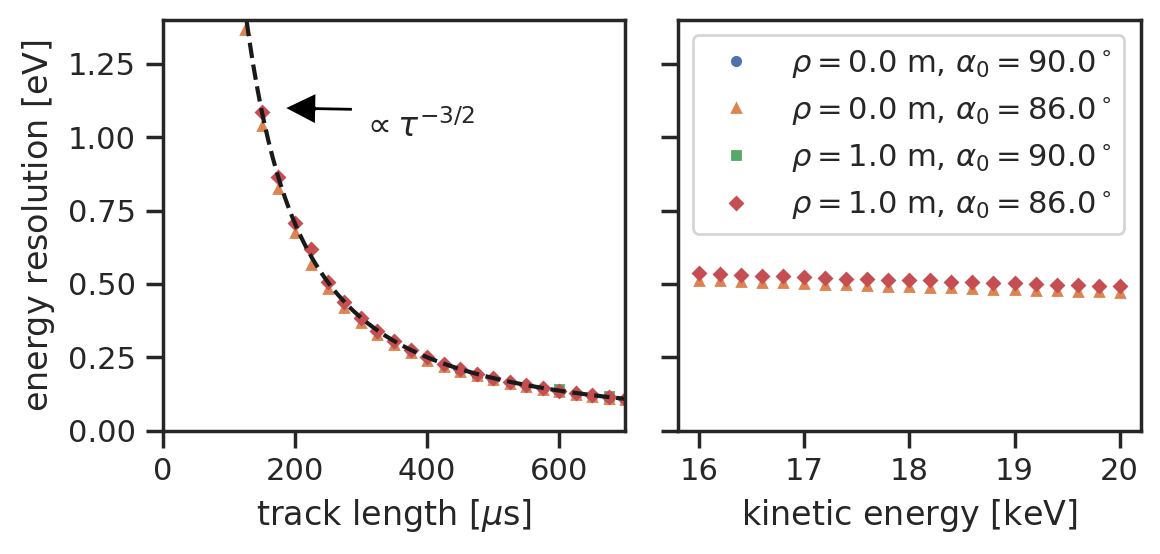}
        \caption{Energy resolution as function of track length (left) and kinetic energy (right) for multiple radial positions and pitch angles. For the left plot a kinetic energy of \SI{18.6}{\kilo \electronvolt} and for the right plot a mean track length of \SI{250}{\micro\second} are used.}
    \label{fig:energy_resolution_vs_track_length}
\end{figure}

In a generic chirp model, as discussed in~\cite{nickPhd}, the \gls{crlb} can be calculated for the variance of the initial frequency, which is~\footnote{Note that in the given reference constant in the equation of the \gls{crlb} is wrong and has to be 192.} 
\begin{equation}
    \mathrm{Var}(\hat{\omega}_0) \geq \slope^2 \mathrm{Var}(\hat{\starttime}) + \frac{192}{\mathrm{\gls{snr}}\cdot \tracklength^2}
\end{equation}
where $\slope$ is slope from \autoref{eq:chirp_rate}.
For low fields where $\slope$ is sufficiently small, the correlation on the start time variance can be neglected. 
It follows that the start frequency resolution is proportional to $\propto \tracklength^{-3}$. 
The event-wise energy resolution can be calculated from the frequency variance by error propagation:
\begin{equation}
    \Delta \ekin = \frac{\sqrt{\mathrm{Var}(\hat{\omega}_0)}(m_0 c^2 + \ekin)^2}{ec^2 B}
    \label{eq:energy_reso_CRLB}
\end{equation}
Since the energy resolution scales like the inverse of the frequency resolution, the energy resolution is expected to scale like $\Delta \ekin \propto \tracklength^{-3/2}$.
In the case of low fields, the start time contribution can be neglected. 

\begin{figure}
    \centering
    \includegraphics[width=\linewidth]{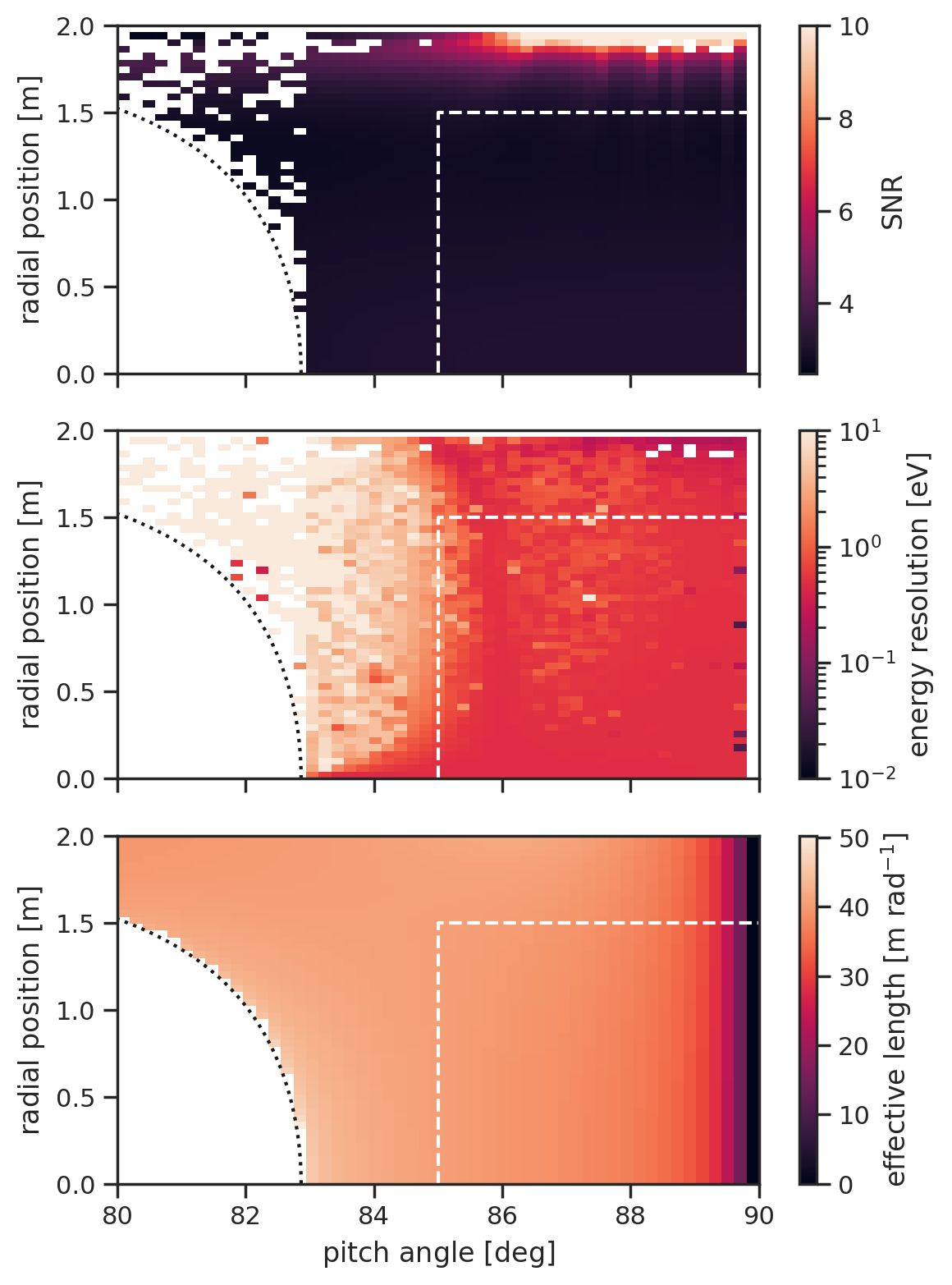}
    \caption{Signal to noise (top), event-wise energy resolution (middle) and effective length  (bottom) as function of pitch angle and radial position for the setup described in \autoref{sec:sensitivity:setup}. The signal to noise and energy resolution are calculated for  $\tracklength=\SI{250}{\micro\second}$ and $\ekin=\SI{18.6}{\kilo \electronvolt}$. Effective length is defined in \autoref{eq:effective_volume}. The black dotted line indicates the boundary between trapped and untrapped electrons. The white dashed line indicates the analysis cuts of pitch angle $>\SI{85}{\degree}$ and radius $<\SI{1.5}{\meter}$. Numerical instabilities occur at large radii and pitch angles close to the boundary of the untrapped region.}
    \label{fig:snr_dE_scan}
    \label{fig:effective_length}
\end{figure}

It is also shown in \autoref{eq:energy_reso_CRLB} that for energies $\ekin \ll m_0 c^2$, the dependence on the kinetic energy itself is very weak. 
While the above chirp model can only be strictly true for an electron with a pitch angle of \SI{90}{\degree}, the expectation is verified by full event simulation for a set of pitch angles and radii, as shown in \autoref{fig:energy_resolution_vs_track_length}.
In general the approximation
\begin{equation}
    \Delta E_\mathrm{reco} (\ekin, \radial, \pitch, \tracklength) \approx 
    \Delta E_\mathrm{reco} (\radial, \pitch, \tracklength_\mathrm{ref}) \cdot (\tracklength/\tracklength_\mathrm{ref})^{-3/2}
    \label{eq:energy_resolution_individual}
\end{equation}
also holds for any radii and pitch angle. 
An example of a $\Delta E_\mathrm{reco}(\radial,\pitch)$ scan is shown in the middle panel of \autoref{fig:snr_dE_scan} for the configuration described in \autoref{sec:sensitivity:setup}.

\subsection{Ensemble Performance Parameters}
\label{sec:performance:ensemble}

While the $\mathrm{\gls{snr}} (\radial,\pitch)$ and $\Delta E(\radial, \pitch)$ are individual event properties and depend on the radial position and pitch angle of the electron, the main interest for a physics analysis are the properties of the full ensemble.

\subsubsection{Background Rate}

The background rate is the rate of background events in a specific region of the analysis spectrum, the region of interest $\Delta E_\mathrm{ROI}$. 
For the matched filter trigger approach (see \autoref{sec:reco:MF}), the noise distribution from pure white noise is given by a $\chi^2$-distribution with two degrees of freedom.  
The false alarm rate $FAR$ can be defined as
\begin{equation}
    FAR = \frac{P_\mathrm{FP}(\gamma)}{\tau'}\,,
    \label{eq:FAR}
\end{equation}
where $\tau'$ is the length of the tested matched filter template, $P_\mathrm{FP}$ is the probability of a false positive, and $\gamma$ is the decision threshold on the matched filter score as defined in \autoref{eq:detection_threshold}. 
\autoref{eq:FAR} is valid only when a single matched filter template is tested. 
If $n_t$ independent matched filter templates are tested, the probability for a false positive is $P_\mathrm{FP}' = 1 - (1-P_\mathrm{FP,1})^{n_t} \approx n_t P_\mathrm{FP,1}$, where the binomial approximation is valid for  $n_t P_\mathrm{FP} \ll 1$.
The region of interest in energy defines the range of templates, thus the background rate can be defined as
\begin{equation}
    b \approx 
    \frac{n_t P_\mathrm{FP,1}(\gamma)}{\tau' \Delta E_\mathrm{ROI}} 
    \approx \mathrm{const}_\mathrm{bgd}\cdot P_\mathrm{FP,1}(\gamma) \,.
    \label{eq:background_rate_simple}
\end{equation}
The number of independent templates differs from the number of tested templates, since similar templates have correlated matched filter scores in a very fine template bank. 
The number of effective independent templates has to be determined by Monte Carlo (MC) simulations and depends linearly on $\Delta E_\mathrm{ROI}$ and $\tau'$. 
Thus the background rate can be calculated from the false alarm rate of a single template and a constant factor $\mathrm{const}_\mathrm{bgd}$.
The background rate does not depend on any signal parameters. 
\subsubsection{Effective Volume}

The effective volume $V_\mathrm{eff}$ is the detection efficiency integrated over the full volume weighted by the probability densities of the electron signal parameters 
\begin{align} V_\mathrm{eff} = \iiint & \epsilon_\mathrm{trap}(\radial,z,\pitchb) \cdot \epsilon_\mathrm{trig}(\radial,\pitch,\tracklength|\gamma) \nonumber \\ & \cdot P(\tracklength) \cdot P(\pitch)\,\mathrm{d}\pitch\,\mathrm{d}\tracklength\,\mathrm{d}V\, 
\label{eq:effective_volume_definition}
\end{align}
which is equivalent to the averaged detection efficiency multiplied by the volume $V_\mathrm{eff} = \langle \epsilon \rangle \cdot V$. 
The total detection efficiency can be decomposed into the efficiency of trapping an electron and the efficiency of triggering on the received signal.
The trapping efficiency $\epsilon_\mathrm{trap}(\radial,z,\pitchb)$ is either $1$ if \autoref{eq:trapped} is fulfilled, or zero if not. 
It depends on the decay position $(\radial,z)$ and the pitch angle at the decay position $\pitchb$, which can be expressed as pitch angle at the center $\pitch$ using \autoref{eq:adiabatic_invariance}.
The trigger efficiency is defined in \autoref{eq:MF_performance} and depends on the decision threshold $\gamma$, which can be determined by the required background rate.
The detection efficiency is assumed to be uneffected by start time, cyclotron phase and kinetic energy.

In Project~8 - Phase~II~\cite{phaseIIprl}, the track length distribution followed an exponential distribution. An exponential distribution is expected for random scattering with other gas atoms, leading to escape from the trap or changed electron properties.
The probability density distribution can be written as $P(\tracklength) = \frac{1}{\langle \tracklength \rangle}  \exp(-\tracklength/\langle \tracklength \rangle)$, where $\langle \tracklength \rangle$ is the mean track length of the population. 
The mean track length $\langle \tracklength \rangle(\density)$ depends on the density of the gas $\density$ in the detection volume. 

The direction of the initial momentum vector of the electron just after the decay is uniformly distributed on a sphere. 
Thus the probability density distribution of the pitch angle of the electron just after the decay is given by $P(\pitchb) = \sin(\pitchb)$. 
Using \autoref{eq:adiabatic_invariance}, the pitch angle at the decay position $\pitchb$ is related to the pitch angle at the trap center $\pitch$ and can be substituted in \autoref{eq:effective_volume_definition} such that the integral is performed over $\mathrm{d}\pitch$. 

Using the cylindrical symmetry of the detector, the volume integral can be written as $\mathrm{d}V = \radial\,\mathrm{d}\radial\,\mathrm{d}\azimuth\,\mathrm{d}z$ and the $\azimuth$ integral can be directly evaluated since no other dependence on $\azimuth$ exists. 
The effective volume integral can be rewritten as
\begin{widetext}
\begin{equation}
 V_\mathrm{eff}(\gamma, \langle \tracklength \rangle) = \int_0^{2\pi}\mathrm{d}\azimuth\, \int_0^\infty \int_0^{\pi/2} \underbrace{\left(\int_0^\infty \epsilon_\mathrm{trig}(\mathrm{\gls{snr}} (\radial,\pitch,\tracklength)|\gamma)P(\tracklength|\langle \tracklength\rangle)\,\mathrm{d}\tracklength\right)}_{\langle \epsilon_\mathrm{trig}\rangle_\tracklength(\radial,\pitch|\gamma,\langle \tracklength\rangle)} \cdot \underbrace{\left(\int_{-\infty}^{\infty} \epsilon_\mathrm{trap}(\radial,z,\pitchb) P(\pitchb)\left|\frac{\mathrm{d}\pitchb}{\mathrm{d}\pitch}\right|\,\mathrm{d}z \right)}_{l_\mathrm{eff}(\radial,\pitch)} \radial\,\mathrm{d}\radial\,\mathrm{d}\pitch \, 
    \label{eq:effective_volume}
\end{equation}
\end{widetext}
where $\langle \epsilon_\mathrm{trig}\rangle_\mathrm{\tracklength}$ is the track length averaged trigger probability and $l_\mathrm{eff}$ is the effective length. 
The trapping condition ensures that the integral boundary conditions are finite for the effective length.
The effective length is a property of the trap and can be calculated purely on the basis of the magnetic field. 
The effective length as a function of radial position and pitch angle for the setup described in \autoref{sec:sensitivity:setup} is shown in \autoref{fig:effective_length}.

\subsubsection{Ensemble Energy Resolution}

The ensemble energy resolution can be calculated by a weighted average, where the weighting factor is proportional to the event rate, which is proportional to the effective volume. Thus the ensemble weighted energy resolution can be defined as
\begin{widetext}
\begin{equation}
 \Delta E_\mathrm{ens}(\gamma, \langle \tracklength \rangle) = \frac{2\pi}{V_\mathrm{eff}}\int_0^\infty \int_0^{\pi/2} \Delta E_\mathrm{reco}(\radial,\pitch,\tracklength_\mathrm{ref}) \cdot   l_\mathrm{eff} (\radial,\pitch) \cdot \label{eq:energy_resolution_ensemble} \left(\int_0^\infty \frac{\tracklength^{-3/2}}{\tracklength_\mathrm{ref}^{-3/2}}\epsilon_\mathrm{trig}(\mathrm{\gls{snr}} (\radial,\pitch,\tracklength)|\gamma)P(\tracklength|\langle \tracklength\rangle)\,\mathrm{d}\tracklength\right)\cdot \radial\,\mathrm{d}\radial\,\mathrm{d}\pitch 
\end{equation}
\end{widetext}
The $V_\mathrm{eff}$ in the denominator ensures that the weighting is properly normalized. 
The track length dependence of the event-wise energy resolution modifies the $\tracklength$ integral compared to the mean trigger efficiency integral in \autoref{eq:effective_volume}. 

The choice of a background rate fixes the decision threshold $\gamma$ and thus determines the corresponding effective volume and energy resolution at the same time.  
In addition, analysis cuts on reconstructed pitch angle and radial position can be introduced, which restrict the integral boundaries in the effective volume and energy resolution calculation. 
As shown in \autoref{fig:effective_length}, the event-wise energy resolution gets significantly worse for pitch angles below \SI{85}{\degree}. 
Therefore, an analysis cut of \SI{85}{\degree} on the pitch angle is introduced. 
Moreover, a radial position $<\SI{1.5}{\meter}$ is required, to avoid the reactive near-field of antennas and allow for gas containment vessels. 
The reactive near-field antenna response and signal simulation is not accurately modeled, but the region used for analysis (\autoref{fig:effective_length}) is not affected by these simplifications.

\begin{figure}[b]
    \centering
    \includegraphics[width=\linewidth]{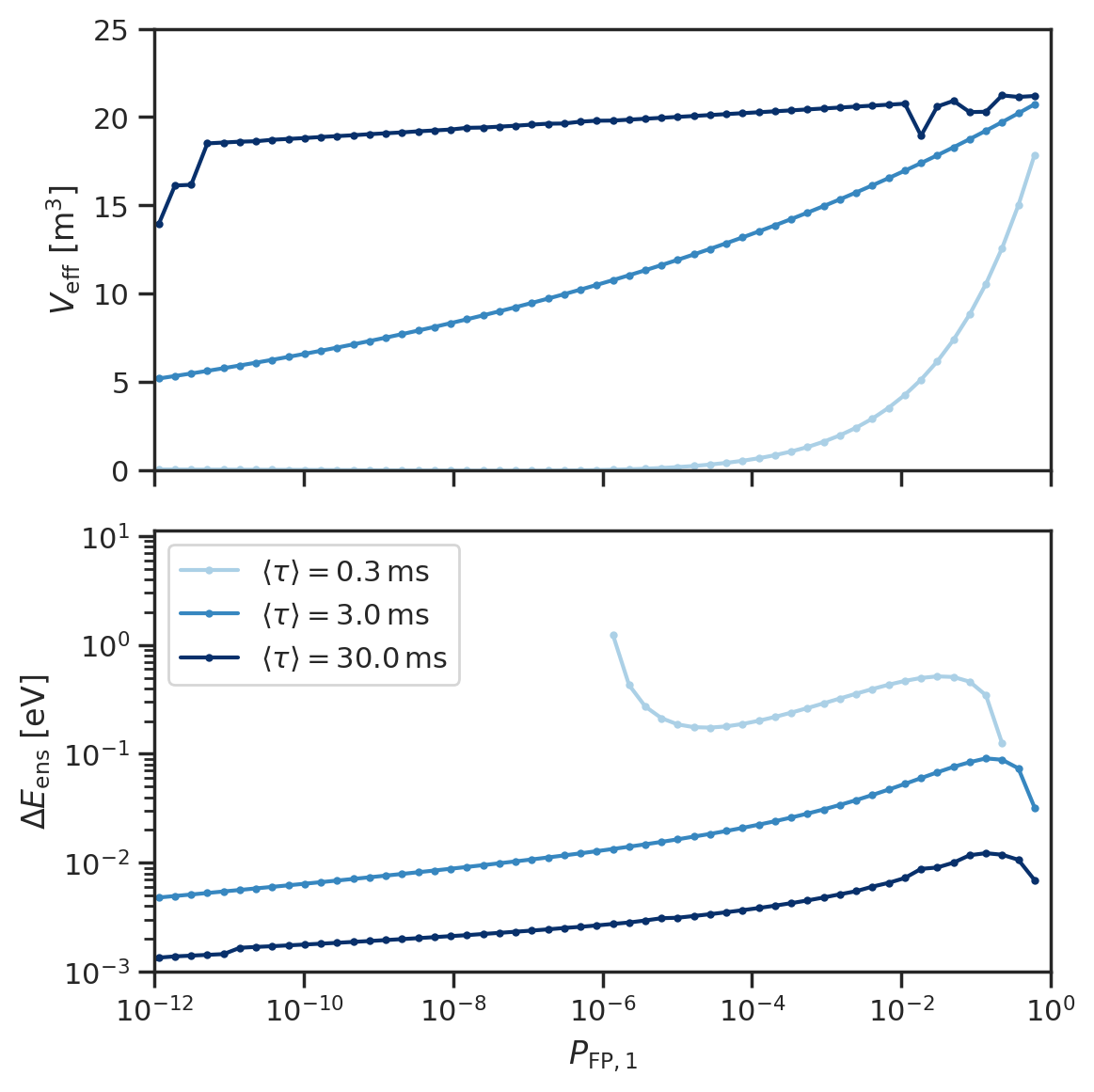}
    \caption{Effective volume (top) and ensemble energy resolution (bottom) as function of false alarm rate for different mean track lengths.  Only events with reconstructed pitch angle $\pitch>\SI{85}{\degree}$ and radius $\radial<\SI{1.5}{\meter}$ are used. Simulation artifacts are visible for $\SI{30}{\milli\second}$ track length at large false alarm rates $>10^{-2}$. }
    \label{fig:ensemble_performance_parameters}
\end{figure}

The effective volume and energy resolution as function of background rate is shown in \autoref{fig:ensemble_performance_parameters} for different mean track lengths $\langle \tracklength \rangle$ and analysis cuts of pitch angle $>\SI{85}{\degree}$ and radius $<\SI{1.5}{\meter}$. 
For long track lengths, the effective volume becomes independent of the chosen false positive rate and thus the decision threshold. 
In this case the detection efficiency approaches 100\% and the effective volume is given by the product of physical volume and trapping efficiency.

\section{Estimating Neutrino Mass Sensitivity}
\label{sec:sensitivity}

\subsection{Sensitivity Estimation Method}
\label{sec:sensitivity:method}

The Project 8 collaboration has developed an analytic model for estimating the sensitivity to neutrino mass based on signal rate, background rate and energy resolution. 
A simple cut and count model is used, where the number of events within an energy window $\Delta E$ below the spectrum endpoint is measured and a limit on the neutrino mass is calculated using Poisson statistics. 
In this estimate it is assumed that the endpoint is known exactly.
In general, the endpoint can be measured by analyzing a larger region than $\Delta E$.
The width of this ``analysis window'' is optimized with respect to the background rate $b$ and energy resolution $\Delta E_{\rm res}$, which is a measure of the full width at half max of the detector response function. 
In this estimate, only the ensemble energy resolution from the event reconstruction as presented in \autoref{sec:recon} is considered. 
A complete description of the model is found in \cite{Formaggio:2021nfz}. 
The analytic model was verified in the analysis of Phase II data~\cite{phaseIIprl,phaseIIprc} and a Monte Carlo study~\cite{bayesian_prc}.

The signal rate in the detection volume in units of \SI{}{\per\second\per\cubic\electronvolt} is given by 
\begin{equation}
    r = \frac{\mathrm{d}^2 N}{\mathrm{d}t\,\mathrm{d}(\varepsilon^3)} = \frac{\density V_\mathrm{eff} \eta_{\SI{1}{\electronvolt}}}{\tau_\mathrm{tritium}}\,
    \label{eq:signal_rate}
\end{equation}
where $\varepsilon$ is an energy interval contiguous with the endpoint, $\density$ is the number density of the tritium gas, $V_\mathrm{eff}$ is the effective volume as described in \autoref{eq:effective_volume}, $\eta_{\SI{1}{\electronvolt}}$ is the branching fraction of decays in the last \SI{1}{\electronvolt} of the tritium beta spectrum in units of \SI{}{\per\cubic\electronvolt} and $\tau_\mathrm{tritium}$ is the mean-life of tritium.
The optimized energy window for the counting experiment is 
\begin{equation}
    \Delta E \simeq \sqrt{\frac{b}{r}+(\Delta E_{\rm res})^2+(\Delta E_{\rm other})^2}\,,
    \label{eq:analysiswindow}
\end{equation}
where $b$ is the background rate per \SI{}{\electronvolt} and $\Delta E_\mathrm{res}$ is the energy resolution as estimated in \autoref{sec:performance:ensemble}. 
The contributions from other effects to the energy resolution $\Delta E_{\rm other}$ are neglected here.
Assuming no signal, the 90\% confidence level upper limit on the neutrino mass can thus be estimated by
\begin{eqnarray}    
    m_\beta \leq  \sqrt{ 1.64 \cdot \frac{2}{3}\sqrt{\frac{\Delta E}{rt} + \frac{b}{r^2 t \Delta E}}},  \label{eq:sig} 
\end{eqnarray}
where $t$ is the amount of time the experiment is run, or livetime.
\autoref{eq:sig} is the  statistical contribution to the neutrino mass limit while systematic effects need to be considered separately and are neglected here.

This framework allows us to estimate the sensitivity to the neutrino mass for a given setup. 
We obtain event-wise parameters \gls{snr} and $\Delta E_{\mathrm{reco}}$  from event simulations, yielding the ensemble parameters $\Delta E_\mathrm{res}$, $b$, and $V_\mathrm{eff}$ as described in \autoref{sec:performance:ensemble}. 
In addition, the sensitivity depends on
$n_\mathrm{t}$, $t$, and $\density$ which determines $\langle \tracklength\rangle$. 

The scaling with livetime is clear from \autoref{eq:sig}; thus, only scenarios with $t=\SI{1}{year}$ are considered. 
As discussed in \autoref{sec:performance:ensemble}, the number of independent templates has to be determined by Monte Carlo simulations. 
Here, it is assumed that $\mathrm{const}_\mathrm{bgd}$ in \autoref{eq:background_rate_simple} is \SI{1}{\per\second\per\electronvolt}. 
The parameters $\density$ and $b$, which in turn fix the decision threshold $\gamma$, can be optimized to achieve the best neutrino mass sensitivity. 

During the optimization procedure of $\density$ and $b$ we require that the number of events within the last \qty{1}{\electronvolt} of the spectrum exceeds 1000 signal events. 
This additional requirement avoids cases of erroneously good sensitivity estimates due to a region of low statistics.

\subsection{Neutrino Mass Sensitivity for an Idealized Antenna Array}
\label{sec:sensitivity:setup}

In this section the neutrino mass sensitivity is calculated for an example detector setup using the method described in \autoref{sec:sensitivity:method}. 
Note that the detector design outlined here is not optimized in any way. 
However, the procedure put forth in this section can be used to evaluate several detector designs and optimize with respect to neutrino mass sensitivity.

The background field used for this setup is \qty{50}{\milli\tesla} pointed along the z-axis. 
At this field an electron at the tritium spectrum endpoint of \qty{18.6}{\kilo \electronvolt} has a cyclotron frequency of $f_\mathrm{cycl}\sim\qty{1.3}{\giga \hertz}$. 
The magnetic field of the electron trap is generated by two circular current loops with a radius of $R_\mathrm{coil}=\qty{2}{\meter}$ located at $z = \pm\qty{20}{\meter}$ with a current of $2500$~amp-turns. 
The large aspect ratio ($L/D=10$) generates a magnetic bottle trap which has a very flat central section and two magnetic field walls at each end. 

The volume is surrounded by \SI{50000}{} dipole antennas. 
The dipole antenna has a peak gain of \SI{3}{dB}, constant gain in the H-plane and a directive gain in the E-plane of  $[\cos( (\pi/2)\cdot\sin\xi ) / \cos\xi ]^2$, where $\xi$ is \SI{0}{\degree} in the direction of peak gain.
Antennas are oriented to look radially inward and the E-plane of the antenna is parallel to the $x$-$y$-plane. 
The antennas are arranged on 400 rings with 125 antennas each, on the lateral surface of the cylindrical volume. A sampling rate of \SI{200}{\mega\hertz} is used to read out each antenna with a thermal noise temperature of \SI{5}{\kelvin}.

\begin{figure}
    \centering
    \includegraphics[width=\linewidth]{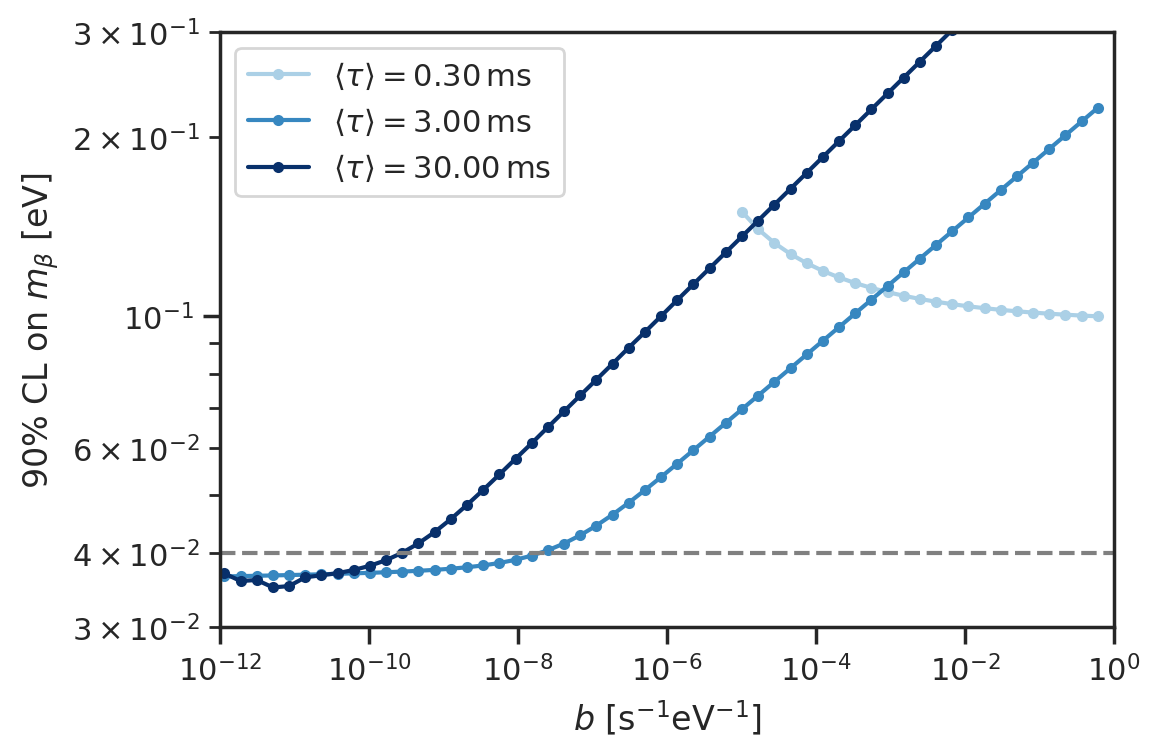}
    \caption{Sensitivity to neutrino mass as a function of background rate for the setup described in \autoref{sec:sensitivity:setup}. Curves are produced by different analysis cuts, which determine the ensemble energy resolution, effective volume and background rate.}
    \label{fig:sensitivity_default}
\end{figure}

The event-wise \gls{snr} and energy resolution are shown in \autoref{fig:effective_length}. 
The effective volume and energy resolution of this setup are shown in \autoref{fig:ensemble_performance_parameters}. 
\autoref{fig:sensitivity_default} shows the neutrino mass sensitivity for this setup as a function of the background rate which are reached for different analysis thresholds. 
It can be seen that a mean track length of \SI{3}{\milli\second}, which corresponds to an atom density of \SI{3.8E16}{\per\cubic\meter}, is needed to achieve a neutrino mass sensitivity of \SI{40}{\milli\electronvolt}.
Longer track lengths reach similar neutrino mass sensitivities but require lower background rates. At low background rates (high analysis cuts), the loss in effective volume is compensated by improvements in energy resolution leading to a nearly plateauing neutrino mass sensitivity. 

\subsection{Impact of Idealizations}

The sensitivity calculated in \autoref{sec:sensitivity:setup} is based on a series of idealizations and thus is a best case scenario. 
In this section we discuss the main idealizations and estimate their impact on the neutrino mass sensitivity. 

\subsubsection{Idealized likelihood reconstruction}

\begin{figure}
    \centering
    \includegraphics[width=\linewidth]{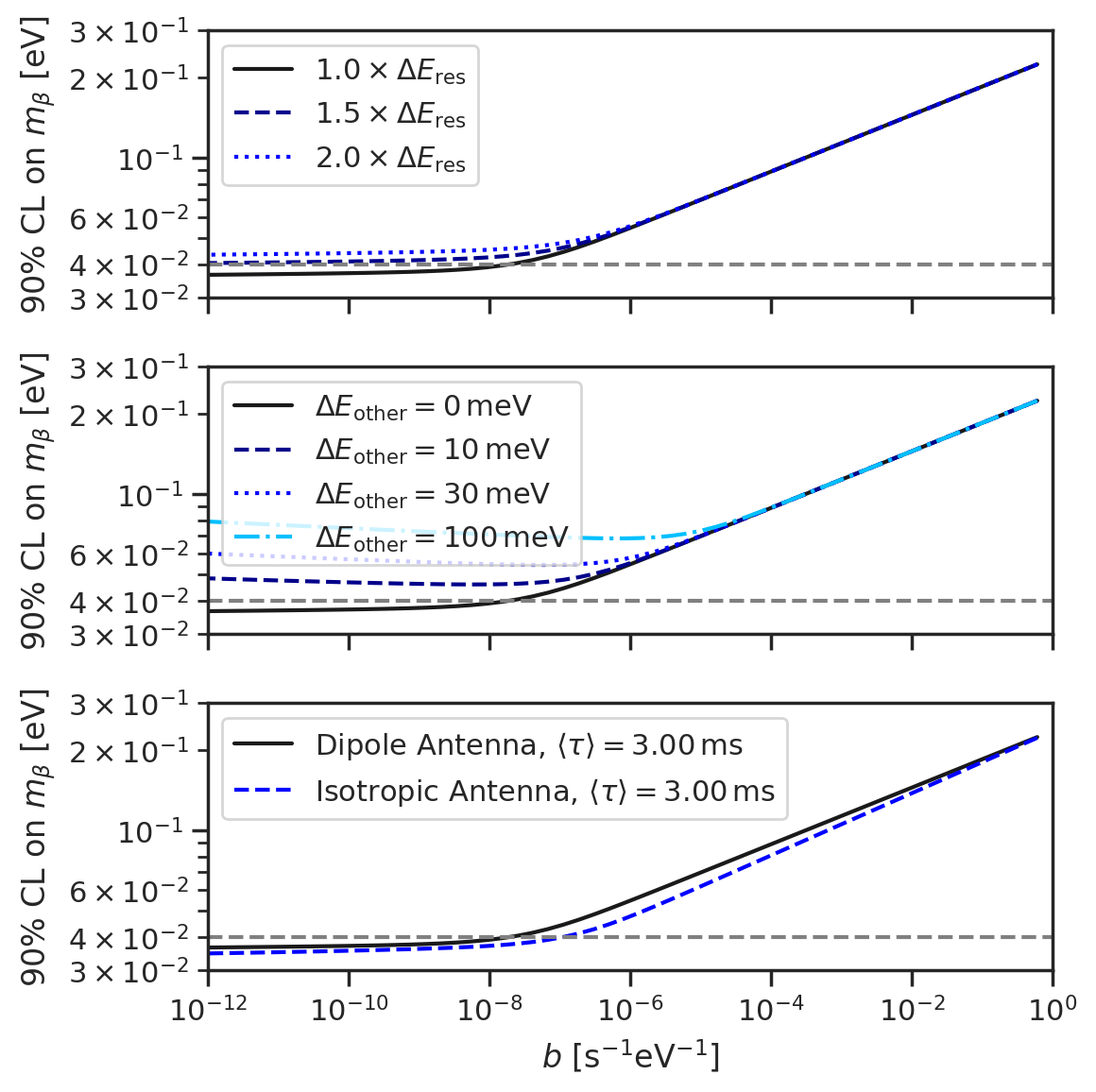}
    \caption{Sensitivity as a function of background rate for different energy resolutions (top), 
    additional contributions to the energy resolution (bottom middle) and  antennas (bottom). Mean track length of \SI{3}{\milli\second} is assumed. The default sensitivity from \autoref{fig:sensitivity_default} is always shown in black here. 
    }
    \label{fig:sensitivity_idealizations}
\end{figure}

The likelihood reconstruction described in \autoref{sec:recon:estimation} is used to estimate the event-wise energy resolution. 
In the likelihood reconstruction, the uncertainties are estimated from the likelihood profile around the true minimum. 
However, a real reconstruction algorithm does not know the true minimum and thus will result in larger event-wise energy resolutions. 
The impact on the neutrino mass sensitivity can be estimated by scaling up the event-wise energy resolution. 
In \autoref{fig:sensitivity_idealizations} (top) the neutrino mass sensitivity is calculated for energy resolutions that are a factor of 1.5 and 2.0 worse then the idealized energy resolution. 
For the background rate in which the default energy resolution reaches \SI{40}{\milli\electronvolt} sensitivity, a 50\% worse energy resolution yields a sensitivity of \SI{43}{\milli\electronvolt}. While this increase seems to be modest, a much stronger background restriction is needed to reach \SI{40}{\milli\electronvolt}, since the curve flattens significantly. 

\subsubsection{Background rate}

In \autoref{sec:performance:ensemble} the dependence of the background rate on the number of independent templates was discussed, and a constant of \qty{1}{\per\electronvolt\per\second} was used for the neutrino mass sensitivity estimates. 
An increase in $\mathrm{const}_\mathrm{bgd}$ can be compensated by requiring a higher analysis threshold. 
Since the neutrino mass sensitivity as function of background rate flattens for sufficiently small background rates, similar sensitivities can be achieved.
This implies that the number of independent templates per energy window and time do not have a strong impact on the neutrino mass sensitivity.

\subsubsection{Additional Contributions to Energy Resolution}

Additional contributions to the energy resolution have numerous origins and a detailed discussion of them is outside the scope of this paper. 
However, the impact on neutrino mass sensitivity can be tested by adding a non-zero $\Delta E_\mathrm{other}$ in \autoref{eq:analysiswindow}. 
\autoref{fig:sensitivity_idealizations} (middle) shows the neutrino mass sensitivity for different amounts of additional energy broadening.
It can be seen that additional energy broadening has to be limited to $\lesssim \SI{10}{\milli\electronvolt}$ to still reach the \SI{40}{\milli\electronvolt} neutrino mass sensitivity, which imposes stringent constraints on a real experiment.

\subsubsection{Dipole Antennas}

The sensitivity was calculated using  idealized dipole antennas. 
While  dipole antennas are some of the simplest antennas, the antenna gain and directivity may influence the performance and the neutrino mass sensitivity. 
To estimate the effect of antenna choice, the idealized dipole antenna is compared to an isotropic gain antenna, which is an even further idealized antenna but serves as a reference for the impact of antenna choice. 
The event-wise \gls{snr} and energy resolution were determined from full event-wise simulations. 
A comparison between neutrino mass sensitivities with dipole and isotropic gain antennas is shown in \autoref{fig:sensitivity_idealizations} (bottom). 
The impact on neutrino mass sensitivity is small. 

\subsubsection{Realization of Setup}

While this section demonstrates how an antenna array can be used to reach a neutrino mass sensitivity of \SI{40}{\milli\electronvolt}, surpassing the range allowed by the inverted mass ordering, we have not included the engineering aspects of the experiment. 
A physical antenna array with $\sim\SI{50000}{}$ antennas seems exceedingly difficult to realize at time of writing. 
With these antennas, the data rates reach $\mathcal{O}(\SI{20}{\tera\byte\per\second})$ at a sampling rate of \SI{200}{\mega\hertz} and 8-bit sampling depth.
The physical detection volume reaches \SI{502}{\cubic\meter} and is 36\% of the size of the KATRIN spectrometer.

\section{Conclusions and Outlook}
\label{sec:conclusion}
Beta spectrum measurements offer a direct kinematic approach to measure neutrino mass through its impact on the shape of the beta spectrum near the endpoint.
However, such a measurement requires a high event rate with minimal background while maintaining high resolution.
While Phase~II of Project~8 was a zero background experiment and showed that a good resolution is possible, it was done at the cost of low efficiency and subsequently low event rate.
In this article, we demonstrated that antenna arrays provide a potential path forward for a large volume \gls{cres} experiment.

Using standard electromagnetic theory, we described  the kinematics and the radiation of magnetically trapped electrons as well as their implementation in simulations.
With the simulation as a guide, we designed and fabricated slotted waveguide antennas optimized for detecting cyclotron radiation.
By making use of these antennas arranged in an inward-facing cylindrical array, we performed room-temperature measurements using a synthetic radiating antenna as a source to benchmark our simulations and validate our reconstruction techniques.
The benchmarked simulations were then used to establish the antenna array performance metrics relevant to neutrino mass measurement.
These metrics were used to estimate the sensitivity to neutrino mass of a hypothetical antenna array.

This comprehensive study provides a reference for the design of antenna arrays for neutrino mass measurements and other CRES-based efforts \cite{Iwasaki:2024voi, PTOLEMY:2025unk, Amad:2024jod}.
While we performed extensive studies for \gls{cres} detection using antenna arrays, cavity resonators were ultimately selected as the detection method for Project~8 due mainly to the requirement to lower the frequency. 
Cavities also offer a practical way to lower the number of channels.
Although a thorough analysis was not conducted for this paper, passively-combined antenna arrays can significantly reduce the number of channels, making them an alternative to cavities.
If other technical challenges result in the infeasibility of resonant cavities for Project 8, returning to antenna arrays is a key alternative strategy.
Regardless of the future of Project 8, this study serves as a benchmark for future antenna array projects aimed at measuring neutrino mass and for \gls{cres} experiments in general.

\gls{cres} is a relatively new beta spectroscopy technique devised and developed for neutrino mass measurement. 
The inherently superior energy resolution and low backgrounds make \gls{cres} an attractive way to perform precision energy measurement of charged particles.
These qualities are being used for other spectroscopy measurements including precision $\beta$-decay measurements for searches for physics beyond the TeV scale~\cite{He6-CRES:2022lev} and x-ray spectroscopy for fundamental physics and applications~\cite{Kazkaz:2019umq}. 
The phenomenology and methodology developed in this article provide a comprehensive guide for antenna array-based \gls{cres} detectors, opening up \gls{cres} as a means of meeting spectroscopy demands far beyond neutrino mass measurement.

\subsection*{Acknowledgements}
This material is based upon work supported by the following sources: the U.S. Department of Energy (DOE) Office of Science, Office of Nuclear Physics, under Award No.~DE-SC0020433 to Case Western Reserve University (CWRU), under Award No.~DE-SC0011091 to the Massachusetts Institute of Technology (MIT), under Field Work Proposal Number 73006 at the Pacific Northwest National Laboratory (PNNL), a multiprogram national laboratory operated by Battelle for the DOE under Contract No.~DE-AC05-76RL01830, under Early Career Award No.~DE-SC0019088 to Pennsylvania State University, under Award No.~DE-SC0024434 to the University of Texas at Arlington, under Award No.~DE-FG02-97ER41020 to the University of Washington, and under Award No.~DE-SC0012654 to Yale University; the National Science Foundation under Grant No.~PHY-2209530 to Indiana University, and under Grant No.~PHY-2110569 to MIT; the Cluster of Excellence “Precision Physics, Fundamental Interactions, and Structure of Matter” (PRISMA+ EXC 2118/1) funded by the German Research Foundation (DFG) within the German Excellence Strategy (Project ID 39083149); the Karlsruhe Institute of Technology (KIT) Center Elementary Particle and Astroparticle Physics (KCETA); the Laboratory Directed Research and Development (LDRD) program at Lawrence Berkeley National Laboratory (LBNL), a national laboratory operated by The Regents of the University of California (UC) for the DOE under Federal Prime Agreement DE-AC02-05CH11231; LDRD 18-ERD-028 and 20-LW-056 at Lawrence Livermore National Laboratory (LLNL), prepared by LLNL under Contract DE-AC52-07NA27344, LLNL-JRNL-2004198; the LDRD Program at PNNL; University of Pittsburgh; and Yale University. This material is based on work supported by the National Science Foundation Graduate Research Fellowship Program under Grant No. DGE-2139841. A portion of the research was performed using Research Computing at PNNL.  We thank the Yale Center for Research Computing for guidance and use of the research computing infrastructure. The design and fabrication of antennas used in this research were supported by the Advanced Prototyping Center at Wright Laboratory at Yale University.	
\bibliography{refs}
\end{document}